\documentclass[fleqn,usenatbib]{mnras}

\usepackage[T1]{fontenc}

\DeclareRobustCommand{\VAN}[3]{#2}
\let\VANthebibliography\thebibliography
\def\thebibliography{\DeclareRobustCommand{\VAN}[3]{##3}\VANthebibliography}

\usepackage{graphicx}	
\usepackage{amsmath}	
\usepackage{amssymb}	
\usepackage{orcidlink}

\usepackage{newtxtext,newtxmath}

\newcommand{\OIII}{[\ion{O}{III}]}
\newcommand{\Ha}{H$\alpha$}
\newcommand{\Hb}{H$\beta$}
\newcommand{\Hg}{H$\gamma$}
\newcommand{\Hd}{H$\delta$}
\newcommand{\Lya}{Ly$\alpha$}
\newcommand{\He}{H$\epsilon$}
\newcommand{\NeIII}{[\ion{Ne}{III}]}
\newcommand{\OII}{[\ion{O}{II}]}

\title[Nebular-dominated galaxies at $1.5 < z < 8.5$]{Possible photometric signatures of nebular-dominated emission in $1.5 < z < 8.5$ JADES galaxies}

\author[J.\@A.\@A.\@ Trussler et al.]{James A.\@ A.\@ Trussler,$^{1}$\thanks{E-mail: james.trussler@cfa.harvard.edu}
Alex J.\@ Cameron,$^{2,3,4}$
Daniel J.\@ Eisenstein,$^{1}$
Harley Katz,$^{5,6}$
Nathan J.\@ Adams,$^{7}$
\newauthor
Duncan Austin,$^{7}$
Andrew J.\@ Bunker,$^{2}$
Stefano Carniani,$^{8}$
Christopher J.\@ Conselice,$^{7}$
Mirko Curti,$^{9}$
\newauthor 
Emma Curtis-Lake,$^{10}$
Kevin Hainline,$^{11}$
Thomas Harvey,$^{7}$
Benjamin D.\@ Johnson,$^{1}$
Qiong Li,$^{7}$
\newauthor 
Tobias J.\@ Looser,$^{1}$
Pierluigi Rinaldi,$^{12}$
Brant Robertson,$^{13}$
Fengwu Sun,$^{1}$
Sandro Tacchella,$^{14,15}$
\newauthor
Christina C.\@ Williams,$^{16}$
Christopher N.\@ A.\@ Willmer,$^{11}$
Chris Willott$^{17}$
and Zihao Wu$^{1}$
\\
$^{1}$Center for Astrophysics $|$ Harvard \& Smithsonian, 60 Garden St., Cambridge MA 02138 USA\\
$^{2}$Department of Physics, University of Oxford, Denys Wilkinson Building, Keble Road, Oxford OX1 3RH, UK\\
$^{3}$Cosmic Dawn Center (DAWN), Copenhagen, Denmark\\
$^{4}$Niels Bohr Institute, University of Copenhagen, Jagtvej 128, DK-2200, Copenhagen, Denmark\\
$^{5}$Department of Astronomy \& Astrophysics, University of Chicago, 5640 S Ellis Avenue, Chicago, IL 60637, USA\\
$^{6}$Kavli Institute for Cosmological Physics, University of Chicago, Chicago IL 60637, USA\\
$^{7}$Jodrell Bank Centre for Astrophysics, University of Manchester, Oxford Road, Manchester M13 9PL, UK\\
$^{8}$Scuola Normale Superiore, Piazza dei Cavalieri 7, I-56126 Pisa, Italy\\
$^{9}$INAF, Osservatorio di Astrofisica e Scienza dello Spazio, Via P. Gobetti 93/3, I-40129 Bologna, Italy\\
$^{10}$Centre for Astrophysics Research, Department of Physics, Astronomy and Mathematics, University of Hertfordshire, Hatfield AL10 9AB, UK\\
$^{11}$Steward Observatory, University of Arizona, 933 N.\@ Cherry Avenue, Tucson, AZ 85721, USA\\
$^{12}$Space Telescope Science Institute, 3700 San Martin Drive, Baltimore, Maryland 21218, USA\\
$^{13}$Department of Astronomy and Astrophysics University of California, Santa Cruz, 1156 High Street, Santa Cruz CA 96054, USA\\
$^{14}$Kavli Institute for Cosmology, University of Cambridge, Madingley Road, Cambridge, CB3 0HA, UK\\
$^{15}$Cavendish Laboratory, University of Cambridge, 19 JJ Thomson Avenue, Cambridge, CB3 0HE, UK\\
$^{16}$NSF National Optical-Infrared Astronomy Research Laboratory, 950 North Cherry Avenue, Tucson, AZ 85719, USA\\
$^{17}$NRC Herzberg, 5071 West Saanich Rd, Victoria, BC V9E 2E7, Canada
}

\date{Accepted XXX. Received YYY; in original form ZZZ}

\pubyear{2026}

\begin{document}
\label{firstpage}
\pagerange{\pageref{firstpage}--\pageref{lastpage}}
\maketitle

\begin{abstract}
The discovery of high-redshift galaxies exhibiting a steep spectral UV downturn potentially indicative of two-photon continuum emission marks a turning point in our search for signatures of star formation following a top-heavy IMF in the early Universe. We develop a photometric search method for identifying further nebular-dominated galaxy candidates, whose nebular continuum dominates over the starlight, due to the high ionising photon production efficiencies $\xi_\mathrm{ion}$ associated with massive star formation. We utilise the extensive medium-band imaging from JADES, which enables the identification of Balmer jumps across a wide range of redshifts ($1.5 < z < 8.5$), through the deficit in rest-frame optical continuum level. As Balmer jumps are a general recombination feature of young starbursts ($\lesssim 3$~Myr), we further demand a high observed $\log\, (\xi_\mathrm{ion, obs}/\mathrm{(Hz\ erg^{-1})}) > 25.60$ to power the strong nebular continuum, together with a relatively non-blue UV slope ($m_\mathrm{F115W}-m_\mathrm{F200W} > -0.4$ at $z=6$) indicating a lack of stellar continuum emission. Our nebular-dominated candidates, constituting ${\sim}11$~per~cent of galaxies at $z \sim 6$ (decreasing to ${\sim}2$~per cent at $z \sim 2$, not completeness-corrected) are faint in the rest-frame optical (median $M_\mathrm{opt} = -17.95$) with extreme line emission (median $\mathrm{EW}_\mathrm{H\alpha, rest} = 1567$~\AA, $\mathrm{EW}_\mathrm{[O\ III] + H\beta,rest} = 2292$~\AA). However, hot \ion{H}{II} region temperatures, collisionally-enhanced two-photon continuum emission, and strong UV lines are expected to accompany top-heavy star formation. Thus nebular-dominated galaxies do not necessarily exhibit the biggest Balmer jumps, nor the largest $\xi_\mathrm{ion, obs}$ or reddest UV slopes. Hence continuum spectroscopy is ultimately required to establish the presence of a two-photon downturn in our candidates, thus advancing our understanding of primordial star formation and AGN. 
\end{abstract}




\section{Introduction} \label{sec:intro}

\emph{JWST} \citep{Gardner2006, Gardner2023} seeks to uncover our cosmic origins, by witnessing the formation of the first stars to shine in the cosmos. Forming out of the primaeval clouds, this primordial star formation is expected to follow a more top-heavy initial mass function (IMF), owing to the lack of metal-line cooling 
\citep[e.g.\@][]{Bromm1999}. In the limit of zero metallicity, Population III galaxies likely exhibit unique spectra and photometry, due to their pristine, massive, hot stars \citep[e.g.\@][]{Bromm2001, Schaerer2002, Schaerer2003, Raiter2010, Zackrisson2011, Mas-Ribas2016, Nakajima2022, Katz2023, Nishigaki2023, Trussler2023, Fujimoto2025}. Characterised by exceptionally strong H and He nebular emission, Pop III galaxies  are thought to have prominent hydrogen lines (e.g.\@ H$\alpha$) and nebular continuum emission, as well as characteristically high equivalent width (EW) \ion{He}{II} emission lines, with a complete lack of metal lines. Indeed, extremely metal-poor (possibly Pop III) galaxy candidates have been identified through their strong \Ha\ (or \Hb) and lack of \OIII\ $\lambda 5007$ emission, in both spectroscopy \citep{Vanzella2023, Vanzella2026, Cai2025, Hsiao2025, Morishita2025, Nakajima2025, Willott2025, Maiolino2026} and photometry \citep{Nishigaki2023, Fujimoto2025}, suggesting very low metallicity. Additionally, Pop III candidates have also been found through high EW \ion{He}{II} emission \citep{Maiolino2024, Wang2024}, indicating a very hard radiation field, possibly powered by pristine star formation.

The search for signatures of a top-heavy IMF in the early Universe can be expanded by looking beyond just Pop III systems. Setting the selection bar so high, by requiring the metallicity to be so exceptionally low, causes galaxies exhibiting top-heavy star formation (i.e.\@ following a top-heavy IMF) in non-pristine environments to be overlooked \citep{Cameron2024, Mowla2024, Upadhyaya2024, Cullen2025, Katz2025}. Venturing past zero metallicity, the iconically strong \ion{He}{II} emission dramatically wanes \citep{Schaerer2003, Raiter2010}, and metal emission lines start to become prominent. Hence these selection criteria are no longer as applicable. Still, the bright hydrogen emission persists, both in terms of the emission lines, and crucially, the nebular continuum emission \citep{Cameron2024, Mowla2024, Katz2025}.

The serendipitous discovery of a steep downturn in the rest-frame UV \emph{JWST}/NIRSpec \citep{Ferruit2022, Jakobsen2022} PRISM spectrum of GS-9422 \citep{Cameron2024}, a galaxy at $z=5.943$, marks a critical turning point in our search for signatures of star formation following a top-heavy IMF in the early Universe. Believed to be attributable to two-photon continuum emission, the nebular continuum must be exceptionally strong relative to the starlight that is powering it, demanding a very high ionising photon production efficiency $\xi_\mathrm{ion}$ \citep{Cameron2024, Katz2025} powered by massive star formation (or an AGN), for the steep decline of the two-photon continuum to be visible over the rising stellar continuum in the vicinity of Ly$\alpha$. With a nebular continuum $f_{\nu,\mathrm{neb}}$ that is believed to dominate over the starlight $f_{\nu, *}$ (i.e.\@ $f_{\nu,\mathrm{neb}} > f_{\nu, *}$), GS-9422 is said to be a nebular-dominated galaxy \citep{Cameron2024, Katz2025}. In contrast, under normal star formation \citep[following a regular IMF, e.g.\@][]{Salpeter1955, Kroupa2001, Chabrier2003} the two-photon continuum downturn is hidden beneath the rise of the much brighter stellar continuum. This two-photon continuum emission \citep{Bottorff2006, Osterbrock2006, Draine2011, Schirmer2016} generally arises due to the recombination of hydrogen, which after cascading down to the 2s state, is forbidden from emitting a single photon (i.e.\@ Ly$\alpha$) to reach the 1s state. Instead, two photons are emitted, whose total energy equals that of Ly$\alpha$, with a continuum of photon--photon energy pairs possible, resulting in the definitive downturn to zero flux density at the Ly$\alpha$ wavelength. In the case of the hot \ion{H}{II} region temperatures expected to accompany star formation following a top-heavy IMF, collisional excitation to the 2s state can become appreciable \citep{Drake1983, Scholz1990, Vrinceanu2014}, boosting the two-photon continuum emission \citep{Raiter2010, Mas-Ribas2016, Katz2025, Schaerer2025}, favouring its detection. Moreover, GS-9422 exhibits a deficit in continuum flux density in the rest-frame optical compared to the rest-frame UV \citep{Cameron2024, Katz2025}. This is attributable to the Balmer jump, a continuum discontinuity at the Balmer limit (3646~\AA), due to the direct recombination of free proton--electron pairs to the $n=2$ state \citep[i.e.\@ free--bound emission,][]{Osterbrock2006, Draine2011, Schirmer2016}. The Balmer jump further indicates that GS-9422 is a young starburst with a prominent nebular continuum, supporting the nebular-dominated scenario \citep{Cameron2024, Katz2025}.

However, the nebular-dominated nature of GS-9422 is under current debate. \citet{Terp2024} suggest that the UV downturn in the spectrum of GS-9422 may be due to damped Ly$\alpha$ absorption, from a massive neutral gas reservoir associated with a $z=5.4$ protocluster \citep[identified by][]{Helton2024, Helton2024b} in the foreground of GS-9422. The offset in redshift between foreground absorber and background galaxy helps explain why GS-9422 exhibits strong Ly$\alpha$ emission despite (supposed) damped Ly$\alpha$ absorption. Indeed, \citet{Heintz2024b} find evidence of foreground DLA absorption by the $z=5.4$ protocluster in the spectra of additional $z > 5.4$ galaxies behind the protocluster. Additionally, \citet{Li2024} find that the emission lines of GS-9422 are compatible with an ionising spectrum powered by a combination of star formation and AGN activity. They find that the UV downturn is well-described by damped Ly$\alpha$ absorption by an absorber near to (i.e.\@ at the same redshift) as such an ionising source. \citet{Tacchella2025} further consider the morphology of GS-9422, finding different spatial emission in \emph{JWST}/NIRCam \citep{Rieke2023} bands probing the rest-frame UV, rest-frame optical, and nebular line emission, which is in contrast to what would be expected (similar morphologies) if the emission of GS-9422 was dominated by the nebular continuum and emission lines. They argue that the off-planar nebular emission (and Ly$\alpha$) is due to an ionisation cone powered by an AGN, while the DLA absorption originates from the galaxy disk,  explaining the wavelength-dependent morphologies and coincidence of strong Ly$\alpha$ emission with notable DLA absorption. Hence GS-9422 is not necessarily powered by star formation following a top-heavy IMF \citep{Li2024, Tacchella2025}. Additionally, it may not be nebular-dominated, with the UV downturn perhaps attributable to DLA absorption \citep{Heintz2024b, Li2024, Terp2024, Tacchella2025} rather than two-photon continuum emission \citep{Cameron2024, Katz2025}.

Given the currently ambiguous nature of GS-9422, further progress requires the identification of additional nebular-dominated candidates to better understand the mechanisms driving the downturn in the rest-frame UV. Indeed, \citet{Katz2025} have comprehensively searched through existing NIRSpec PRISM catalogs, finding four high-redshift sources (in addition to GS-9422) consistent with exhibiting nebular-dominated emission, having both a prominent UV downturn and a Balmer jump in their spectra. Two sources are galaxies, with 2198\_7807 at $z=5.387$, and 1210\_5217 at $z=4.888$ \citep[both in the GOODS-S field,][]{Giavalisco2004}, with the nebular-dominated emission possibly powered by massive star formation. Two are distinct regions (2561\_17467 and 2756\_301) within the Cosmic Gummy Worm \citep{Vanzella2022, Lin2023}, a  gravitationally lensed arc at $z=3.99$, where the spectra are thought to be purely nebular emission, due to a possible spatial offset (made possible by the lensing) between the \ion{H}{II} region (captured by the NIRSpec slit) and the star formation that is powering it (off slit). Additionally, \citet{Mowla2024} find a prominent UV downturn and substantial Balmer jump in the NIRSpec PRISM spectrum of a a highly-magnified star cluster within the Firefly Sparkle, a strongly-lensed galaxy at $z = 8.296$, which they therefore believe may be nebular-dominated. Moreover, \citet{Katz2025} outline the definitive features associated with nebular-dominated emission, these being the two-photon downturn, high observed ionising photon production efficiencies $\xi_\mathrm{ion, obs}$, red UV slopes, a Balmer jump, and high Balmer line EWs (e.g.\@ \Ha\ or \Hb). 

The current search for nebular-dominated candidates has been limited to serendipitous discoveries in existing NIRSpec PRISM catalogs \citep{Cameron2024, Mowla2024, Katz2025}, the sources being chanced upon (though the JADES prioritisation of UV-bright sources did favour the selection of GS-9422), their possible nebular-dominated nature a welcome surprise rather than being expected. To accelerate our understanding of nebular-dominated emission, we therefore develop a photometric search method for systematically identifying nebular-dominated galaxy candidates from photometric data. Key to this is the deep, extensive medium-band imaging (F162M, F182M, F210M, F250M, F300M, F335M, F410M, F430M, F460M, F480M) in the GOODS-S and GOODS-N fields from JADES \citep{Rieke2023, Bunker2024, Hainline2024, D'Eugenio2025, Eisenstein2025, Eisenstein2026}, JEMS \citep{Williams2023}, FRESCO \citep{Oesch2023}, PANORAMIC \citep{Williams2025} and BEACON \citep{Morishita2025b}, which enables Balmer-jump galaxy candidates to be identified across a wide range of redshifts ($1.5 < z < 8.5$), through their deficit in continuum flux density in the rest-frame optical. As Balmer jumps are a general recombination feature of all young starbursts ($\lesssim 3$~Myr), and thus do not necessarily demand star formation following a top-heavy IMF, we apply additional nebular-dominated selection criteria to identify nebular-dominated galaxy candidates. We require a large observed ionising photon production efficiency $\xi_\mathrm{ion, obs}$ to power the strong nebular continuum emission, as well as a relatively non-blue UV slope ($m_\mathrm{F115W}-m_\mathrm{F200W} > -0.4$ at $z=6$) indicating a lack of stellar continuum emission. We examine the continuum and emission-line properties of our Balmer-jump and nebular-dominated candidates, tracing the evolution of Balmer jumps in the galaxy population across cosmic time.  

This article is structured as follows. In Section~\ref{sec:data}, we discuss the data used in our analysis. In Section~\ref{sec:emission}, we discuss nebular-dominated emission, outlining how it demands high stellar ionising photon production efficiencies $\xi_\mathrm{ion, *}$ powered by star formation following a top-heavy IMF, and its dependence on \ion{H}{II} region temperature and the collisional enhancement of two-photon continuum emission. In Section~\ref{sec:selection}, we discuss our selection procedure for identifying Balmer-jump and nebular-dominated galaxy candidates. In Section~\ref{sec:statistics}, we discuss the continuum and emission-line properties of our Balmer-jump and nebular-dominated candidates, as well as the redshift-evolution of Balmer jumps in the galaxy population. Finally, we conclude in Section~\ref{sec:conclusions}. We assume a \citet{Planck2020} cosmology throughout and adopt the AB magnitude system \citep{Oke1983}.

\section{Data} \label{sec:data}

We use HST+NIRCam imaging data in the GOODS-S and GOODS-N fields \citep{Giavalisco2004}. Of primary interest is the medium-band imaging, as it enables the identification of Balmer-jump galaxy candidates through their drop in continuum flux density in the rest-frame optical compared with the UV. Hence we restrict our analysis to the GOODS-S and GOODS-N footprints where medium-band imaging is available. Across the various contributing programs, the medium-band filters covered are: F162M, F182M, F210M, F250M, F300M, F335M, F410M, F430M, F460M, F480M. The extensive medium-band filter set thus allows Balmer-jump galaxy candidates to be be identified across a wide range of redshifts ($1.5 < z < 8.5$). The bulk of the medium-band depth and area stems from JADES-related programs. Namely the JADES GTO \citep{Bunker2020, Rieke2020, Rieke2023, Bunker2024, Hainline2024, D'Eugenio2025, Eisenstein2026}, the JADES Origins Field \citep{Eisenstein2025}, JEMS \citep{Williams2023} and OASIS (Looser et al.\@, in prep.). This is supplemented by data from FRESCO \citep{Oesch2023}, PANORAMIC \citep{Williams2025} and BEACON \citep{Morishita2025b}. We further utilise the NIRCam wide-band imaging (F070W, F090W, F115W, F150W, F200W, F277W, F356W, F444W) from the previously-mentioned programs, as well as HST/ACS imaging in F435W, F606W, F775W, F814W, F850LP \citep{Giavalisco2004, Beckwith2006, Grogin2011, Koekemoer2011, Ellis2013, Illingworth2013, Whitaker2019}. Owing to the various programs contributing to the JADES DR5 footprint, the range of exposure times results in substantial variation in image depth across the GOODS-S and GOODS-N fields. We note that the 5$\sigma$ aperture-corrected point source depths in NIRCam filters with 0.2~arcsec diameter circular apertures is usually between 28.5--31.0~AB~mag \citep[see][]{Johnson2026, Robertson2026}. We refer the reader to the JADES Data Release 5 (DR5) papers \citep{Johnson2026, Robertson2026, Wu2026} for an extensive discussion of the depths and areas in the various NIRCam filters, as well as the full details of the data reduction and cataloguing procedures.

Briefly, the data reduction and cataloguing process builds on the procedure implemented in earlier JADES data releases. We reduce NIRCam imaging data using a modified version of the \emph{JWST} reduction pipeline, incorporating 1/f noise correction, background subtraction, and alignment onto Gaia WCS \citep{Johnson2026}. Notably, we subtract wisps using newly-developed wisp templates \citep{Wu2026} and remove persistence. Source detection is performed on an inverse-variance-weighted sum of the F277W, F335M, F356W, F410M, F430M, F444W, F460M, F480M filter images \citep{Robertson2026}. The parameters governing source deblending have been modified to allow for improved deblending in crowded fields. Images are also visually inspected to impose further source deblending and to mask diffraction spikes from bright stars. Aperture photometry is performed using various aperture sizes, we utilise 0.3~arcsec diameter circular apertures in this analysis (CIRC2 in the photometric catalog). Uncertainties in the circular aperture photometry are determined from the rms in flux densities measured in empty apertures in the vicinity of the source of interest, which is added in quadrature to the Poisson noise from the source itself. The circular aperture photometry is aperture-corrected using corrections based off the simulated STPSF \citep{Perrin2012, Perrin2014} point spread functions for each filter, using the method described by \citet{Ji2024}. `Total' source flux densities are determined using a modified procedure for performing Kron elliptical aperture photometry \citep{Robertson2026}. We use the ratio of these total flux densities to the circular aperture-based measurements, to scale the aperture-based line fluxes and continuum flux densities into total line luminosities and absolute magnitudes, taking into account also the source redshift. 

Photometric redshifts are determined \citep{Hainline2026} using the EAZY \citep{Brammer2008} SED-fitting code, following a similar procedure to previous JADES data releases \citep{Hainline2024}. The full NIRCam photometry, together with HST/ACS F435W, F606W, F775W, F814W, F850LP photometry in 0.2~arcsec diameter (i.e.\@ different to the 0.3~arcsec diameters we use in our analysis) circular apertures are fit in the SED-fitting process, adopting a minimum error of 5~per~cent in each band. As before \citep{Hainline2024}, the EAZY fit is performed using a modified template set, consisting of the original EAZY templates plus seven additional templates that were created to better span the observed colour space of galaxies in the JAGUAR simulations \citep{Williams2018}. The \citet{Asada2025} IGM prescription to account for damped Ly$\alpha$ absorption is now newly implemented in the EAZY fitting process. We generally use the EAZY best-fit (i.e.\@ minimum $\chi^2$) photometric redshifts in our analysis, though we adopt the JADES DR4 \citep{Curtis-Lake2025, Scholtz2025} spectroscopic redshifts whenever these are available. 

We construct our high-redshift sample ($1.5 < z < 8.5$) by applying the following quality cuts to the full photometric catalog. We require the source to be ${>} 5\sigma$ detected in all wide-bands fully redward of Ly$\alpha$. We further require the source to be ${<}3 \sigma$ detected (i.e.\@ non-detected) in all bands fully blueward of the Ly$\alpha$ break (for high-redshift galaxies) or fully blueward of the Lyman break (for low-redshift galaxies where the Ly$\alpha$ break is less strong). The appropriate redshifts for this dropout requirement are determined by establishing when the bandpass-averaged flux density in a given filter drops out by $> 5$~mag, assuming a flat $f_\nu$ spectrum with the \citet{Inoue2014} IGM prescription. To account for very bright dropout galaxies in the deep imaging, we alternatively require the magnitude difference $\Delta m_\mathrm{break}$ between the dropout filter and first filter fully redward of Ly$\alpha$ to be $\Delta m_\mathrm{break} > 4.5$. Finally, we require the source to be ${>}5\sigma$ detected in the medium-band filter used to probe the rest-frame optical continuum level. We apply a reduced $\chi_\mathrm{red} ^2 < 6.25$ cut for most galaxies, to ensure the EAZY fits provide a reasonable description of the data. We do not apply the $\chi_\mathrm{red} ^2$ cut for the nebular-dominated galaxy candidates, as it is possible that the EAZY templates may not provide a particularly good fit (as they do not include nebular-dominated templates), despite correctly identifying the redshift (given the prominent photometric excess caused by the emission lines, plus Ly$\alpha$ break). As mentioned previously, the image depths vary across the GOODS-S and GOODS-N fields, and between filters. We do not apply a magnitude or stellar mass limit cut in our subsequent analysis. Photometry is corrected for Galactic dust extinction using the \citet{Schlafly2011} $\mathrm{E(B-V)}$ dust map, adopting the \citet{Fitzpatrick1999} dust attenuation law with total-to-selective extinction ratio $R_\mathrm{V}=3.1$.

We fit the HST+NIRCam photometry of select sources with Bagpipes \citep{Carnall2018}, adopting a minimum error of 5~per~cent in each band. This is for display purposes only, to accompany the photometry shown in our SED figures. We use the v2.2 BPASS \citep{Stanway2018} stellar templates with a 300~$\mathrm{M}_\odot$ upper mass threshold. We fix the redshift to the best-fit EAZY photometric redshift (or JADES DR4 spectroscopic redshift, where available). We adopt uniform priors on all fitted parameters, with the logarithm of the stellar mass $\log_{10} (M_*/\mathrm{M}_\odot) \in [1, 15]$, the logarithm of the metallicity $\log_{10} (Z/\mathrm{Z}_\odot) \in [\log_{10}0.005, 0]$, logarithm of the ionisation parameter $\log_{10} U \in [-4, -1]$ and dust attenuation $A_\mathrm{V} \in [0, 5]$ \citep[with the same total $A_\mathrm{V}$ for emission inside/outside birth clouds, assuming the dust attenuation law from][]{Calzetti2000}. For the star-formation history (SFH), we generally use a single burst model, with burst age $a \in [1, 5]$~Myr. However, for Balmer-jump galaxy candidates with very blue slopes a double power-law SFH may yield a better fit, where the falling power-law slope $\log_{10} \alpha \in [-2, 3]$, rising power-law slope $\log_{10} \beta \in [-2, 3]$, and time of peak star formation $\tau \in [0, \mathrm{age_{univ}}(z)]$. 

We also use other data to characterise the observed properties of previously known nebular-dominated galaxy candidates. We utilise the 0.32~arcsec diameter circular aperture photometry tabulated in the UNCOVER \citep{Bezanson2024} + MegaScience \citep{Suess2024} data release photometric catalog. This is to photometrically determine the Balmer jump $\Delta m_\mathrm{jump}$ and observed ionising photon production efficiency $\xi_\mathrm{ion, obs}$ for the two distinct regions (2561\_17467 and 2756\_301) in a $z=3.99$ gravitationally lensed arc that may exhibit nebular-dominated emission in their NIRSpec PRISM spectra \citep{Katz2025}.

\begin{figure*}
\centering
\includegraphics[width=.6\linewidth]{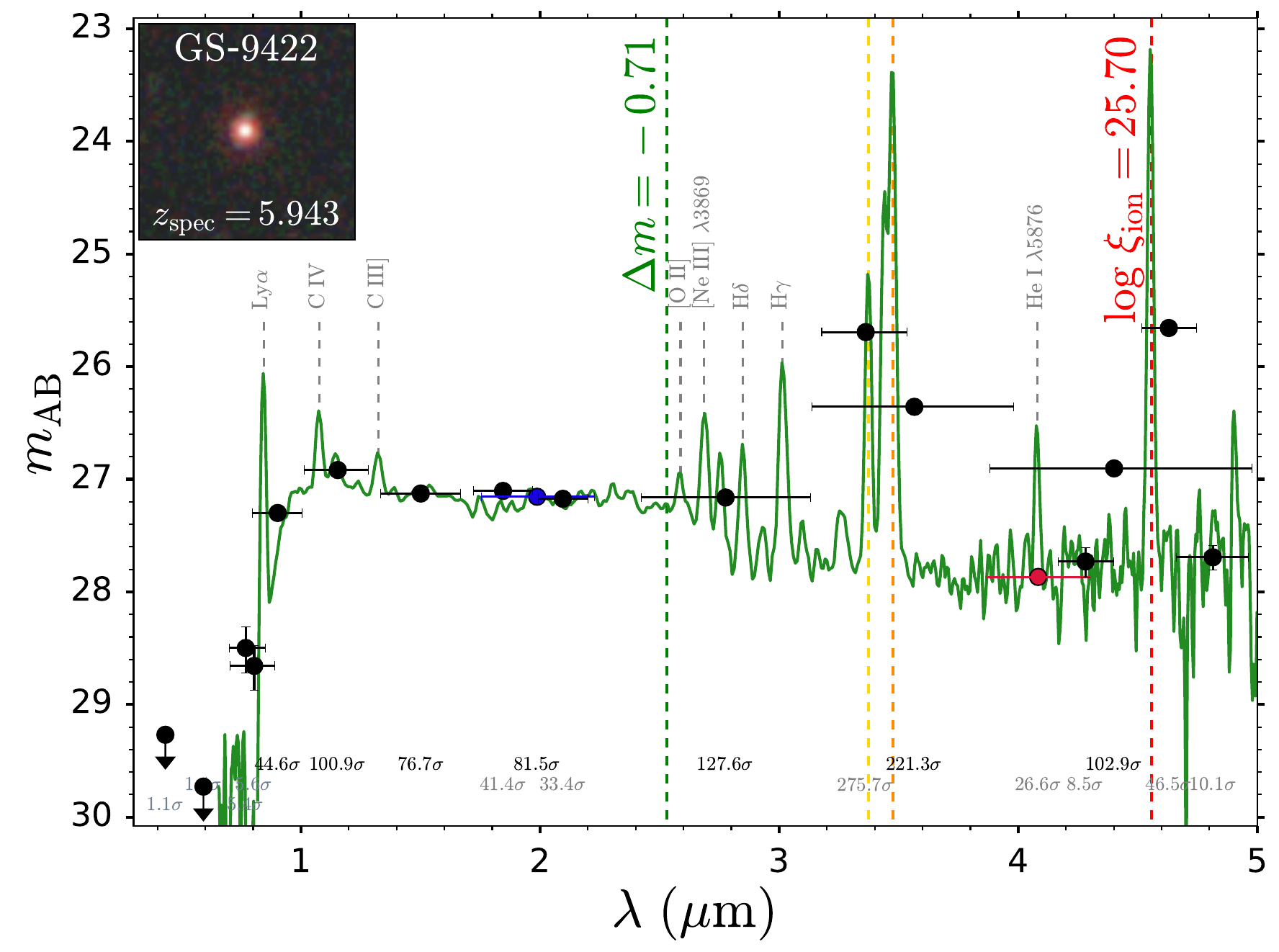}
\caption{The NIRSpec PRISM spectrum (dark green), HST+NIRCam photometry (black) and $2\times 2$~arcsec RGB (F444W, F200W, F115W) cutout for GS-9422, the nebular-dominated galaxy candidate reported by \citet{Cameron2024}. Vertical dashed lines indicate the location of the Balmer limit at 3646~\AA\ (green), \Hb\ (yellow), \OIII\ $\lambda 5007$ (orange) and \Ha\ (red). Grey vertical dashed lines and labels denote other notable rest-frame UV/optical emission lines. The SNR in 0.3~arcsec diameter circular apertures is denoted at the bottom for each filter. Non-detections are depicted as downward arrows located at the 3$\sigma$ upper limit. The spectrum of GS-9422 exhibits a prominent downturn at ${\sim}1.1$~\textmu m, possibly due to two-photon continuum emission (or perhaps damped Ly$\alpha$ absorption), suggesting nebular-dominated emission \citep{Cameron2024, Katz2025}. GS-9422 also exhibits a discontinuity (at 3646~\AA\ rest-frame, ${\sim}2.5$\textmu m observed-frame)  in its continuum levels between the rest-frame UV and optical: the Balmer jump, a general recombination feature of all young ($\lesssim 3$~Myr) starbursts. Medium-band photometry, narrow enough to fit between the prominent rest-frame optical emission lines (\OIII\ and \Ha), reveals the low-lying rest-frame optical continuum level (measured with F410M, red, though F430M and F480M could also be adopted), being $\Delta m_\mathrm{jump} = -0.71$~mag fainter than the rest-frame UV (measured with F200W, dark blue). The photometric boost caused by the strong \Ha\ emission in F444W, together with the 1500~\AA\ continuum measurement (via F115W), establish the high ionising photon production efficiency $\log\, (\xi_\mathrm{ion, obs} /\mathrm{(Hz\ erg^{-1})})= 25.70$ in this source, supporting the nebular-dominated scenario.}
\label{fig:GS-9422}
\end{figure*}

We use the GS-9422 NIRSpec PRISM spectrum from the JADES public data release 4 \citep{Curtis-Lake2025, Scholtz2025}. We also use v4 PRISM spectra from the DAWN JWST Archive \citep{deGraaff2025, Heintz2025} for the four additional nebular-dominated galaxy candidates reported by \citet{Katz2025}: 1210\_5217, 2198\_7807, 2561\_17467 and 2756\_301. These spectra are from programs 1210 (i.e.\@ JADES Deep-HST), 2198 \citep{Barrufet2025}, 2561 \citep[i.e.\@ UNCOVER,][]{Bezanson2024} and 2756 (PI: Chen).

\section{Nebular-dominated emission} \label{sec:emission}

In this section we discuss nebular-dominated emission. In this work, we define a galaxy to be nebular-dominated if its nebular continuum $f_{\nu, \mathrm{neb}}(\lambda)$ exceeds its stellar continuum $f_{\nu,*}(\lambda)$ at all wavelengths beyond 1615~\AA, corresponding to the wavelength where the two-photon continuum peaks (in $f_\nu$), i.e.\@ $f_{\nu, \mathrm{neb}}(\lambda) > f_{\nu, *}(\lambda)$ for $\lambda > 1615~$\AA. This definition is motivated by the fact that even in Pop III models \citep[e.g.\@][]{Zackrisson2011} the nebular continuum becomes weaker than the stellar continuum slightly redward of \Lya\ (with the nebular-/stellar-dominated transition wavelength depending on Pop III IMF and starburst age), due to the steeply declining two-photon continuum near \Lya. With this nebular-dominated galaxy definition, the two-photon downturn should be evident in continuum spectroscopy. We examine the most notable nebular-dominated galaxy candidate, GS-9422, in Section~\ref{subsec:gs-9422}. We describe the components and behaviour of the nebular continuum in Section~\ref{subsec:nebcont}. Finally, we compare the relative normalisation of the nebular vs.\@ stellar continuum in Section~\ref{subsec:nebular_vs_stellar}.

\subsection{GS-9422} \label{subsec:gs-9422}

We begin by examining the most notable known nebular-dominated galaxy candidate, GS-9422 \citep{Cameron2024, Katz2025}, in more detail. We show its NIRSpec PRISM spectrum (dark green), HST+NIRCam photometry (black) and RGB cutout \citep[$2\times2$ arcsec, R: F444W; G: F200W; B: F115W, generated using Trilogy,][]{Coe2012} in Fig.~\ref{fig:GS-9422}. Vertical dashed lines correspond to the location of the Balmer jump (3646~\AA, green), H$\beta$ (yellow), [\ion{O}{III}] $\lambda 5007$ (orange) and H$\alpha$ (red), respectively. Grey vertical dashed lines and labels denote other notable rest-frame UV/optical emission lines. The SNRs in 0.3~arcsec diameter circular apertures are listed at the bottom for each filter. 

A clear downturn in the rest-frame UV continuum level is visible immediately redward of Ly$\alpha$ in the PRISM spectrum, dropping sharply at wavelengths below ${\sim}1.1$\textmu m (1615~\AA\ rest-frame). This steep continuum drop is thought to be due to either two-photon continuum emission \citep[i.e.\@ the nebular-dominated scenario,][]{Cameron2024, Katz2025} or damped Ly$\alpha$ absorption \citep{Heintz2024b, Li2024, Terp2024, Tacchella2025}. While a prominent feature in the spectrum of GS-9422, there is no clear signature of the UV downturn in the photometry, notably in the F090W dropout filter. Here the strong Ly$\alpha$ emission hides the subtle photometric deficit that the UV downturn would otherwise leave on the F090W dropout photometry. We discuss the challenges in identifying the imprint of two-photon continuum emission on photometry in more detail in Appendix~\ref{app:deficit}.

In addition, GS-9422 also exhibits a discontinuity in its continuum levels between the rest-frame UV and optical \citep{Cameron2024, Katz2025}. The drop in flux density at 3646~\AA\ rest-frame is the Balmer jump, a general recombination feature of all young starbursts ($\lesssim 3$~Myr). While the rest-frame UV level is well-traced using wide bands (e.g.\@ F200W, coloured blue), they are generally not satisfactory for tracing the rest-frame optical continuum level in starbursts. This is because the broad bands are sufficiently wide that they almost always have their bandpass-averaged flux densities photometrically boosted by strong line emission to be well-above the actual continuum level. Here F444W is boosted by \Ha, F356W is boosted by \OIII\ + \Hb, and F277W is boosted by the complex of weaker emission lines (\Hg, \Hd, \He, \NeIII, \OII, etc.\@). Medium bands, with their narrower spectral range, are able to fit inside the gap (for certain redshift ranges) between the strong optical lines (e.g.\@ \OIII\ and \Ha), thus providing a cleaner measurement of the rest-frame optical continuum level, such as the F410M (coloured red), F430M and F480M filters in the case of GS-9422. Even then, the rest-frame optical continuum level can still be overestimated due to the photometric boost caused by the weaker He~I $\lambda 5876$ line, seen at ${\sim}4.1$\textmu m in the spectrum. By comparing the continuum levels in the F200W and F410M filters, we measure a Balmer jump $\Delta m_\mathrm{jump} = \mathrm{F200W} - \mathrm{F410M} = -0.71 \pm 0.05$ (noted in Fig.~\ref{fig:GS-9422}) for GS-9422.

Moreover, the measurement of the optical continuum level (via F410M) and photometric boost caused by \Ha\ (via F444W, or possibly F460M) enables the determination of the \Ha\ flux, a tracer for the ionising photon production rate. Combined with the continuum level at 1500~\AA\ (via F115W), the observed ionising photon production efficiency $\xi_\mathrm{ion,obs}$ can be determined. The inferred $\log\, (\xi_\mathrm{ion, obs} /\mathrm{(Hz\ erg^{-1})}) = 25.70$ for GS-9422 is rather high \citep[compared to the typical value of 25.29 at $z=6$,][]{Simmonds2024b}, suggesting a lot of recombination emission compared to the starlight that is powering it, hinting at its possible nebular-dominated nature \citep{Cameron2024, Katz2025}. We will discuss this key observational proxy for nebular-dominated emission in more detail in Section~\ref{subsubsec:xi_selection}.

\subsection{Nebular continuum} \label{subsec:nebcont}

\begin{figure}
\centering
\includegraphics[width=\linewidth]{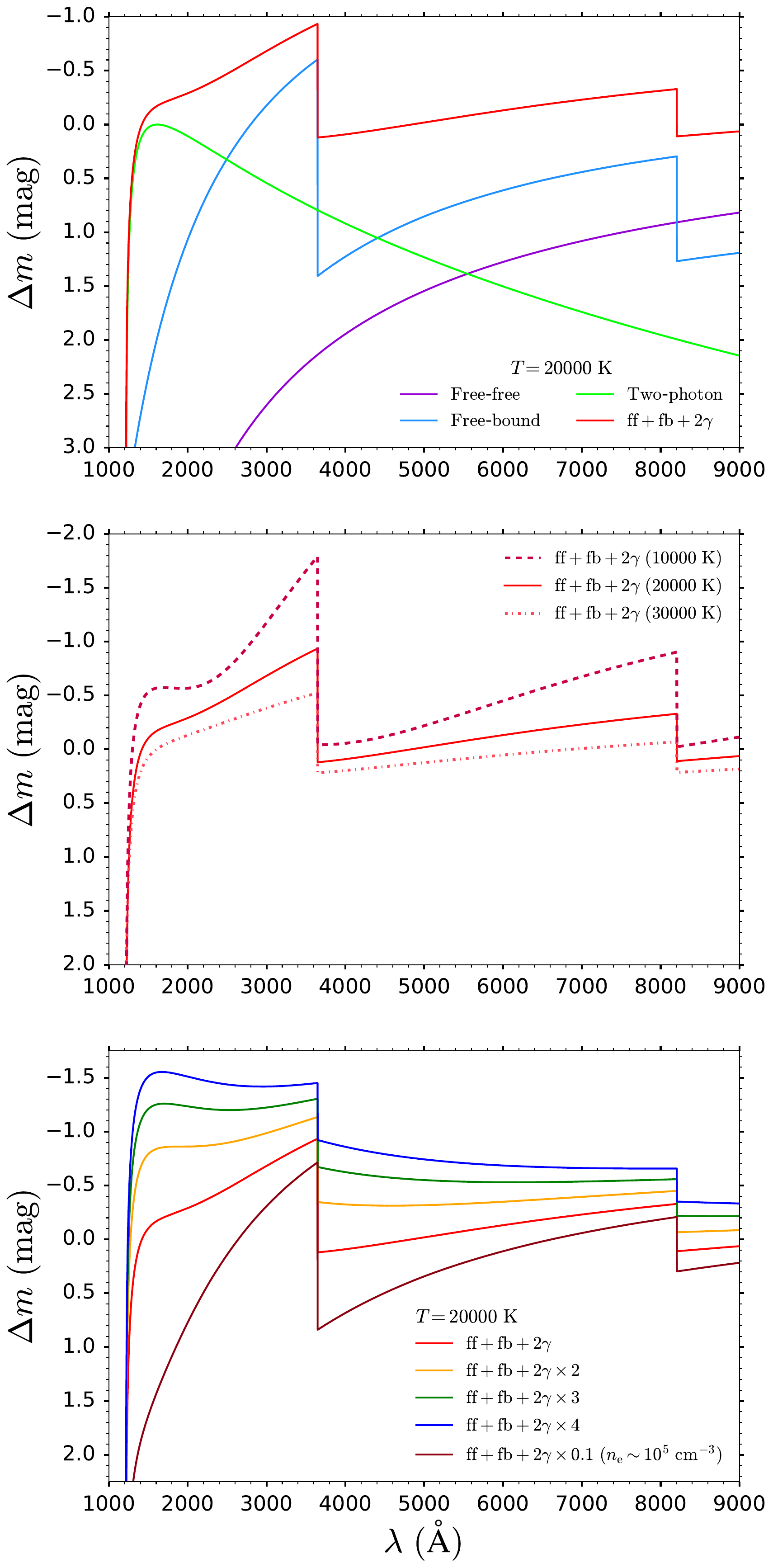}
\caption{Top panel: Various components of the nebular continuum emission, assuming pure hydrogen gas at $T=20000~\mathrm{K}$, generated using PyNeb \citep{Luridiana2015}. Shown are the two-photon continuum (light green), free--bound emission (blue) which is responsible for the Balmer jump at 3646~\AA\ and Paschen jump at 8206~\AA, free--free emission (purple), and the total nebular continuum (ff + fb + $2\gamma$, red). All components are normalised by the peak value of the two-photon continuum at 1615~\AA. Middle panel: The temperature dependence of the total nebular continuum, with ff + fb + $2\gamma$ at $T = 10000~\mathrm{K}$ (dashed), $T = 20000~\mathrm{K}$ (solid) and $T = 30000~\mathrm{K}$ (dot-dashed). The Balmer jump decreases in $\left | \Delta m\right |$ with increasing temperature, so top-heavy, metal-poor star formation (resulting in higher \ion{H}{II} region temperatures) is not necessarily powering the largest Balmer jumps. Bottom panel: The total nebular continuum with varying amounts of collisional enhancement (caused by high \ion{H}{II} region temperatures and non-negligible neutral \ion{H}{I} fractions) of two-photon continuum emission. Shown are regular two-photon continuum emission ($2\gamma \times 1$, red), collisionally-enhanced two-photon emission with $2\gamma \times [2, 3, 4]$ (orange, green, blue, respectively) and collisionally-suppressed two-photon emission with $2\gamma \times 0.1$ due to high density $n_\mathrm{e} \sim 10^{5}~\mathrm{cm}^{-3}$ gas (dark red).}
\label{fig:nebular_continuum}
\end{figure}

We show the various components of the nebular continuum in the top panel of Fig.~\ref{fig:nebular_continuum}. These were generated using PyNeb \citep{Luridiana2015}, assuming pure hydrogen gas at $T = 20000~\mathrm{K}$ \citep[which is comparable to the $T = 20156$~K \ion{H}{II} region temperature inferred for GS-9422,][]{Katz2025}. The two-photon continuum (light green) rises steeply with decreasing wavelength from the optical to the ultraviolet, plateaus in $f_\nu$ at 1615~\AA, before precipitously declining towards zero at $\lambda_\mathrm{Ly\alpha} = 1216~$\AA. The other nebular components displayed in Fig.~\ref{fig:nebular_continuum} are all normalised by the peak value of the two-photon continuum emission at $T = 20000~\mathrm{K}$ (so that $\Delta m = 0$ corresponds to the continuum flux density $f_\nu$ being the same as the two-photon peak).

The free--bound continuum emission (blue) is marked by discontinuities, with direct recombinations to $n=2$ causing the Balmer jump at 3646~\AA, and direct recombinations to $n=3$ causing the Paschen jump at 8206~\AA. Outside the discontinuities, the free--bound continuum goes approximately as $f_\nu \propto e^{-hc/\lambda kT}$ \citep{Brown1970, Ercolano2006}. This has three notable consequences. First, that the free--bound emission is always rising (apart from at the discontinuities) with increasing wavelength, i.e.\@ it has a red colour (though can be blue when bracketing discontinuities, like the Balmer jump). Thus, if the full spectrum is blue, this must either be because of the two-photon continuum (if the source is nebular-dominated), or the starlight / AGN accretion disk emission. Second, the slope of the free--bound emission flattens with increasing wavelength, simply due to the $1/\lambda$ behaviour. So while the free--bound emission notably drops in the rest-frame UV, its decline is less prominent in the optical and even less so in the NIR. Third, the free--bound emission flattens with increasing temperature due to the $1/T$ behaviour. The free--free emission (also referred to as bremsstrahlung, purple) follows the same exponential dependence \citep{Brown1970, Ercolano2006, Osterbrock2006, Draine2011}, so has the same properties.

The combined free--free + free--bound + two-photon continuum emission (ff + fb + $2\gamma$) is also shown (red). The introduction of additional continuum components reduces the magnitude shift $\Delta m$ associated with the Balmer jump, as the increased normalisation now causes a given linear shift (the Balmer jump discontinuity) to correspond to a smaller multiplicative change. Even at $T = 20000~\mathrm{K}$, the total nebular continuum still has a relatively red UV slope, which acts to flatten the characteristically blue slopes of young stellar populations, resulting in a combined nebular+stellar spectrum with a less steep UV slope \citep[e.g.\@][]{Dunlop2013, Byler2017, Topping2022, Cullen2024, Austin2025}

The dependence of the total nebular continuum on ISM temperature is shown in the middle panel of Fig.~\ref{fig:nebular_continuum}, at $T = [10000, 20000, 30000]~\mathrm{K}$ (dashed, solid, dot-dashed, respectively). Increasing the temperature causes the Balmer jump to decrease in terms of a magnitude shift $\Delta m$. This is because the discontinuity at 3646~\AA\ is driven by proton--electron pairs recombining with exactly zero kinetic energy to the $n=2$ state. The greater the temperature, the smaller the fraction of recombinations taking place with zero kinetic energy. Moreover, the normalisation of the free--free emission drops less strongly ($\propto T^{-1/2}$) compared to the free--bound emission ($\propto T^{-3/2}$), further contributing to the decline in the Balmer jump. Thus top-heavy (higher average ionising photon energy contributing more ISM heating), metal-poor (less effective ISM cooling) star formation may not necessarily be powering the largest Balmer jumps, due to the hotter expected \ion{H}{II} region temperatures. Additionally, the rest-frame UV (and optical) slope of the total nebular continuum flattens with increasing temperature, due to the gentler decline of the free--bound emission.

\begin{figure*}
\centering
\includegraphics[width=\linewidth]{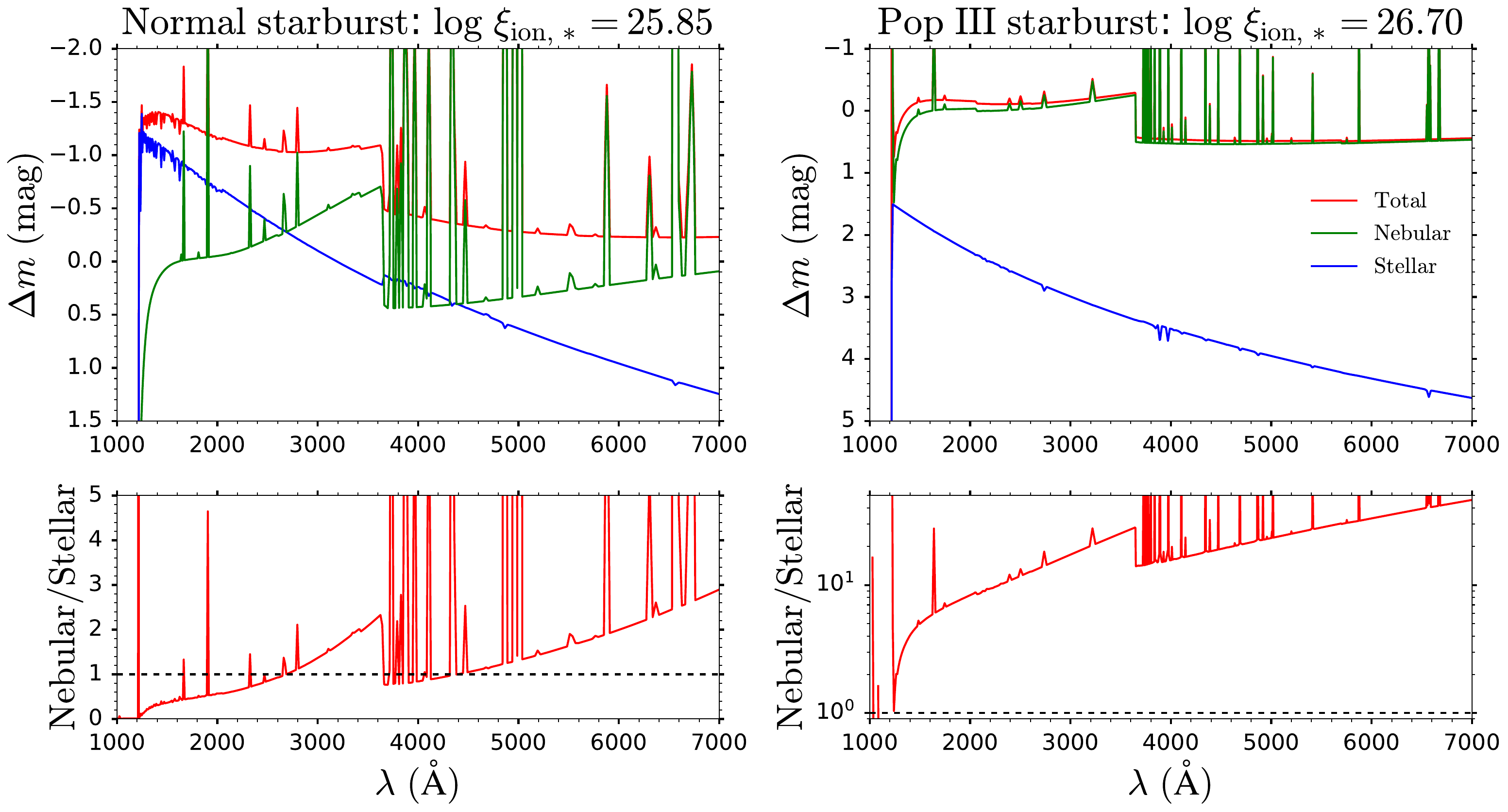}
\caption{Left panel: For normal starbursts \citep[following a regular IMF, e.g.\@][]{Salpeter1955, Kroupa2001, Chabrier2003} with $\log\, (\xi_\mathrm{ion, *} /\mathrm{(Hz\ erg^{-1})}) \approx 25.85$ such as the 1~Myr old starburst shown generated with Bagpipes \citep{Carnall2018}, the normalisation of the nebular continuum (green) is too low for the steep decline of the two-photon continuum emission to be seen over the rise of the stellar continuum (blue) in the total spectrum (red). Right panel: It is only with star formation following a top-heavy IMF \citep[here a 1~Myr old Pop III.1 starburst from][]{Zackrisson2011}, where the ionising photon production efficiency is much higher (here $\log\, (\xi_\mathrm{ion, *} /\mathrm{(Hz\ erg^{-1})}) \approx 26.70$) , so that the normalisation of the nebular continuum becomes substantial relative to the ionising starlight that is powering it, that the two-photon turnover is very prominent in the total spectrum: the galaxy is nebular-dominated \citep{Cameron2024, Katz2025}.}
\label{fig:nebular_stellar}
\end{figure*}

Under normal \ion{H}{II} region conditions ($T \sim 10000~\mathrm{K}$), the two-photon continuum emission arises solely due to recombination, following the cascade of hydrogen to the 2s state. However, if the \ion{H}{II} region temperature is sufficiently high (possibly powered by top-heavy, metal-poor star formation / AGN activity), collisional excitation of ground state hydrogen to the $n=2$ state occurs at an appreciable rate \citep{Drake1983, Scholz1990, Vrinceanu2014}, opening up another channel for two-photon continuum emission \citep{Raiter2010, Mas-Ribas2016, Katz2025, Schaerer2025}. For example, at $T = 10000~\mathrm{K}$, the $n = 1 \rightarrow 2$ collisional excitation rate coefficient $k_{12} \approx \alpha _B$, the case-B recombination rate coefficient \citep{Drake1983, Scholz1990, Draine2011}. Assuming a neutral fraction ($=n_\mathrm{HI}/n_\mathrm{H}$) of $10^{-4}$, the collisional excitation rate ($n_\mathrm{HI}n_\mathrm{e}k_{12}$) to $n=2$ is four orders of magnitudes lower than the recombination rate ($n_\mathrm{p}n_\mathrm{e}\alpha _B$) to $n=2$, so negligible. In contrast, at $T = 30000~\mathrm{K}$ $k_{12} \approx 5000\alpha _B$, so collisional excitation enhances the two-photon continuum emission by 50~per~cent. So in addition to the \ion{H}{II} region temperature, the ionisation structure (neutral fraction) of the \ion{H}{II} region is important in determining the collisional enhancement of two-photon continuum emission.

In addition to the total nebular continuum arising purely from recombination (ff + fb + $2\gamma$, red lines in Fig.~\ref{fig:nebular_continuum}), we therefore also show the total nebular continuum when the two-photon continuum is enhanced through collisional excitations, with $2\gamma \times [2, 3, 4]$ shown in orange, green and blue, respectively. At $T=20000~\mathrm{K}$ ($T=30000~\mathrm{K}$), these collisional enhancements correspond to neutral fractions $n_\mathrm{HI}/n_\mathrm{H} = [2, 4, 6] \times 10^{-3}$  ($[2, 4, 6]\times 10^{-4}$), respectively. We note that collisional enhancements up to $2\gamma \times 3$ are seen in the Pop III models of \citet{Raiter2010}. In the bottom panel of Fig.~\ref{fig:nebular_continuum}, we see that the enhanced two-photon continuum emission reduces the Balmer jump and makes the UV continuum slope less red (i.e.\@ more blue). Comparing to the relatively flat UV slope for GS-9422 in Fig.~\ref{fig:GS-9422}, this implies that there is substantial collisional enhancement of the two-photon continuum emission in this source (provided the UV downturn is not due to a DLA). 

The two-photon continuum emission can also be suppressed, provided that the electron density approaches the critical density \citep[${\sim} 10^4~\mathrm{cm}^{-3}$,][]{Draine2011} of the 2s state, through $l$-changing collisions to the 2p state, resulting in enhanced Ly$\alpha$ emission \citep{Cameron2024, Katz2025, Schaerer2025}. Assuming $n_\mathrm{e}{\sim} 10^5~\mathrm{cm}^{-3}$, the two-photon continuum emission is reduced to $0.1\times$ the regular value (shown in dark red), resulting in very red nebular continuum emission \citep{Katz2025}, dominated by the free--free and free--bound components.

\subsection{Nebular vs.\@ stellar} \label{subsec:nebular_vs_stellar}

Recombination powers both nebular line emission and nebular continuum emission. Like e.g.\@ \Ha\ emission, the nebular continuum emission is always present in recombination spectra. However, unlike \Ha\ emission, the nebular continuum emission is generally not very prominent, even for young starbursts. This is because the normalisation of the nebular continuum is usually small compared to the starlight that is powering it, especially in the rest-frame ultraviolet around 1500~\AA. Since the nebular continuum is (mostly) powered by recombination (and thus photoionisation), this implies that the ionising photon production rate is relatively low compared to the level of the non-ionising stellar continuum, i.e.\@ the stellar ionising photon production efficiency $\xi_\mathrm{ion,*}$ is too small for the nebular continuum to dominate \citep{Cameron2024, Katz2025}. We show this more clearly in the left panel of Fig.~\ref{fig:nebular_stellar}, showcasing the stellar (blue), nebular (green) and total (nebular+stellar, red) spectrum of a 1~Myr old starburst generated using Bagpipes \citep{Carnall2018}. While the nebular continuum can be prominent immediately blueward of the Balmer jump (due to the sudden increase in nebular continuum level), and at increasingly longer optical wavelengths (due to the divergence in flat/red nebular slopes vs.\@ blue stellar slopes), it is negligible around the vicinity of the two-photon continuum turnover (1615~\AA). Thus, in the UV, the nebular continuum mostly serves to make the slope of the total continuum less blue \citep[i.e.\@ increasing $\beta$,][]{Dunlop2013, Topping2022, Cullen2024, Austin2025}.

Hence under regular circumstances, with starbursts following a normal IMF \citep[e.g.\@][]{Salpeter1955, Kroupa2001, Chabrier2003}, the decrease in nebular continuum due to the two-photon continuum turnover is hidden beneath the rise of the much brighter stellar continuum, as the nebular continuum is too faint because of the low $\log\, (\xi_\mathrm{ion, *} /\mathrm{(Hz\ erg^{-1})}) \approx 25.85$ \citep[for a 1~Myr old starburst in Bagpipes,][]{Carnall2018}. Seeing the two-photon continuum turnover therefore demands a much larger normalisation of the nebular continuum, indicating a very high $\xi_\mathrm{ion,*}$ \citep{Cameron2024, Katz2025}. We show this scenario in the right panel of Fig.~\ref{fig:nebular_stellar}, adopting a 1~Myr old Pop III.1 starburst from \citet{Zackrisson2011}. Here the typical mass of Pop III stars is ${\sim}100~\mathrm{M}_\odot$, resulting in $\log\, (\xi_\mathrm{ion, *} /\mathrm{(Hz\ erg^{-1})}) \approx 26.70$. Powered by the photoionisation of the massive, hot stars, now the normalisation of the nebular continuum is much higher, the UV turnover is very prominent in the total spectrum, and the incident starlight is fainter than the nebular continuum at all wavelengths: the galaxy is nebular-dominated \citep{Cameron2024, Katz2025}.

In addition to the ionising photon production efficiency, the normalisation of the nebular continuum is further affected by additional parameters. The \ion{H}{II} region temperature affects the shape of the continuum, with increasing temperature flattening the slope in the UV (see Fig~\ref{fig:nebular_continuum}). Moreover, the collisional enhancement of two-photon continuum emission naturally favours its detection. On the other hand, high electron density suppresses the two-photon continuum emission. Finally, the ionising photon escape fraction further sets the level of the nebular continuum. Non-unity covering fractions \citep[ionisation-bounded case, e.g.\@][]{Zackrisson2013}, a lack of \ion{H}{I} gas \citep[density-bounded case, e.g.\@][]{Zackrisson2013, Cullen2025, McClymont2025} or absorption of ionising photons by dust grains in the \ion{H}{II} region \citep{Tacchella2022, Tacchella2023} all reduce the normalisation of the nebular continuum emission.

\begin{figure*}
\centering
\includegraphics[width=\linewidth]{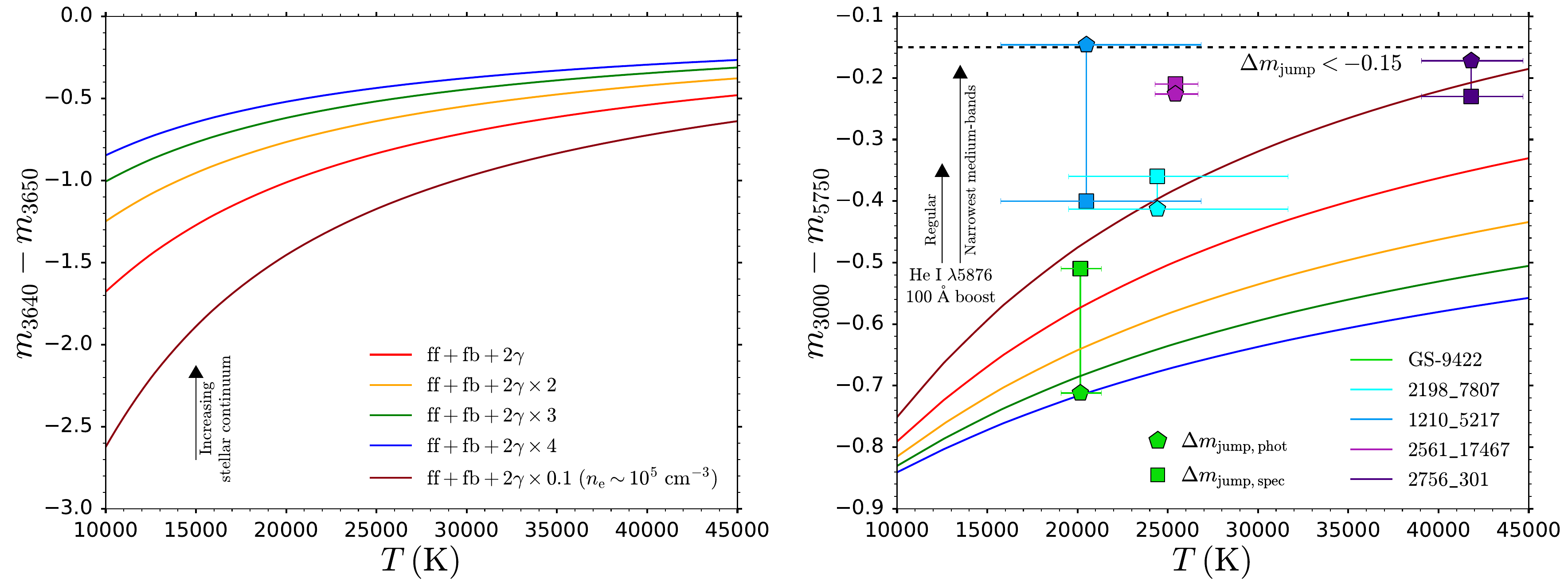}
\caption{Left panel: The Balmer jump, as traced by the magnitude difference $m_{3640}-m_{3650}$ between the continuum levels immediately blueward and redward of the Balmer limit at 3646~\AA, shown as a function of \ion{H}{II} region temperature for different levels of collisionally-enhanced two-photon continuum emission (various colours). We assume only the nebular continuum, including the H and \ion{He}{I} contribution \citep[using PyNeb,][]{Luridiana2015}, but excluding any stellar continuum component (which would reduce the magnitude shift, arrow). Right panel: Low-resolution PRISM spectroscopy and photometry demand longer wavelength baseline measurements $m_\mathrm{3000}-m_\mathrm{5750}$, to avoid blending of emission lines and continuum. Here the continuum level in the rest-frame optical $m_\mathrm{5750}$ is measured using a medium-band filter that lies between \OIII\ $\lambda 5007$ and \Ha. \ion{He}{I} $\lambda 5876$ emission (maximally ${\sim}100~$\AA\ EW), can cause a photometric boost in the medium band, causing the Balmer jump to be underestimated by up to 0.16--0.32~mag (depending on filter spectral resolution $R$, arrows). Photometric (pentagons) and spectroscopic (squares) measurements of the Balmer jump in the spectroscopically-identified nebular-dominated galaxy candidates reported by \citet{Katz2025}: GS-9422 (green), 2198\_7807 ($z=5.347$, cyan), 1210\_5217 ($z=4.888$, blue), 2561\_17467 ($z=3.99$, light purple), 2756\_301 ($z=3.99$, dark purple). Motivated by the smallest Balmer jumps seen in current observations of nebular-dominated galaxy candidates \citep{Cameron2024, Katz2025}, as well as Pop III models \citep{Zackrisson2011, Nakajima2022}, we set our minimal Balmer jump threshold (dashed horizontal line) as $\Delta m_\mathrm{jump} < -0.15$.}
\label{fig:jump_temperature}
\end{figure*}

We note that nebular-dominated emission does not necessarily imply exceptionally high ionising photon production efficiencies powered by massive, hot stars \citep{Cameron2024, Katz2025, Tacchella2025}. AGN accretion disks can also power strong nebular continuum emission. In principle, this would demand hot disk temperatures (so that $\xi_\mathrm{ion,BH}$ is high), through accretion onto low mass black holes and/or at high accretion rates \citep{Wilkins2025}. However, two-photon continuum emission will never be visible in the broad line region, due to suppression by $l$-changing collisions in the high density gas ($n_\mathrm{e} \sim 10^{10}~\mathrm{cm}^{-3} \gg 10^{4}~\mathrm{cm}^{-3} = n_\mathrm{crit}$) that is well-above the critical density of the 2s state. Thus only two-photon continuum emission from the narrow line region will be visible. Provided that the accretion disk and BLR are fully obscured, and the contribution from starlight in the host galaxy is negligible, only the nebular continuum emission will be visible \citep{Tacchella2025}, so the two-photon continuum emission should dominate independent of the temperature of the AGN accretion disk. Additionally, if the AGN has faded, but the NLR emission has not yet, then again the nebular continuum should dominate \citep{Cameron2024}. Finally, if there is a spatial separation between the photoionisation source (stars/AGN) and the \ion{H}{II} region, such that the nebular emission can be isolated, the spectrum will be nebular-dominated \citep{Cameron2024, Katz2025}. This can come about by e.g.\@ AGN photoionisation of giant \ion{H}{II} regions in the haloes of galaxies \citep{Lintott2009, D'Eugenio2025b}, or by strong gravitational lensing of \ion{H}{II} regions that are being externally radiated by starlight \citep{Katz2025}.

\section{Selection} \label{sec:selection}

\subsection{Balmer-jump selection} \label{subsec:bj_selection}

We start our search for nebular-dominated candidates by applying a Balmer-jump selection, the principle being that galaxies exhibiting a Balmer jump will be young starbursts, some subset of which may be undergoing star formation following a top-heavy IMF and exhibit nebular-dominated emission. The identification of the Balmer jump ensures that the nebular continuum contributes considerably to the rest-frame optical, so it is possible that for some of these systems the nebular continuum may also dominate in the rest-frame UV.

\subsubsection{Balmer jump} \label{subsubsec:jump}

As shown in the left panel of Fig.~\ref{fig:jump_temperature}, the Balmer jump $m_{3640} - m_{3650}$, as traced by the immediate discontinuity blueward and redward of 3646~\AA, decreases in absolute value with increasing temperature and with increasing two-photon continuum contribution. Here we assume only the nebular continuum, including the H and \ion{He}{I} contribution \citep[using PyNeb,][]{Luridiana2015}, but excluding any stellar continuum component. For regular star formation, the stellar continuum is relatively substantial, further reducing the magnitude shift $\Delta m$ associated with the Balmer jump (see Fig.~\ref{fig:nebular_stellar}). Measuring the Balmer jump in this manner would require high-resolution spectroscopy, to avoid blending the higher order Balmer lines and e.g.\@ \OII\ emission with the rest-frame optical continuum level. In the case of low-resolution PRISM spectroscopy or photometry, a longer wavelength baseline is needed to avoid this blending of emission lines and continuum (see Fig.~\ref{fig:GS-9422}). 

Hence we show a more practical measure of the Balmer jump $m_{3000} - m_{5750}$ in the right panel of Fig.~\ref{fig:jump_temperature}. Here 3000~\AA\ corresponds to the typical wavelength probed by the first NIRCam wide-band filter fully blueward of the Balmer limit, and 5750~\AA\ corresponds roughly to the midpoint of the wavelength range between \OIII\ $\lambda5007$ and \Ha, which is relatively devoid of strong emission lines, hence offering a measure of the rest-frame optical continuum level via medium-band photometry. Interestingly, the magnitude shift $\Delta m$ associated with our measure of the Balmer jump now actually increases with increasing two-photon continuum contribution. Due to the longer wavelength baseline, this measure is sensitive to both the immediate discontinuity (left panel), as well as the UV--optical slope. The smaller discontinuity is more than compensated for by the flatter slopes in the UV and optical.

Depending on the source redshift, the \ion{He}{I} $\lambda 5876$ line may contribute a photometric boost to the medium-band filter used to assess the rest-frame optical continuum level. While weak \citep[maximally ${\sim}100~$\AA\  rest-frame equivalent width,][]{Zackrisson2011, Nakajima2022}, it can still lead to an overestimation of the continuum level (and thus an underestimation of the Balmer jump) by up to ${\sim}0.16$~mag for most medium-band filters, and at most ${\sim}0.32$~mag for the narrowest filters (F430M, F460M).

Balmer-jump measurements using NIRCam photometry (pentagons) and NIRSpec PRISM spectroscopy (squares) are shown for the nebular-dominated candidates identified by \citet{Katz2025} in Fig.~\ref{fig:jump_temperature} (the displayed temperatures are from their fitting to the PRISM spectra). These candidates are GS-9422 (green), 2198\_7807 ($z=5.347$, cyan), 1210\_5217 ($z=4.888$, blue), and two regions of a $z=3.99$ strongly gravitationally lensed arc each thought to exhibit pure nebular continuum: 2561\_17467 (light purple) and 2756\_301 (dark purple). The photometric Balmer jump measurement corresponds to the magnitude difference $m_\mathrm{b,UV} - m_\mathrm{r,opt}$ between the first filter fully blueward of the Balmer jump and the appropriate medium-band filter between \OIII\ $\lambda 5007$ and \Ha\ (to be discussed in more detail in the next section, though we determine the Balmer jump indirectly for 2198\_7807, see Appendix~\ref{app:indirect}). The spectroscopic Balmer jump measurements are obtained by determining the bandpass-averaged flux densities in the same filters using the NIRSpec PRISM spectrum, so accounting for the photometric boost by \ion{He}{I} $\lambda5876$ if in the filter.  The candidates exhibit a wide range in Balmer jump strengths, with photometric measurements between $-0.712 \leq \Delta m_\mathrm{jump} \leq -0.146$. The key point we wish to make is that nebular-dominated galaxies, possibly powered by star formation following a top-heavy IMF, need not necessarily have the largest Balmer jumps. Although they are thought to lack a substantial stellar continuum component (thus increasing the Balmer jump), the massive, metal-poor star formation could lead to high \ion{H}{II} region temperatures (reducing the Balmer jump). Indeed, using the \citet{Nakajima2022} Pop III models, we find that the Balmer jump measured in this way can be as small as $-0.15$ mag, due to the \ion{He}{I} $\lambda 5876$ photometric boost and a red UV slope. 

Motivated by the smallest Balmer jumps seen in current observations of nebular-dominated galaxy candidates \citep{Cameron2024, Katz2025}, as well as nebular-dominated (i.e.\@ Pop III) models \citep{Zackrisson2011, Nakajima2022}, we thus set our minimal Balmer jump threshold as $\Delta m_\mathrm{jump} < -0.15$. Our selection of nebular-dominated candidates via the preliminary Balmer-jump cut may still not be complete, due to photometric errors on the Balmer jump measurement (especially driven by the medium-band continuum measurement, which can cause galaxies to scatter out of the selection window, e.g. 2198\_7807 upon reprocessing of the NIRCam data). 

We note that there is not always good agreement between the photometric and spectroscopic Balmer jump measurements. This may stem from slightly different extraction apertures for the photometry or spectroscopy or it may be attributable to the NIRSpec flux calibration and/or noise in the rest-frame optical part of the PRISM spectrum where the SNR is lower. In particular, we note that for GS-9422 (Fig.~\ref{fig:GS-9422}) the photometric and spectroscopic measurements of the continuum appear very similar. However, F410M, which we use in our Balmer-jump measurement, is slightly boosted by \ion{He}{I} $\lambda 5876$. After accounting for this, the spectroscopic Balmer-jump measurement is ${\sim}0.20$~mag larger than the photometric measurement (see Fig.~\ref{fig:jump_temperature}).

\subsubsection{Balmer jump colour selection} \label{subsubsec:colour_selection}

We determine the Balmer jump by using a medium-band filter to measure the rest-frame optical continuum level $m_\mathrm{r, opt}$ between \OIII $\lambda5007$ and \Ha. We adopt this spectral range as it is relatively devoid of prominent emission lines, with the exception of \ion{He}{I} $\lambda5876$ \citep[$\mathrm{EW_{rest}}$ maximally ${\sim}100$~\AA,][]{Zackrisson2011, Nakajima2022}. In principle the spectral range between \Ha\ and [\ion{S}{III}] $\lambda 9069$ could also be adopted, with the continuum level even accessible to wide-band filters \citep{Nishigaki2023}. The spectral range between the Balmer limit (3646~\AA) and \Hb\ is unsatisfactory, as the dense network of (weak) emission lines means that a medium-band measurement (and certainly a wide-band measurement) will always be substantially photometrically boosted (in the case of a star-forming galaxy) above the actual continuum level. However, we note that the continuum between \Hg\ and \Hb\ is accessible by the narrowest medium-band filters (F250M, F430M, F460M, F480M), enabling the Balmer jump to be measured photometrically out to $z=9.6$.  

The redshift range of applicability for each medium-band filter is set by when \Ha\ redshifts out of the spectral range (yielding the redshift lower limit) and when \OIII\ $\lambda 5007$ redshifts into the spectral range (yielding the redshift upper limit) of the filter. We define the filter spectral range by the wavelength interval where the throughput is more than 2.5~per~cent of the maximum throughput. We set this very low throughput threshold to ensure that the continuum measurement is not affected by \OIII\ $\lambda 5007$ or \Ha\ line emission, even in the case of maximally strong emission ($\mathrm{EW_{rest}} \sim 4000$~\AA).

Given the narrow nature of the medium bands, it is possible that multiple bands simultaneously reside in the wavelength interval between \OIII $\lambda 5007$ and \Ha. Hence the redshift ranges of applicability somewhat overlap. To ensure a unique filter choice for a given redshift, we prioritise the filter with the greatest sensitivity and area coverage, in this case F335M and F410M from the deep extensive JADES imaging.

We determine the rest-frame UV continuum level $m_\mathrm{b, UV}$ using the first filter fully blueward of the Balmer limit (again using the 2.5~per~cent of maximum throughput threshold to define the filter spectral range). Generally a wide-band filter is used. Though if the data quality is comparable in a medium-band filter, with the noise $\sigma_\mathrm{med}$ at most $1.25\times$ that in the wide band $\sigma_\mathrm{wide}$, the medium-band filter is adopted to allow for a UV continuum measurement closer to the Balmer limit. 

The Balmer jump $\Delta m_\mathrm{jump}$ is thus defined as the magnitude difference between the continuum levels blueward and redward of the Balmer limit:
\begin{equation} \label{eq:jump}
\Delta m_\mathrm{jump} = m_\mathrm{b, UV} - m_\mathrm{r, opt}.
\end{equation}
\noindent Negative $\Delta m_\mathrm{jump}$ values correspond to a lower-lying rest-frame optical continuum level, indicating a Balmer jump and/or blue UV--optical slope, while positive values indicate an elevated optical continuum level, indicating a Balmer break and/or red UV--optical slope. As motivated earlier (the smallest Balmer jump seen empirically in current known nebular-dominated galaxy candidates, and in current models of Pop III galaxies), our Balmer-jump galaxy candidates satisfy:
\begin{equation} \label{eq:jump_cut}
\Delta m_\mathrm{jump} < -0.15.
\end{equation}

\subsubsection{Balmer-jump galaxy candidates} \label{subsubsec:bj_candidates}

\begin{figure*}
\centering
\includegraphics[width=.475\linewidth] {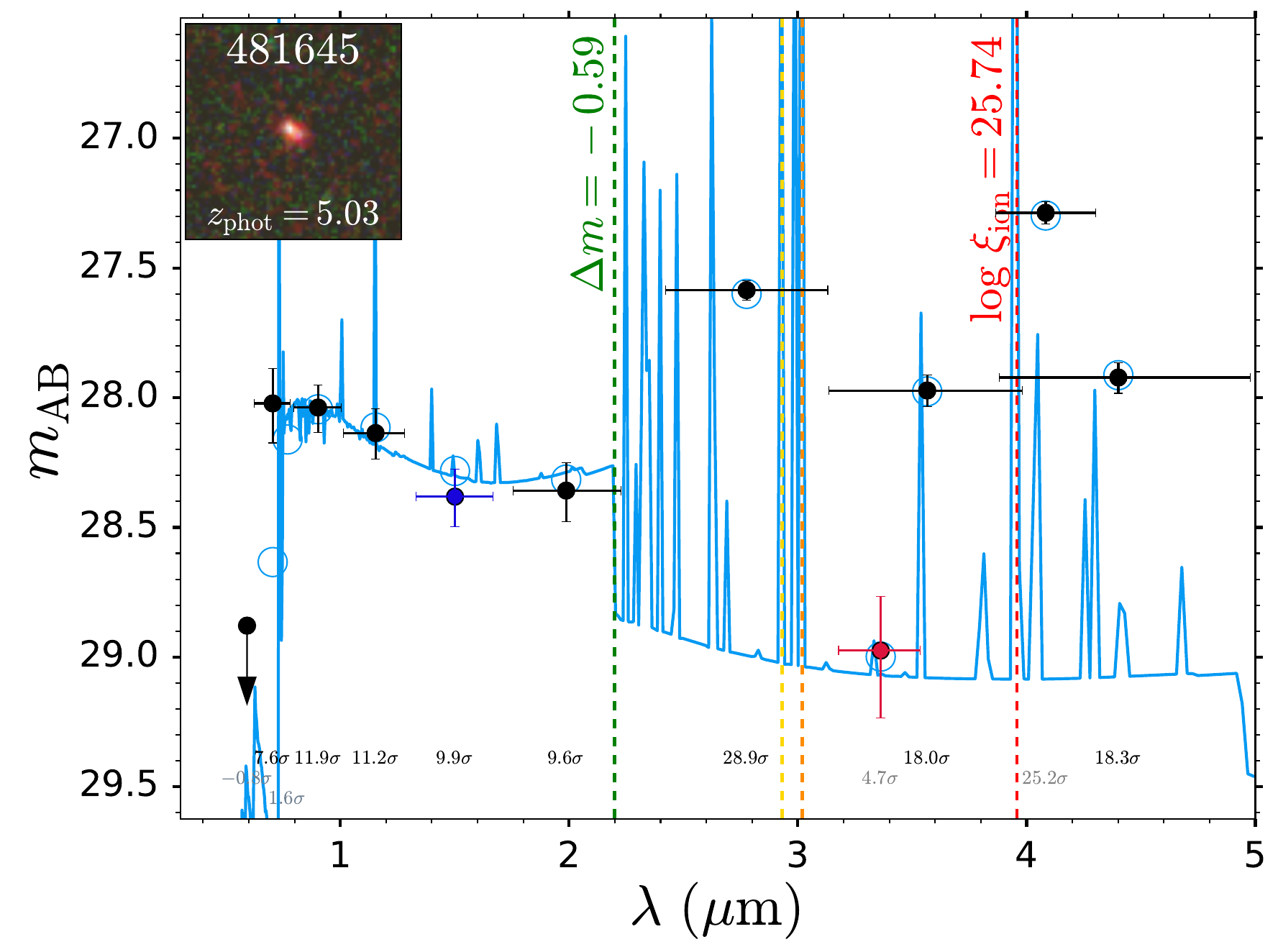} \hfill
\includegraphics[width=.475\linewidth]{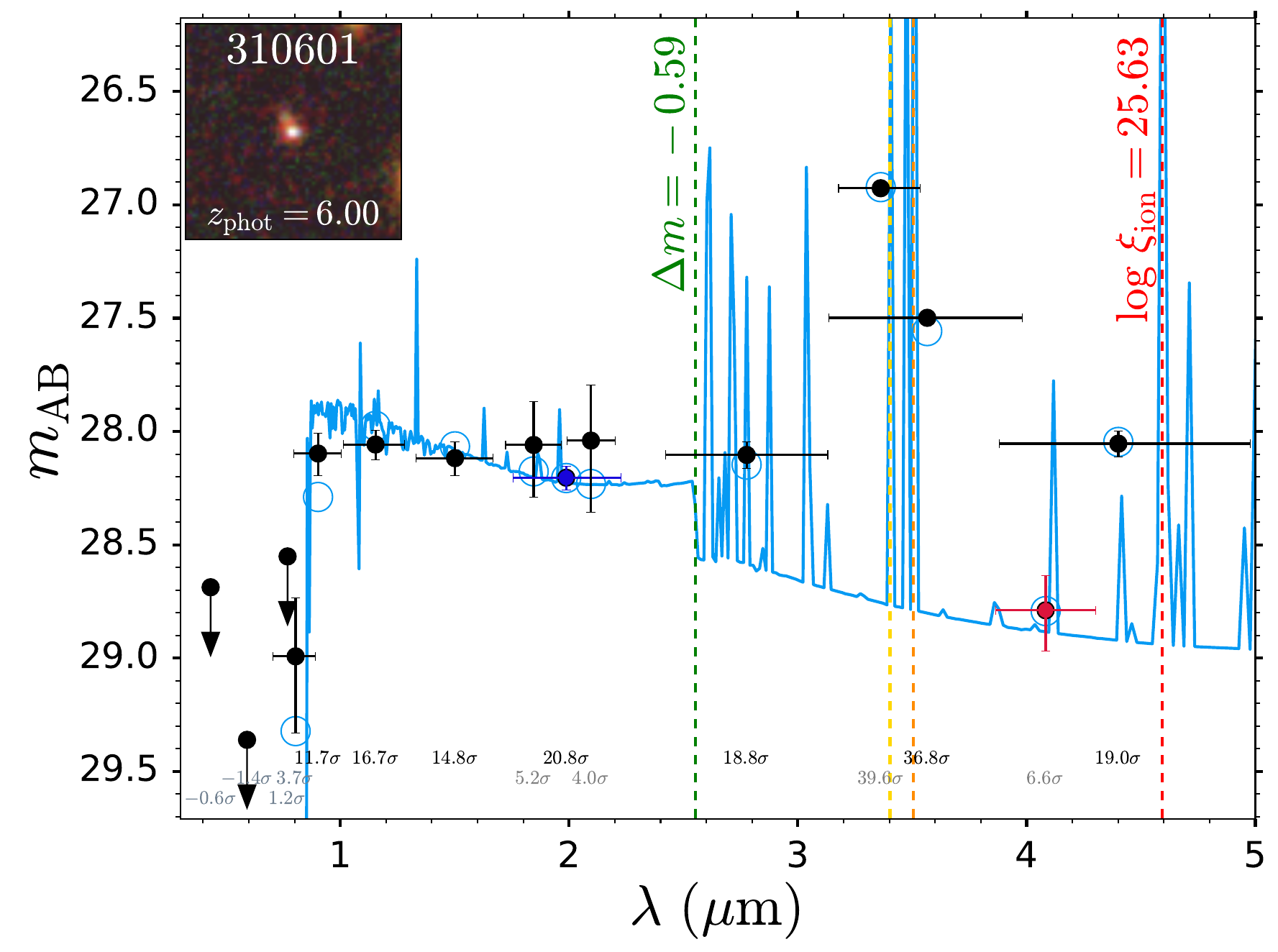} \\[4.5ex]
\includegraphics[width=.475\linewidth] {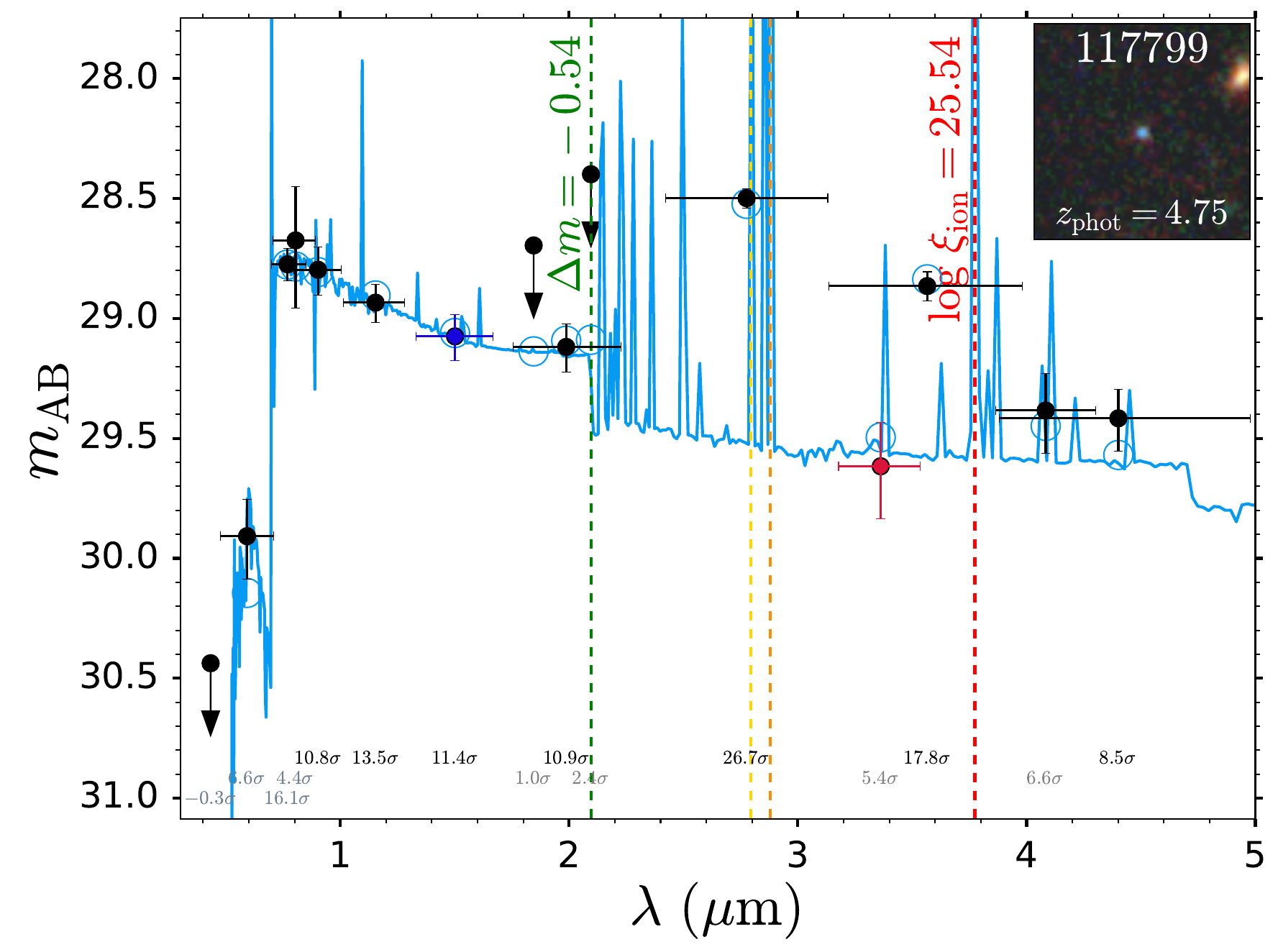} \hfill
\includegraphics[width=.475\linewidth]{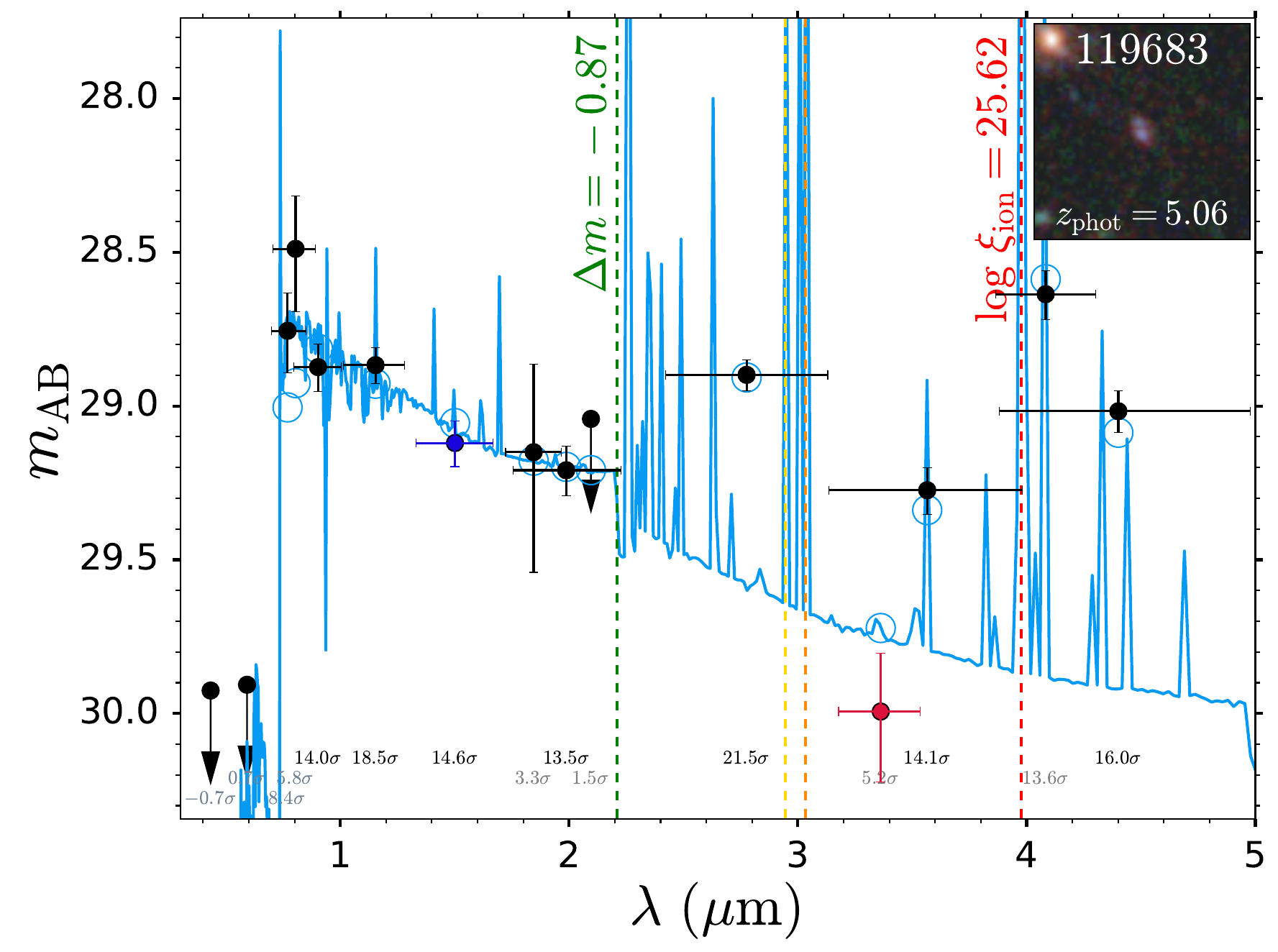} \\[4.5ex]
\includegraphics[width=.475\linewidth] {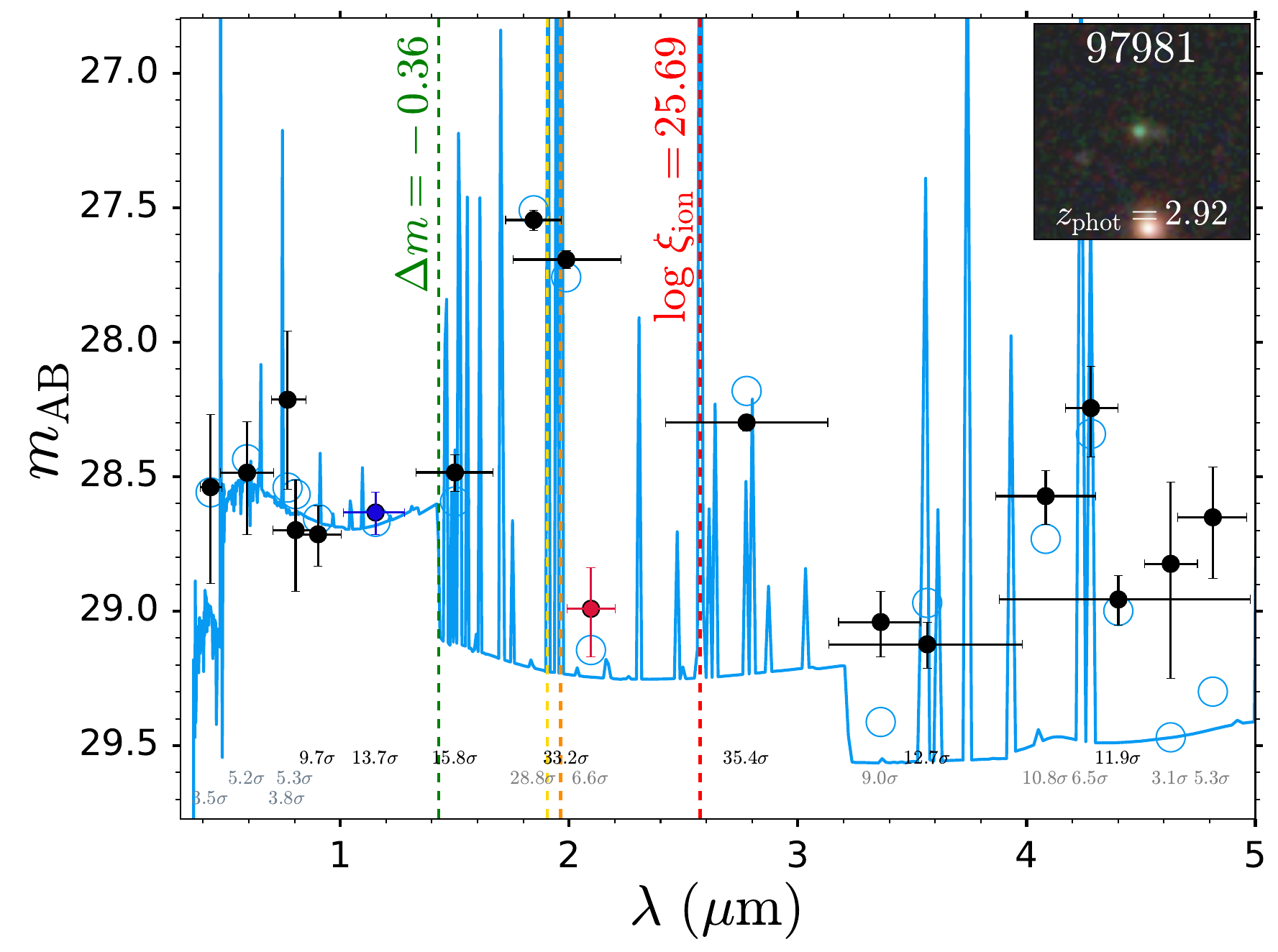} \hfill
\includegraphics[width=.475\linewidth]{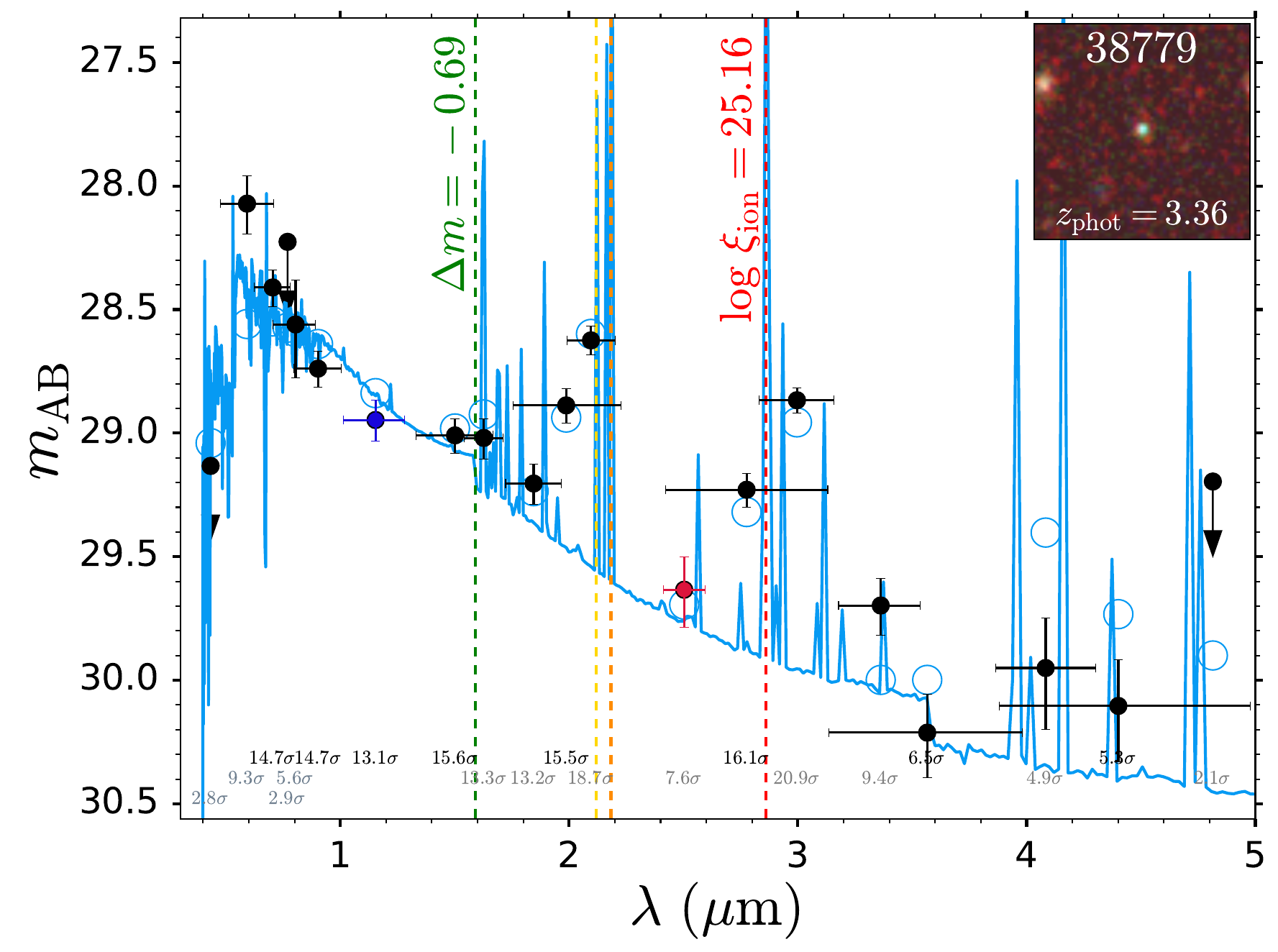} 
\caption{SEDs of Balmer-jump galaxy candidates, selected as having UV--optical colours $\Delta m_\mathrm{jump} < -0.15$. Bagpipes fits to the full photometry are shown in light blue. As in Fig.~\ref{fig:GS-9422}, the RGB (F444W, F200W, F115W) cutout spans $2\times 2$~arcsec. Left panels: Candidates where the UV--optical flux density gap is assessed to mostly be due to the Balmer jump (through visual inspection of the Bagpipes fit). Right panels: Candidates where the UV--optical flux density gap is due to a combination of a Balmer jump and blue slope (top and middle), and a blue slope (bottom).} 
\label{fig:bj_candidates}
\end{figure*}

\begin{figure*}
\centering
\includegraphics[width=\linewidth]{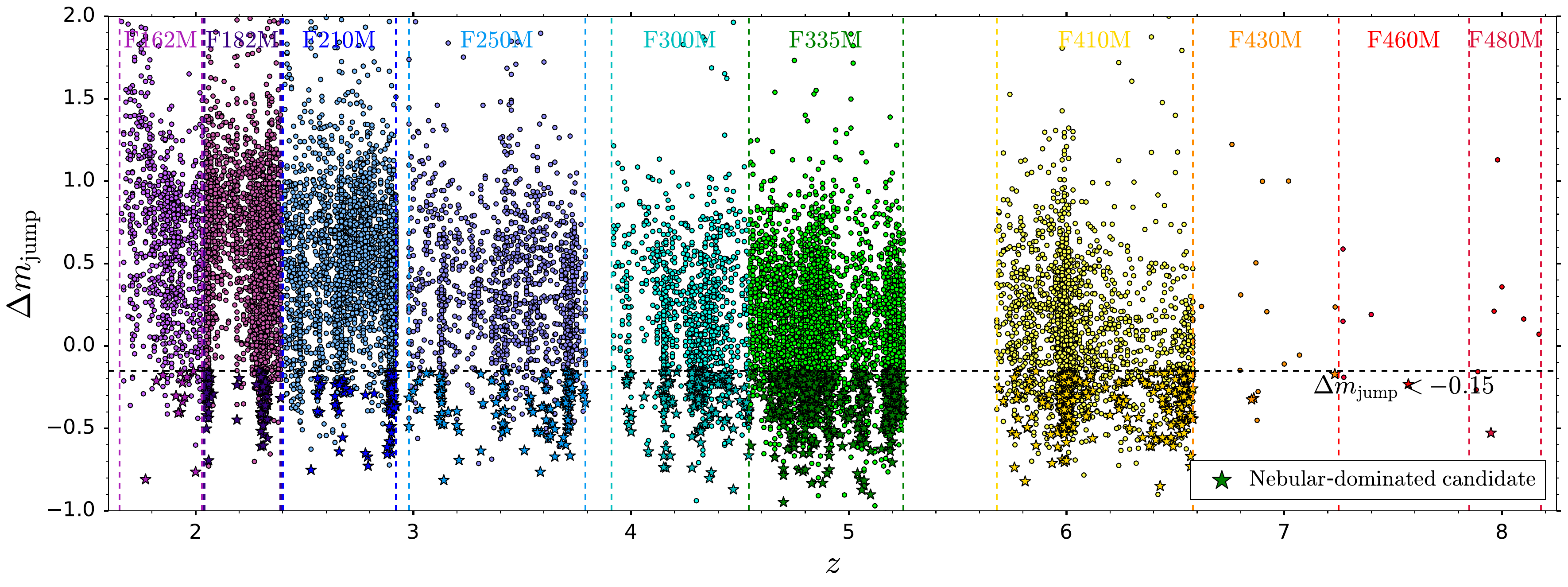}
\caption{The distribution of sources in the $\Delta m_\mathrm{jump}$--$z$ plane, colour-coded by the medium-band filter used to measure the rest-frame optical continuum level. Vertical dashed lines indicate the redshift range of applicability for each medium band. Note that the depth and area covered by the various medium bands varies, hence why there are e.g.\@ a lack of sources associated with the redshifts spanned by the shallower, limited coverage of the F430M, F460M and F480M filters. Balmer-jump candidates are below the horizontal dashed line, with $\Delta m_\mathrm{jump} < -0.15$. Galaxies that further satisfy our nebular-dominated criteria (high $\xi_\mathrm{ion, obs}$ and non-blue UV slopes) are nebular-dominated candidates (stars).}
\label{fig:jump_redshift}
\end{figure*}

We show example Balmer-jump galaxy candidates selected via our Balmer jump colour selection procedure in Fig.~\ref{fig:bj_candidates}. The filters probing the continuum levels blueward $m_\mathrm{b,UV}$ and redward $m_\mathrm{r,opt}$ of the Balmer jump are coloured blue and red, respectively. The Balmer jump $\Delta m_\mathrm{jump}$ is also denoted (green). The Bagpipes fit to the full HST+NIRCam photometry is shown in light blue. 

We stress that these are Balmer-jump candidates, identified through the gap in flux density between the UV and optical, using just two filters. It is possible that this gap is attributable to an immediate discontinuity (the Balmer jump), a blue UV--optical slope, or a combination of the two. Hence not all of our candidates likely actually exhibit a true Balmer jump. From the context of the full HST+NIRCam photometry, the Bagpipes fits suggest that the sources in the left panels have Balmer jumps (assessed via visual inspection), while the UV--optical gap for the sources in the right panels is due to a combination of a Balmer jump and blue slope (top and middle), and a blue slope (bottom).

We show the full distribution of sources in the $\Delta m_\mathrm{jump}$--$z$ plane in Fig.~\ref{fig:jump_redshift}, colour-coded by the medium-band filter used to probe the rest-frame optical continuum level. Galaxies below the horizontal dashed line are Balmer-jump candidates (2646 in total) with $\Delta m_\mathrm{jump} < -0.15$. Galaxies that further satisfy our nebular-dominated selection (discussed in the next section) are nebular-dominated galaxy candidates (displayed as stars).

\subsection{Nebular-dominated selection} \label{subsec:nd_selection}

Having identified Balmer-jump candidates through their gap in continuum flux density between the UV and optical, we now apply further selection criteria to identify possible nebular-dominated galaxy candidates. In principle, one would aim to seek the imprint of the two-photon downturn on the photometry, being the defining feature of nebular-dominated emission. In practice, this is challenging, as we outline in Appendix~\ref{app:deficit}. Thus we instead select based off two further nebular-dominated features: a high $\xi_\mathrm{ion, obs}$ (Section~\ref{subsubsec:xi_selection}) and a relatively non-blue UV slope (Section~\ref{subsubsec:uv_selection}). We showcase example nebular-dominated galaxy candidates in Section~\ref{subsubsec:nd_candidates}.

\subsubsection{$\xi _\mathrm{ion, obs}$ selection} \label{subsubsec:xi_selection}

As discussed earlier, for the nebular continuum to dominate, a high stellar ionising photon production efficiency $\xi_\mathrm{ion,*}$ is required \citep{Cameron2024, Katz2025}. Hence we apply a $\xi_\mathrm{ion,obs}$ cut to identify possible nebular-dominated candidates. 

From the medium band tracing the rest-frame optical continuum level $m_\mathrm{r,opt}$, together with the wide band whose bandpass-averaged flux density $m_\mathrm{H\alpha}$ is boosted by \Ha, we can determine the \Ha\ flux $F_\mathrm{H\alpha}$. This can be crudely estimated using the approximate formula $\Delta m = -2.5\log_{10}(1 + \mathrm{EW_{rest}(1+z)/\Delta \lambda})$, where $\Delta m = m_\mathrm{H\alpha} - m_\mathrm{r, opt}$, and $\Delta \lambda$ is the width of the wide-band filter. However, this formula does not take the filter throughput into account, nor the fact that the NIRCam detectors count electrons, rather than measure energy, so introducing unnecessary systematics. We therefore use the relation:

\begin{equation} \label{eq:boost}
\langle f_{\lambda,\mathrm{H\alpha}} \rangle = \frac{\int f_\lambda (\lambda)\lambda T(\lambda)\mathrm{d}\lambda}{\int \lambda T(\lambda)\mathrm{d}\lambda} = 
\langle f_{\lambda,\mathrm{r, opt}}\rangle + \frac{F_\mathrm{H\alpha}\lambda_\mathrm{H\alpha, obs}T(\lambda_\mathrm{H\alpha, obs})}{\int \lambda T(\lambda)\mathrm{d}\lambda},
\end{equation}

\noindent where $f_\lambda = f_{\lambda, \mathrm{cont}} + f_{\lambda, \mathrm{H\alpha}}$ is the combined continuum+\Ha\ flux density per unit wavelength interval, $\langle f_\lambda \rangle$ is the bandpass-averaged flux density per unit wavelength interval, $T(\lambda)$ is the throughput of the wide-band filter, and $\lambda_\mathrm{H\alpha, obs}$ is the observed wavelength of \Ha. Given the relatively flat optical slope for nebular-dominated emission (see Fig.~\ref{fig:nebular_continuum}), we assume that the source spectrum is flat in $f_\nu$, i.e.\@ $\langle f_{\lambda,\mathrm{r, opt}}\rangle = \langle f_{\nu,\mathrm{r, opt}}\rangle \frac{c}{\lambda^2_\mathrm{pivot,H\alpha}}$, where $\lambda_\mathrm{pivot,H\alpha}$ is the pivot wavelength of the wide-band filter covering \Ha. Using the best-fit photometric redshift we subsequently determine the \Ha\ luminosity $L_\mathrm{H\alpha}$. We estimate the ionising photon production rate $\dot{N}_\mathrm{ion} = 7.28\times 10^{11} L_\mathrm{H\alpha}$ assuming the standard case-B conversion rate at 10000~K \citep{Osterbrock2006}. We adopt a model-independent approach (as EAZY/Bagpipes may not provide an adequate description of nebular-dominated galaxies) for estimating the monochromatic luminosity density at 1500~\AA, $L_{\nu, 1500}$, using the bandpass-averaged flux density $\langle f_\nu \rangle$ in the first wide-band filter fully redward of Ly$\alpha$. This procedure yields $\xi_\mathrm{ion,obs} = \dot{N}_\mathrm{ion}/L_{\nu,1500}$.

Now the 1500~\AA\ monochromatic\ luminosity $L_{\nu,1500}$ is the sum of the stellar $L_{\nu,1500,*}$ and nebular $L_{\nu,1500,\mathrm{neb}}$ components, i.e.\@ $L_{\nu,1500} = L_{\nu,1500,*} + L_{\nu,1500,\mathrm{neb}}$ \citep[see also the discussion in][]{Schaerer2025}. With $\xi_\mathrm{ion, *} = \dot{N}_\mathrm{ion} / L_{\nu,1500,*}$ and defining $\xi_\mathrm{ion, neb} = \dot{N}_\mathrm{ion} / L_{\nu,1500,\mathrm{neb}}$, we have that $\xi_\mathrm{ion, obs} = \dot{N}_\mathrm{ion} / (L_{\nu,1500,*} + L_{\nu,1500,\mathrm{neb}}) \leq \min(\xi_\mathrm{ion, *},\ \xi_\mathrm{ion, neb})$.  This distinction is important because one observationally measures $\xi_\mathrm{ion, obs}$ and not $\xi_\mathrm{ion, *}$. In normal circumstances the nebular component is negligible ($L_{\nu,1500,\mathrm{neb}} \ll L_{\nu,1500,\mathrm{*}}$, see Fig.~\ref{fig:nebular_stellar}), so the total continuum is only slightly larger than the stellar continuum, with $\xi_\mathrm{ion,obs}$ only slightly smaller than $\xi_\mathrm{ion, *}$, i.e.\@ $\xi_\mathrm{ion,obs} \lessapprox \xi_\mathrm{ion, *}$. However, in the nebular-dominated scenario the nebular component can be substantial ($L_{\nu,1500,\mathrm{neb}} \gg L_{\nu,1500,*}$), so the total continuum is only slightly larger than the nebular continuum, with $\xi_\mathrm{ion,obs}$ only slightly smaller than $\xi_\mathrm{ion, neb}$, i.e.\@ $\xi_\mathrm{ion,obs} \lessapprox  \xi_\mathrm{ion, neb} \ll \xi_\mathrm{ion, *}$. 

\begin{figure*}
\centering
\includegraphics[width=.7\linewidth]{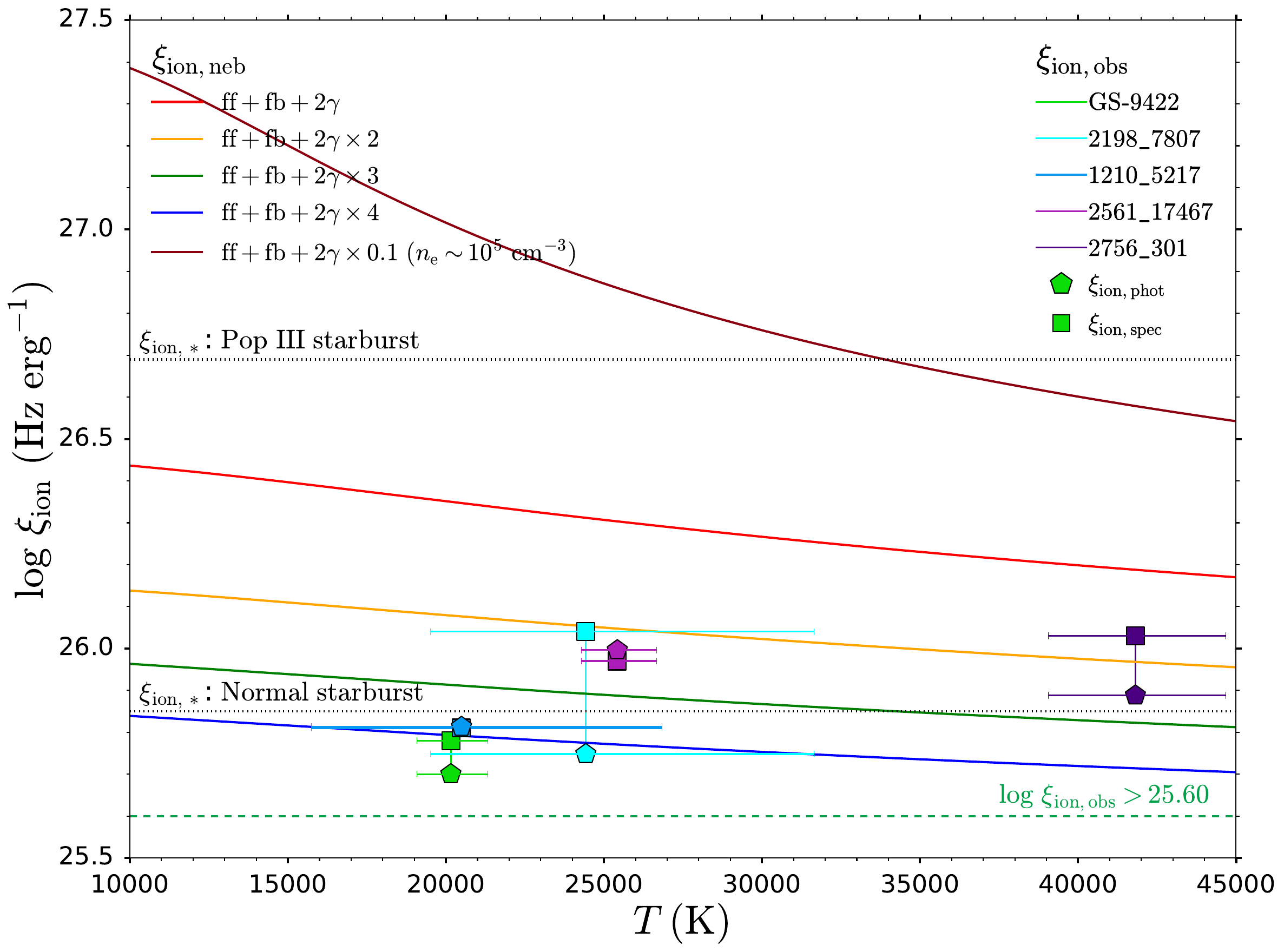}
\caption{For normal starbursts (1~Myr old starburst from Bagpipes shown, dotted horizontal line), due to the relatively low stellar ionising photon production efficiency $\xi_\mathrm{ion, *} = \dot{N}_\mathrm{ion}/L_{\nu, 1500, *}$, the stellar continuum at 1500~\AA\ $L_{\nu, 1500, *}$ is substantially brighter than the nebular continuum $L_{\nu, 1500, \mathrm{neb}}$, so the observed ionising photon production efficiency $\xi_\mathrm{ion,obs} = \dot{N}_\mathrm{ion} / L_{\nu, 1500}  \lessapprox \xi_\mathrm{ion_*}$. With star formation following a top-heavy IMF \citep[here a 1~Myr old Pop III.1 starburst from][]{Zackrisson2011}, due to the much larger $\xi_\mathrm{ion, *}$, the nebular continuum dominates and $\xi_\mathrm{ion, obs} \lessapprox \xi_\mathrm{ion, neb} = \dot{N}_\mathrm{ion}/L_{\nu, 1500, \mathrm{neb}}$. Given the high \ion{H}{II} region temperatures and collisionally-enhanced two-photon continuum emission expected to accompany star formation following a top-heavy IMF, the $\xi_\mathrm{ion, obs}$ (photometric: pentagons; spectroscopic: squares) for nebular-dominated galaxy candidates \citep{Katz2025} is not substantially larger than that expected for normal starbursts. Hence we set our nebular-dominated selection threshold: $\log\, (\xi_\mathrm{ion, obs} /\mathrm{(Hz\ erg^{-1})})> 25.60$ (green dashed horizontal line).}
\label{fig:xi_ion_temperature}
\end{figure*}

As shown in Fig.~\ref{fig:xi_ion_temperature}, for normal starbursts with a regular IMF \citep[e.g.\@][]{Salpeter1955, Kroupa2001, Chabrier2003}, $\log\, (\xi_\mathrm{ion, *} /\mathrm{(Hz\ erg^{-1})})$ ($\approx 25.85$) is substantially smaller than $\log\, \xi_\mathrm{ion, neb}$ with regular two-photon continuum emission ($2\gamma\times 1$, $\approx 26.4$), i.e.\@ the stellar continuum at 1500~\AA\ is approximately 3--4$\times$ larger than the nebular continuum, so the total continuum and thus $\log\, (\xi_\mathrm{ion, obs} /\mathrm{(Hz\ erg^{-1})})$ ($\approx 25.75$) is only slightly smaller than $\log\, \xi_\mathrm{ion, *}$. For a starburst following a top-heavy IMF, $\log\, (\xi_\mathrm{ion, *} /\mathrm{(Hz\ erg^{-1})})$ \citep[$\approx 26.70$ for a Pop III.1 IMF from][]{Zackrisson2011} can be substantially larger than $\log\, \xi_\mathrm{ion, neb}$, especially if the two-photon continuum emission is collisionally enhanced, i.e.\@ the nebular continuum dominates and $\log\, \xi_\mathrm{ion, obs}$ is only slightly smaller than $\log\, \xi_\mathrm{ion, neb}$. 

In other words, $\xi_\mathrm{ion,obs}$ does not become arbitrarily large with an increasingly more top-heavy IMF (i.e.\@ does not indefinitely scale with increasing $\xi_\mathrm{ion,*}$), but instead saturates at the nebular value $\xi_\mathrm{ion,neb}$ \citep[for more details, see also the helpful discussions in][]{Katz2025, Schaerer2025}.  With the higher \ion{H}{II} region temperatures expected to accompany top-heavy, metal-poor star formation, as well as the collisionally-enhanced two-photon continuum emission (both of which decrease $\xi_\mathrm{ion,neb}$), the limiting nebular $\log\, (\xi_\mathrm{ion, obs} /\mathrm{(Hz\ erg^{-1})})$ value ($\approx 26.0$, depending on temperature and collisional enhancement) is not substantially larger than the $\log\, \xi_\mathrm{ion, obs}$ values expected for normal starbursts ($\approx 25.75$), as also discussed in \citet{Katz2025} and \citet{Schaerer2025}. Hence observationally, one unfortunately deduces relatively comparable $\xi_\mathrm{ion, obs}$ for these two extremely different scenarios. It is only in the case of star formation following a top-heavy IMF that is radiating into high density gas ($n_\mathrm{e} \gg 10^4~\mathrm{cm}^{-3}$, dark red in Fig.~\ref{fig:xi_ion_temperature}) that $\xi_\mathrm{ion, obs}$ can become arbitrarily large. 

We show the spectroscopic \citep[pentagon, reported in][]{Katz2025} and photometric $\xi_\mathrm{ion, obs}$ (square, derived using the procedure outlined above) values for GS-9422 (light green) in Fig.~\ref{fig:xi_ion_temperature}, as well as the other nebular-dominated spectroscopic candidates reported by \citet{Katz2025}. Due to the possibly considerable collisional enhancement of two-photon continuum emission in GS-9422, the spectroscopic (25.78) and photometric (25.70) $\log\,(\xi_\mathrm{ion, obs} /\mathrm{(Hz\ erg^{-1})})$ values are not exceptionally large (though still high compared to most galaxies). Motivated by this, we set our nebular-dominated selection threshold: 
\begin{equation} \label{eq:xi_ion}
\log\, (\xi_\mathrm{ion, obs} /\mathrm{(Hz\ erg^{-1})}) > 25.60.
\end{equation}
\noindent We allow for some buffer with respect to the measured photometric value to account for photometric measurement uncertainties, as well as accommodating even stronger collisional enhancement of the two-photon continuum emission. Thus our nebular-dominated $\xi_\mathrm{ion, obs}$ cut is readily achievable by normal starbursts (especially if dust attenuation is present, inflating the $\xi_\mathrm{ion, obs}$ values), and is not unique to top-heavy starbursts \citep{Simmonds2023, Simmonds2024, Simmonds2024b, Boyett2024, Rinaldi2024, Saxena2024, Begley2025, Laseter2025, Papovich2026}\footnote{See also the tabulated list of $\xi_\mathrm{ion, *}$ (stellar continuum only) and $\xi_\mathrm{ion, obs}$ (stellar+nebular) values for a range of IMFs in \citet{Schaerer2025}.}. Hence we cannot guarantee that our nebular-dominated candidates will actually exhibit a two-photon continuum turnover in their spectra. 

We note that we do not correct our $\xi_\mathrm{ion,obs}$ values for possible dust attenuation effects. This is because the inferred dust correction may depend on the IMF assumed. SED-fitting is often used to estimate the amount of dust reddening from the observed photometry \citep[e.g.\@][]{Shivaei2018, Pahl2025, Pahl2026}, based on how much redder the continuum appears relative to the dust-free scenario. As we shall discuss in Section~\ref{subsubsec:uv_selection}, the unreddened UV slopes for young starbursts following a regular IMF \citep[e.g.\@][]{Salpeter1955, Kroupa2001, Chabrier2003} are generally bluer than for nebular-dominated galaxies. Hence the amount of dust reddening inferred to match a given observed UV slope will be greater than in the nebular-dominated case. Since we are considering the possibility of the nebular-dominated scenario, but likely cannot adequately distinguish between dust-reddened starbursts and nebular-dominated galaxies using photometry alone, we apply no dust correction to the $\xi_\mathrm{ion, obs}$ values. The development of nebular-dominated spectral templates would facilitate such dust corrections in future work.

\begin{figure*}
\centering
\includegraphics[width=\linewidth]{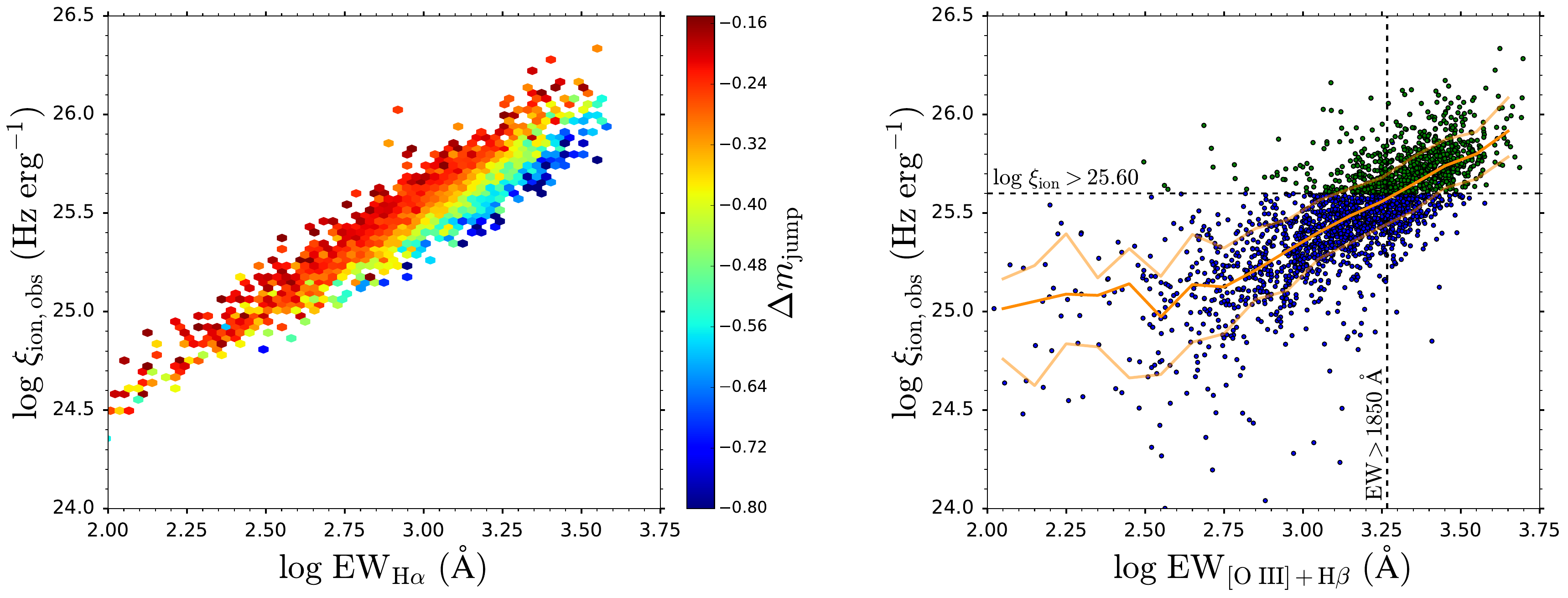}
\caption{Left panel: the observed ionising photon production efficiency $\xi_\mathrm{ion, obs}$ is clearly correlated with the \Ha\ equivalent width $\mathrm{EW_{H\alpha}}$(with both parameters determined empirically from the photometry) for our sample of Balmer-jump candidates at $z < 6.6.$ The scatter in the relation is mostly explained by variations in the Balmer jump $\Delta m_\mathrm{jump}$ (colour coding). Right panel: We find no such clear trend to describe the scatter in the correlation between $\xi_\mathrm{ion, obs}$ and the \OIII\ $\lambda\lambda 4959,5007$ + \Hb\ equivalent width $\mathrm{EW_{[O\ III] + H\beta}}$ (so colour-coding by Balmer jump not shown), which is perhaps to be expected given the additional dependence of \OIII\ emission on metallicity and ionisation parameter. At $z > 6.6$, \Ha\ is redshifted out of the NIRCam spectral range, so we instead use $\mathrm{EW_{[O\ III] + H\beta}}$ as a proxy for $\xi_\mathrm{ion, obs}$ in our nebular-dominated selection. As $z < 6.6$ Balmer-jump candidates start appreciably populating the nebular-dominated regime ($\log\,(\xi_\mathrm{ion, obs} /\mathrm{(Hz\ erg^{-1})}) > 25.60$, green) at $\mathrm{EW_{[O III] + H\beta}} \gtrsim 1850$~\AA, we use this equivalent width cut at $z > 6.6$.}
\label{fig:xi_ion_ew}
\end{figure*}

At $z > 6.6$, \Ha\ is redshifted out of the NIRCam spectral range, so $\xi_\mathrm{ion, obs}$ cannot be estimated following the above procedure. Instead, we rely on using the combined equivalent width $\mathrm{EW_{[O\ III] + H\beta}}$ of \OIII\ $\lambda\lambda 4959,5007$ and \Hb\ as a tracer of $\xi_\mathrm{ion, obs}$ \citep{Chevallard2018, Tang2019, Izotov2021, Boyett2024, Simmonds2024, Laseter2025}. As shown in Fig.~\ref{fig:xi_ion_ew} (left panel), there is a clear correlation between $\xi_\mathrm{ion, obs}$ and $\mathrm{EW_{H\alpha}}$ \citep[see also][]{Tang2019, Izotov2021, Laseter2025} for Balmer-jump galaxies, with the scatter mostly being driven by the range of relative continuum levels in the UV and optical, traced by the Balmer jump $\Delta m_\mathrm{jump}$. For a given \Ha\ luminosity, increasing the Balmer jump reduces the optical continuum level, thereby increasing $\mathrm{EW_{H\alpha}}$ for a given $\xi_\mathrm{ion, obs}$. However, we find no such clear trend to describe the scatter in the correlation between $\xi_\mathrm{ion, obs}$ and $\mathrm{EW_{[O\ III] + H\beta}}$ (right panel), where the solid orange and light lines correspond to the median and (16,84) percentiles, respectively. This should not come as a surprise, as the \OIII\ emission \citep[which tends to dominate over \Hb,][]{Matthee2023, Meyer2024} introduces additional dependencies on metallicity and ionisation parameter, complicating the scatter compared to what was seen with \Ha. 

Here we estimate $\mathrm{EW_{[O\ III] + H\beta}}$ following a similar procedure to \Ha\ (using Eq.~\ref{eq:boost}), now using the wide-band filter boosted by \OIII\ $\lambda 5007$, also accounting for the contributions from \OIII\ $\lambda 4959$ and \Hb\ \citep[assuming \OIII\ $\lambda 5007 / \mathrm{H}\beta = 5$, which is comparable to the line ratio typically seen in high-redshift galaxies,][]{Matthee2023, Meyer2024}. We note that a line ratio needs to be assumed to account for and weight correctly the filter throughputs at the \OIII\ and \Hb\ lines, though generally the ratio assumed has a very negligible effect on the equivalent width determined, unless the lines are near a filter edge.

From the observed correlation between $\mathrm{EW_{H\alpha}}$ and $\xi_\mathrm{ion, obs}$, $\mathrm{EW_{H\alpha}}\approx 1200$~\AA\ roughly corresponds to a median $\log\, (\xi_\mathrm{ion, obs} /\mathrm{(Hz\ erg^{-1})}) = 25.60$ (without further accounting for the $\Delta m_\mathrm{jump}$ dependence). Assuming a line flux ratio $\mathrm{H\alpha} / \mathrm{H\beta} = 2.86$, and a flat $f_\nu$ spectrum, this corresponds to $\mathrm{EW_{H\beta}} = 230$~\AA. Further assuming a typical flux ratio \OIII\ $\lambda 5007 / \mathrm{H}\beta = 5$, our nebular-dominated $\xi_\mathrm{ion, obs}$ threshold corresponds $\mathrm{EW_{[O\ III] + H\beta}} = 1850$~\AA. Hence we select nebular-dominated candidates at $z > 6.6$ with:
\begin{equation} \label{eq:ew}
\mathrm{EW_{[O\ III] + H\beta}} > 1850~\textrm{\AA}.
\end{equation}
\noindent At the 1850~\AA\ threshold, we expect (assuming no redshift evolution) $\approx$50~per~cent of these to actually exhibit $\log\, (\xi_\mathrm{ion, obs} /\mathrm{(Hz\ erg^{-1})}) > 25.60$. We acknowledge that extremely metal-poor galaxies (with weak \OIII\ emission and thus low $\mathrm{EW_{[O\ III] + H\beta}}$) that actually have large $\xi_\mathrm{ion, obs}$ would not be selected through this procedure, due to our assumption of \OIII\ $\lambda 5007 / \mathrm{H}\beta = 5$ in determining the EW threshold. We also adopt this procedure to identify nebular-dominated galaxy candidates when \Ha\ resides in the gap between NIRCam wide-band filters. 

We close by briefly noting some challenges with correctly estimating $ \xi_\mathrm{ion, obs}$. Firstly, $L_\mathrm{H\alpha}$ can be underestimated due to the photometric boost by \ion{He}{I} $\lambda 5876$ causing the rest-frame optical continuum level to be overestimated (${\sim}10$~per~cent effect). On the other hand, we assume that \Ha\ solely contributes to the photometric excess seen in its accompanying wide-band filter. However, helium and metal (i.e.\@ [\ion{N}{II}] and [\ion{Ar}{III}]) lines can also contribute, so \Ha\ can be overestimated (${\sim}10$~per~cent). Additionally, the higher \ion{H}{II} region temperatures expected for metal-poor star formation following a top-heavy IMF, demand a larger $L_\mathrm{H\alpha}$ to $\dot{N}_\mathrm{ion}$ conversion factor (${\sim}10$~per~cent), increasing $\xi_\mathrm{ion, obs}$ \citep{Katz2025}. However, just like collisional excitations can enhance the two-photon continuum emission, so too can they enhance \Ha\ emission, through excitation to the $n=3$ state \citep[though at a lesser rate than to $n=2$,][]{Raiter2010}, causing the photoionisation rate to be overestimated (${\sim}30$~per~cent). Finally, strong UV lines (see Fig.~\ref{fig:GS-9422}) can contribute to a photometric excess in the wide-band filter used to determine $L_{\nu,1500}$, causing $\xi_\mathrm{ion, obs}$ to be underestimated (${\sim}20$~per~cent, which might explain the slight discrepancy in the spectroscopic and photometric $ \xi_\mathrm{ion, obs}$ values seen for GS-9422 in Fig.~\ref{fig:xi_ion_temperature}). We will discuss the implications of these strong UV lines in more detail in the next section.

\subsubsection{UV colour selection} \label{subsubsec:uv_selection}

\begin{figure}
\centering
\includegraphics[width=\linewidth]{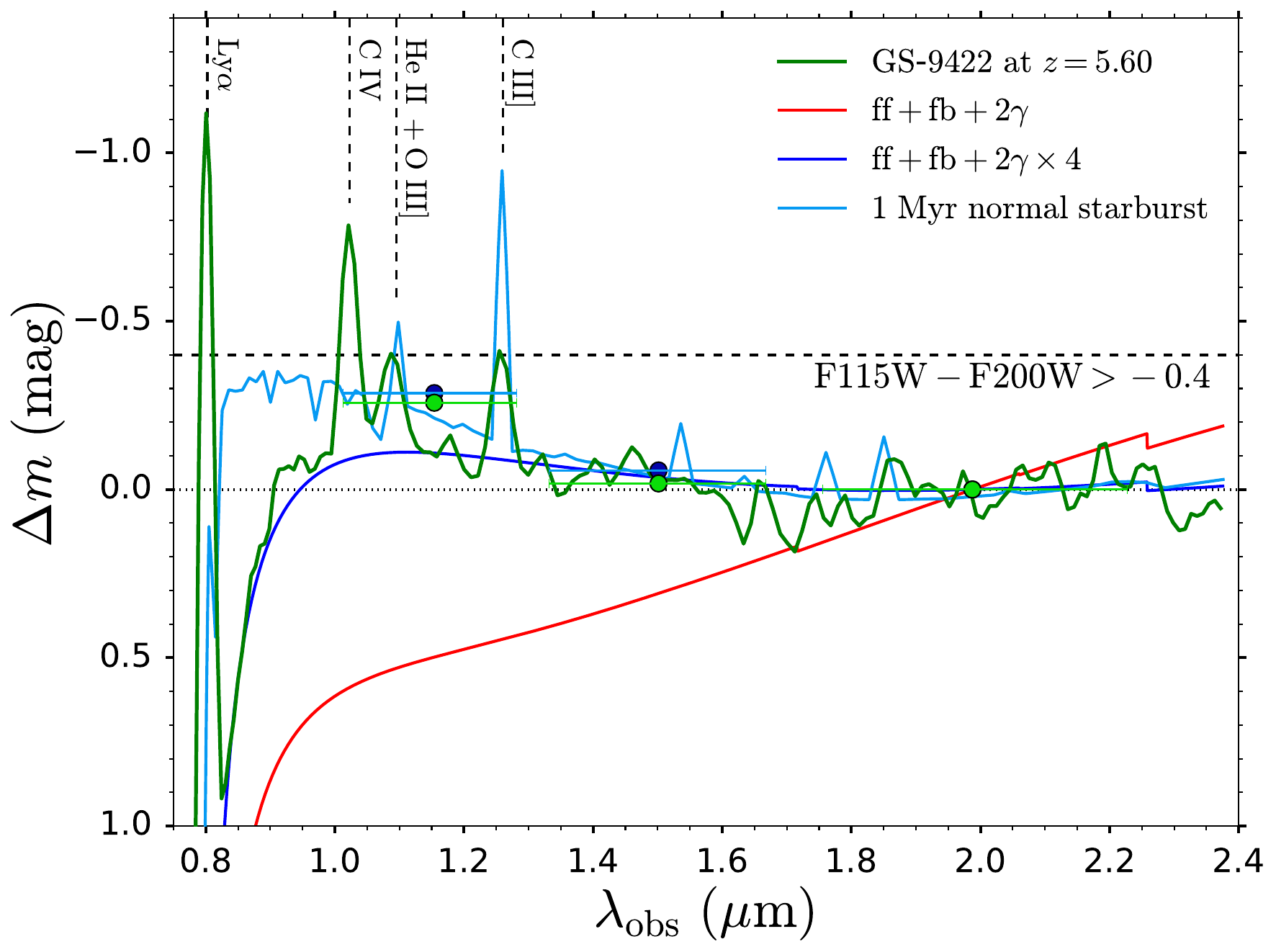}
\caption{The UV slope of a 1~Myr old normal starburst (generated using Bagpipes, light blue) is blue, while the UV slope of a nebular-dominated galaxy \citep[at $T=20156$~K, the inferred temperature for GS-9422,][]{Katz2025} with regular two-photon emission (ff + fb + $2\gamma$, red) is red. This distinction wanes with increasing collisional enhancement of the two-photon continuum emission ($2\gamma \times 4$, dark blue). Moreover, strong UV lines (\ion{C}{IV}, \ion{He}{II}, \ion{O}{III}] and \ion{C}{III}]) are expected \citep{Nakajima2018} for top-heavy starbursts (such as GS-9422, redshifted to $z=5.60$, dark green), which contribute a photometric excess, yielding photometry (light green) that is very similar to that of a normal starburst (dark blue). Thus we apply a relatively conservative UV colour cut (Eq.~\ref{eq:uv_colour}) to select nebular-dominated candidates (serving as more of a consistency check). At $z = 5.60$, this cut is $\mathrm{F115W} - \mathrm{F200W} > -0.4$ (horizontal dashed line).}
\label{fig:uv_colour_selection}
\end{figure}

In addition to demanding a high $\xi_\mathrm{ion, obs}$, we further require our nebular-dominated candidates to have a relatively non-blue UV slope. The reason for this is that under normal star formation \citep[i.e.\@ regular IMF,][]{Salpeter1955, Kroupa2001, Chabrier2003}, the UV slope is rather blue due to the rising stellar continuum that dominates, especially at UV wavelengths well below the Balmer limit, see the 1~Myr starburst (light blue) generated with Bagpipes shown in Fig.~\ref{fig:uv_colour_selection}. On the other hand, the UV continuum for a nebular-dominated galaxy \citep[here at $T = 20156$~K, the inferred temperature for GS-9422,][]{Katz2025} with no collisional enhancement of the two-photon emission (ff + fb + 2$\gamma$, red) is red. With collisional enhancement of the two-photon continuum emission, the UV slope of the nebular continuum flattens (see Fig.~\ref{fig:nebular_continuum}), as the fall of the free--bound and free--free emission is compensated for by the rise of the two-photon continuum emission \citep{Cameron2024}. However, in the case of considerable collisionally-enhanced two-photon continuum emission (e.g.\@ ff + fb + 2$\gamma \times 4$, dark blue), the nebular UV slope becomes comparably blue to that of a normal starburst, weakening the distinction between regular and nebular-dominated galaxies. Note that the continuum of GS-9422 (dark green), here redshifted to $z=5.60$, is rather well-described as having considerable collisional enhancement ($2\gamma \times 4$). 

In addition, GS-9422 exhibits strong UV emission lines \citep{Cameron2024}: \ion{C}{IV} $\lambda\lambda 1548,1550$, \ion{He}{II} $\lambda 1640$ + \ion{O}{III}] $\lambda\lambda 1661, 1666$ and \ion{C}{III}] $\lambda 1907, 1909$. The \ion{C}{IV} and \ion{He}{II} + \ion{O}{III}] lines likely contribute to the photometric excess (${\sim}0.2$~mag) seen in the F115W filter in Fig.~\ref{fig:GS-9422}. By placing GS-9422 at $z=5.60$, the \ion{C}{III}] line further contributes to the photometric excess, resulting in UV photometry (light green, Fig.~\ref{fig:uv_colour_selection}) that is almost indistinguishable from that of a normal starburst (dark blue). Such high EW UV lines are expected in the case of star formation following a top-heavy IMF in a non-pristine environment \citep{Nakajima2018}. Hence the same star formation following a top-heavy IMF that powers the high normalisation of the nebular continuum, also possibly powers strong UV lines and collisional enhancement of the two-photon continuum, which results in hiding the otherwise characteristically red UV slope of nebular-dominated emission \citep{Katz2025} in photometric data. 

\begin{figure*}
\centering
\includegraphics[width=.475\linewidth] {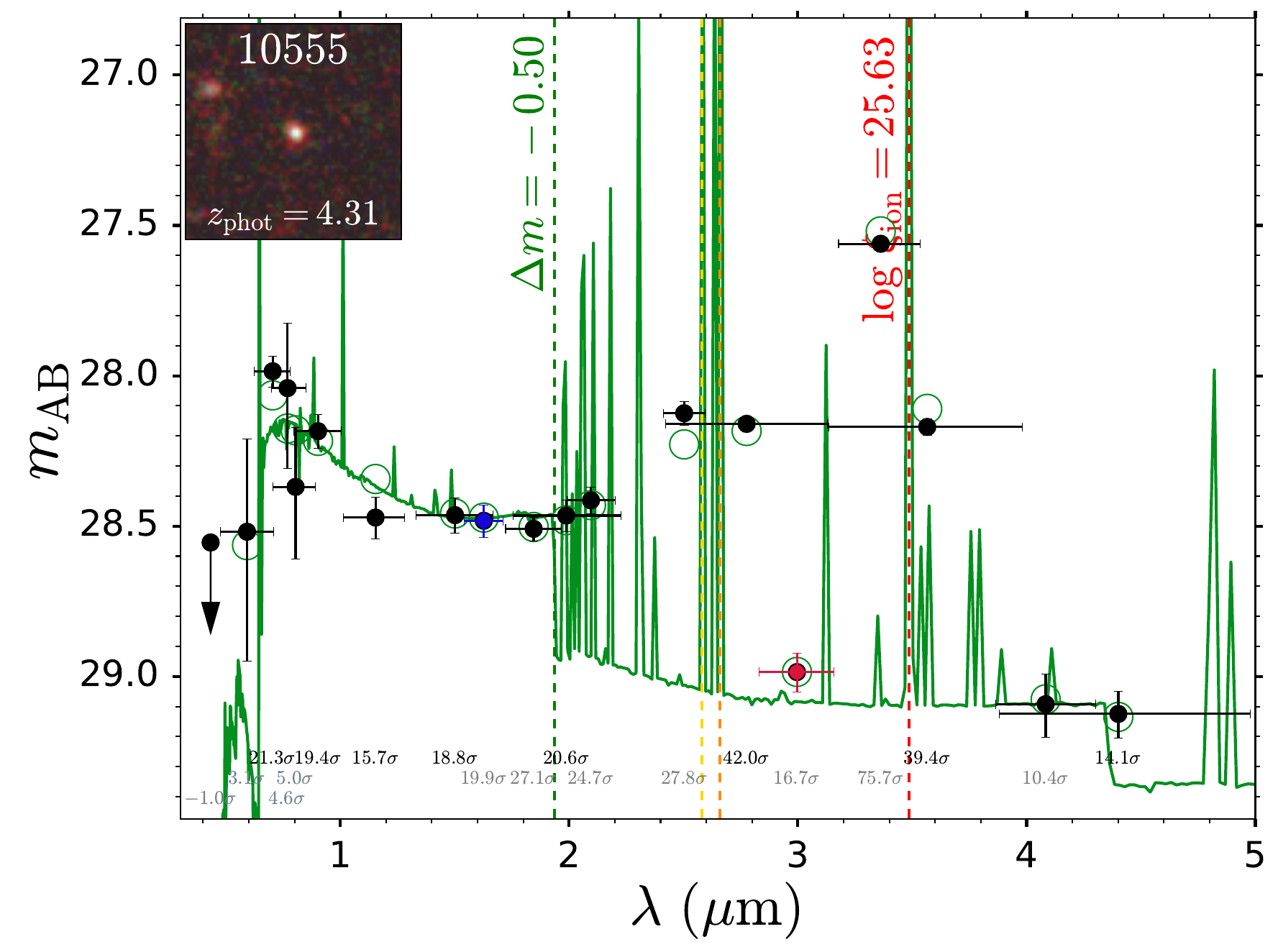} \hfill
\includegraphics[width=.475\linewidth]{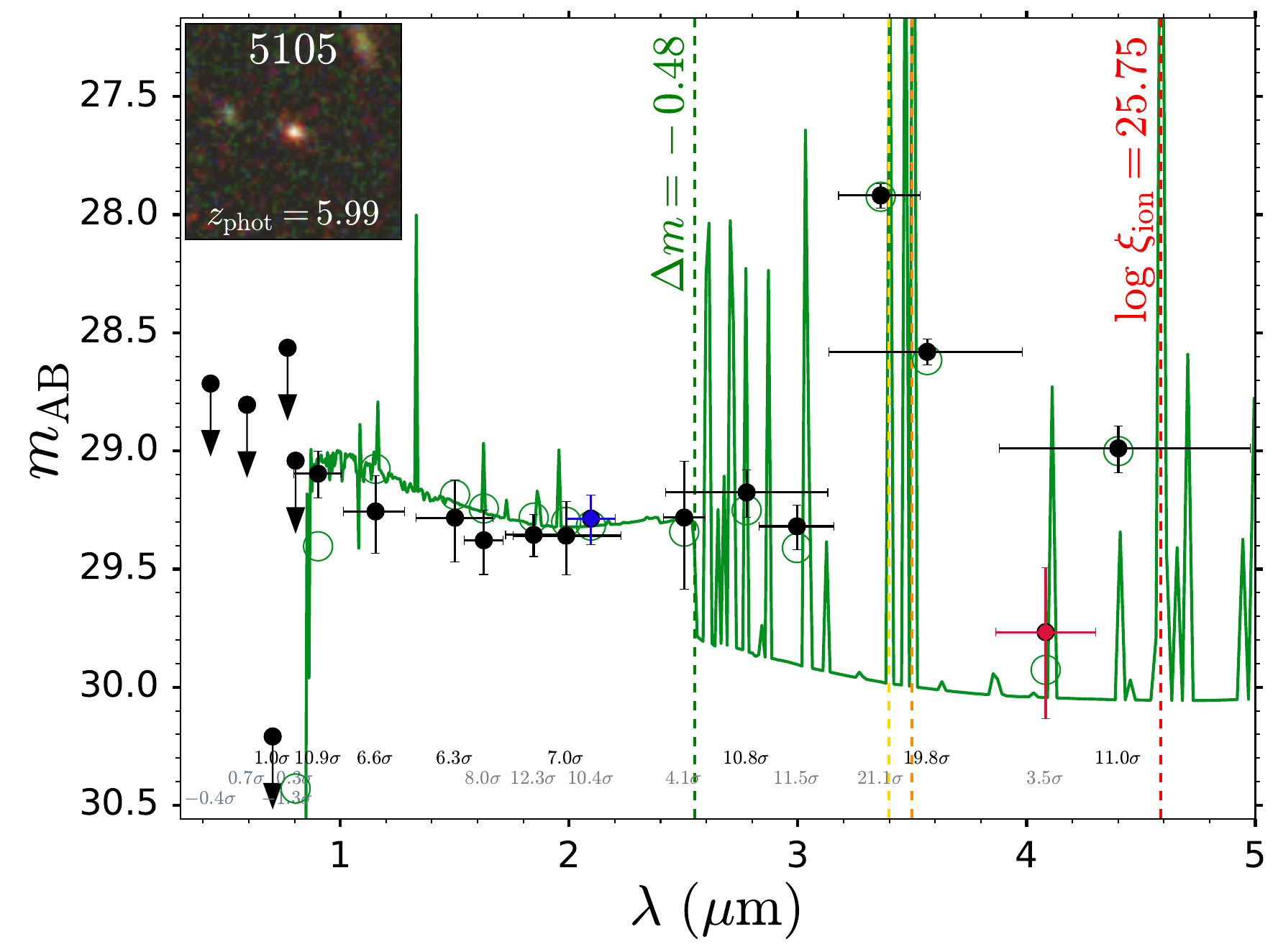} \\[4.5ex]
\includegraphics[width=.475\linewidth] {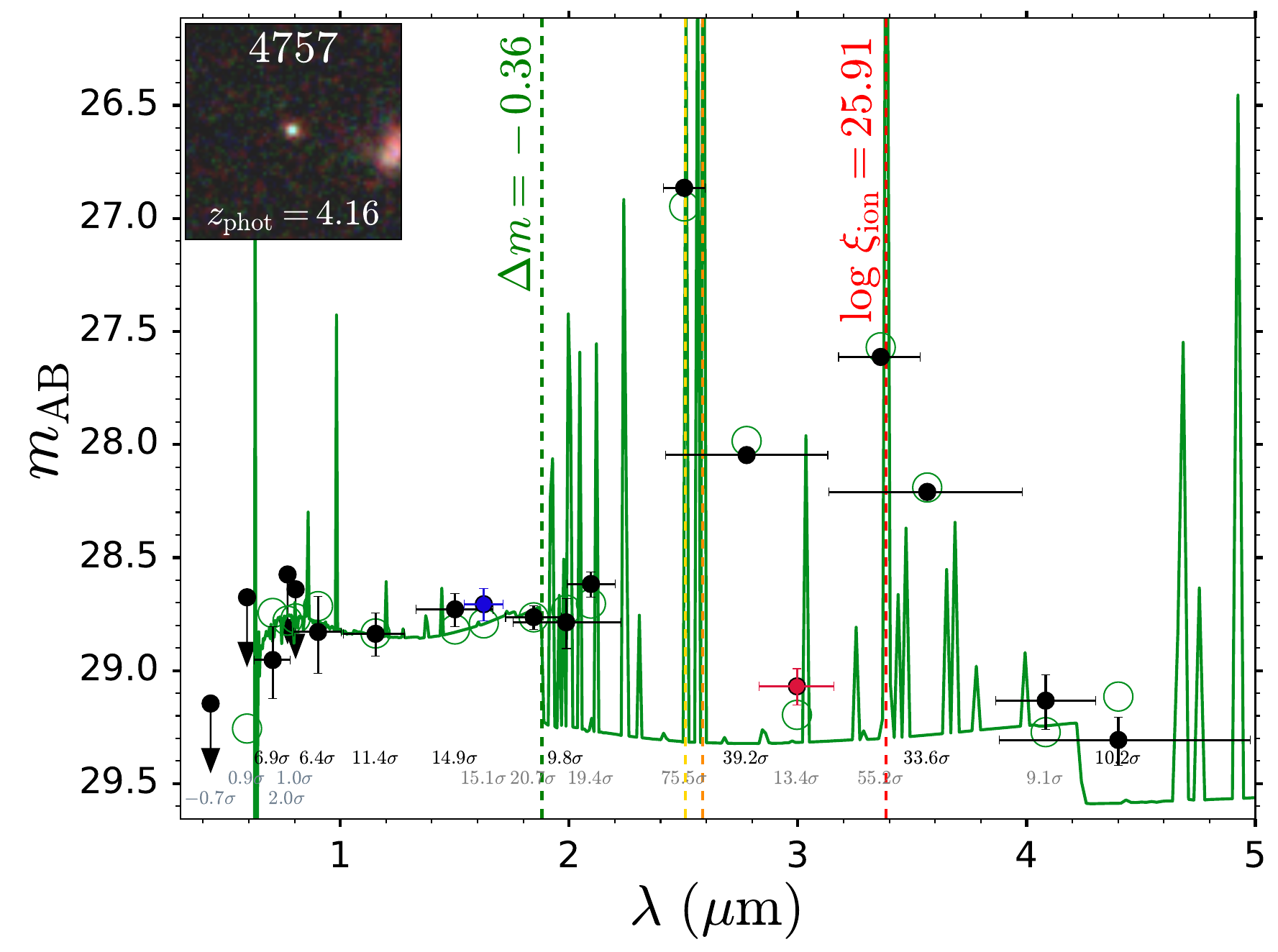} \hfill
\includegraphics[width=.475\linewidth]{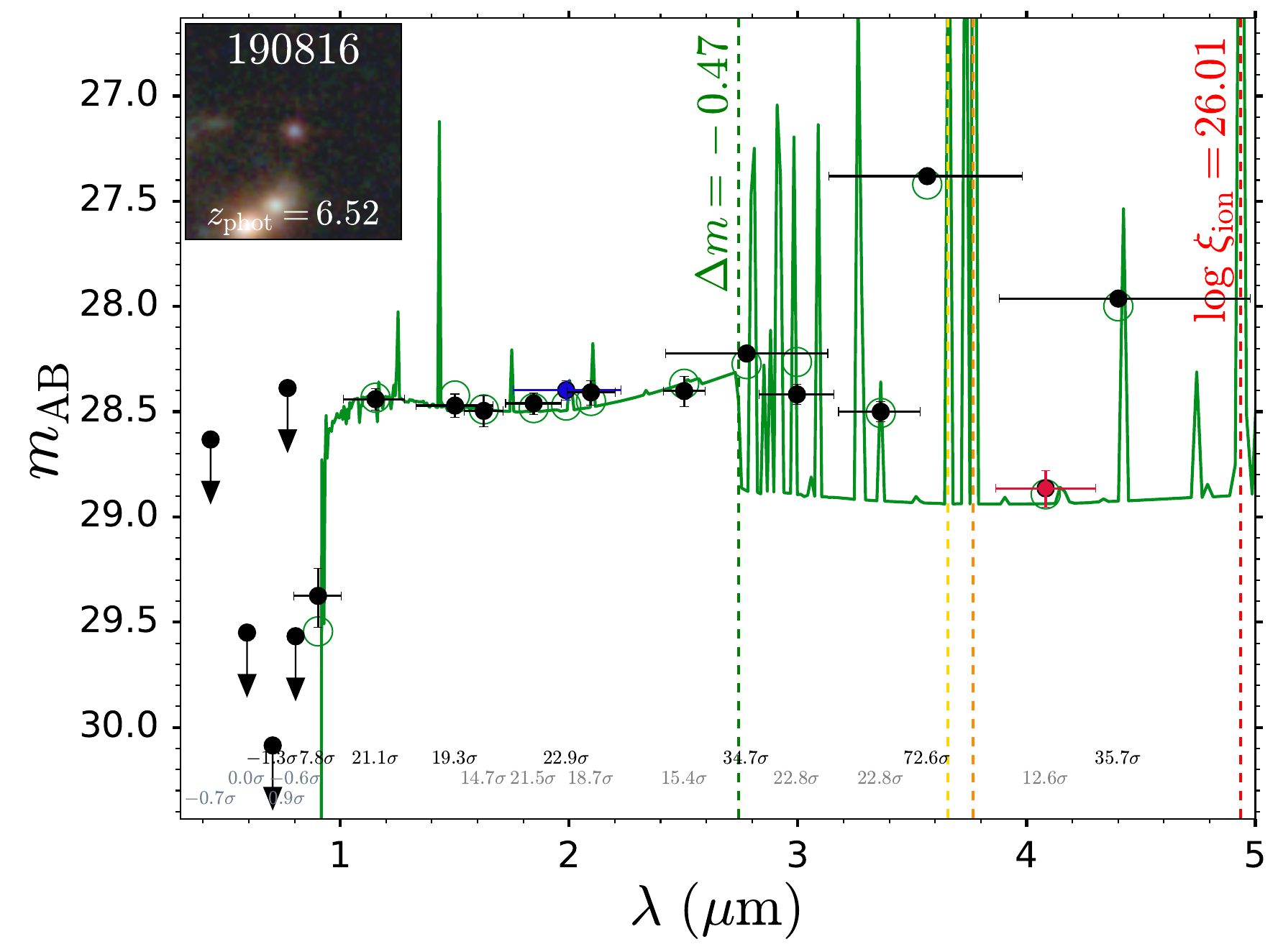}
\caption{SEDs of nebular-dominated galaxy candidates. These are Balmer-jump candidates, that further satisfy our nebular-dominated selection criteria: a high observed ionising photon production efficiency with $\log\, (\xi_\mathrm{ion, obs} /\mathrm{(Hz\ erg^{-1})}) > 25.60$ (Eq.~\ref{eq:xi_ion}) to power the strong nebular continuum emission, and relatively non-blue UV colour (Eq.~\ref{eq:uv_colour}) indicating a lack of stellar continuum emission. Note that the Bagpipes fits (using normal stellar populations, dark green) to the HST+NIRCam photometry provide a reasonable description of the data, indicating that these galaxies do not necessarily have to be starbursts following a top-heavy IMF with nebular-dominated emission. Deep follow-up continuum spectroscopy with the NIRSpec PRISM is needed to establish the presence or absence of the two-photon continuum turnover in their spectra, thus determining whether they are nebular-dominated galaxies or dust-reddened starbursts.}
\label{fig:nd_candidates}
\end{figure*}

It is for this reason that we apply a relatively conservative UV colour cut, that demands the UV colour to be less blue than some threshold. We require the magnitude difference between the first $m_\mathrm{UV, 1}$ and third wide-band $m_\mathrm{UV, 3}$ filters fully redward of Ly$\alpha$ to be greater than $-0.4$: 
\begin{equation} \label{eq:uv_colour}
m_\mathrm{UV, 1} - m_\mathrm{UV, 3} > -0.4.
\end{equation}
\noindent So at e.g.\@ $z=5.60$, this amounts to $\mathrm{F115W} - \mathrm{F200W} > -0.4$. The $-0.4$ colour threshold is motivated by the example of GS-9422 redshifted to $z=5.60$, introducing some buffer to account for possibly even stronger UV lines and/or photometric error on the colour measurement. The $\xi_\mathrm{ion,obs}$ cut (Eq.~\ref{eq:boost}) is the relatively more demanding nebular-dominated selection criterion, while the UV colour cut (Eq.~\ref{eq:uv_colour}) serves as more of a consistency check.

\subsubsection{Nebular-dominated galaxy candidates}\label{subsubsec:nd_candidates}

We show example nebular-dominated galaxy candidates in Fig.~\ref{fig:nd_candidates}, arranged in order of increasing $\xi_\mathrm{ion, obs}$ (noted in red in the individual panels), going from left to right, top to bottom. Bagpipes fits to the HST+NIRCam photometry are shown in dark green. The sources exhibit notable Balmer jumps, strong emission lines, high $\xi_\mathrm{ion, obs}$ and relatively flat UV slopes, following our selection procedure. Thus they are compatible with being nebular-dominated galaxies \citep{Cameron2024, Katz2025}, given our various cuts, though we stress that they do not necessarily have to be nebular-dominated. To accommodate e.g.\@ the range of possible Balmer jump strengths, $\xi_\mathrm{ion, obs}$ values and UV colours, we have had to be quite general, adopting relatively relaxed selection criteria (Eq.~\ref{eq:jump_cut}, Eq.~\ref{eq:xi_ion}, Eq.~\ref{eq:uv_colour}). Hence it is possible that when fitting the full photometry (rather than the select data points comprising our cuts) with tailored nebular-dominated templates, that not all of our candidates persist. 

\begin{table*}
\begin{center}
\begin{tabular}{|c|c|c|c|c|c|c|c|c|c|c|} 
\hline
ID & R.\@A.\@ & Dec.\@ & $z$ & $\Delta m_\mathrm{jump}$ & $\log\, \xi_\mathrm{ion,obs}$ & $m_\mathrm{UV,1} - m_\mathrm{UV,3}$ & $M_\mathrm{UV}$ & $M_\mathrm{opt}$ & $\mathrm{EW_{[O\ III]+H\beta}}$ & $\mathrm{EW_{H\alpha}}$ \\
 & & & & (mag) & (Hz~erg$^{-1}$) & (mag) & (mag) & (mag) & (\AA) & (\AA) \\
 \hline
1225368 & $189.40965$ & $62.26255$ & $6.85^{+0.02}_{-0.03}$ & $-0.32^{+0.18}_{-0.22}$ & - & $-0.20^{+0.11}_{-0.12}$ & $-19.56^{+0.10}_{-0.08}$ & $-19.16^{+0.19}_{-0.16}$ & $2728^{+761}_{-565}$ & - \\
9723 & $53.03720$ & $-27.89547$ & $6.57^{+0.01}_{-0.04}$ & $-0.28^{+0.13}_{-0.15}$ & $25.78^{+0.11}_{-0.12}$ & $-0.01^{+0.18}_{-0.16}$ & $-17.87^{+0.19}_{-0.13}$ & $-17.70^{+0.14}_{-0.10}$ & $2048^{+470}_{-311}$ & $1551^{+540}_{-390}$ \\
407498 & $52.99844$ & $-27.83942$ & $6.53^{+0.02}_{-0.05}$ & $-0.17^{+0.05}_{-0.04}$ & $25.67^{+0.04}_{-0.04}$ & $-0.08^{+0.04}_{-0.04}$ & $-19.84^{+0.03}_{-0.02}$ & $-20.13^{+0.04}_{-0.04}$ & $1567^{+106}_{-87}$ & $1161^{+146}_{-138}$ \\
144012 & $53.10025$ & $-27.75862$ & $6.53^{+0.03}_{-0.06}$ & $-0.37^{+0.10}_{-0.11}$ & $25.77^{+0.07}_{-0.05}$ & $-0.17^{+0.09}_{-0.06}$ & $-19.45^{+0.05}_{-0.05}$ & $-18.98^{+0.12}_{-0.08}$ & $2828^{+382}_{-277}$ & $1909^{+559}_{-311}$ \\
354358 & $53.04097$ & $-27.92838$ & $6.50^{+0.03}_{-0.25}$ & $-0.28^{+0.11}_{-0.14}$ & $25.67^{+0.09}_{-0.08}$ & $-0.12^{+0.12}_{-0.10}$ & $-19.06^{+0.08}_{-0.07}$ & $-18.89^{+0.11}_{-0.09}$ & $1552^{+315}_{-225}$ & $1352^{+394}_{-308}$ \\
473105 & $53.14579$ & $-27.93137$ & $6.36^{+0.06}_{-0.22}$ & $-0.23^{+0.09}_{-0.10}$ & $25.86^{+0.03}_{-0.03}$ & $-0.11^{+0.07}_{-0.07}$ & $-19.22^{+0.03}_{-0.03}$ & $-18.77^{+0.06}_{-0.07}$ & $1308^{+167}_{-143}$ & $1954^{+281}_{-234}$ \\
1164318 & $189.37816$ & $62.32236$ & $6.00^{+0.02}_{-0.02}$ & $-0.29^{+0.08}_{-0.09}$ & $25.80^{+0.04}_{-0.03}$ & $-0.11^{+0.08}_{-0.06}$ & $-19.56^{+0.05}_{-0.04}$ & $-19.01^{+0.07}_{-0.07}$ & $2865^{+235}_{-263}$ & $1791^{+218}_{-219}$ \\
333108 & $53.17585$ & $-27.81998$ & $6.00^{+0.01}_{-0.03}$ & $-0.23^{+0.10}_{-0.13}$ & $25.73^{+0.07}_{-0.06}$ & $-0.10^{+0.09}_{-0.08}$ & $-18.11^{+0.06}_{-0.05}$ & $-17.74^{+0.14}_{-0.07}$ & $2515^{+498}_{-268}$ & $1444^{+429}_{-266}$ \\
112267 & $53.10333$ & $-27.79974$ & $5.99^{+0.01}_{-0.05}$ & $-0.36^{+0.05}_{-0.04}$ & $25.77^{+0.02}_{-0.02}$ & $-0.10^{+0.03}_{-0.03}$ & $-20.13^{+0.02}_{-0.01}$ & $-19.72^{+0.04}_{-0.03}$ & $2781^{+137}_{-123}$ & $1794^{+131}_{-109}$ \\
412941 & $53.13490$ & $-27.71650$ & $5.99^{+0.02}_{-0.03}$ & $-0.42^{+0.13}_{-0.17}$ & $25.71^{+0.09}_{-0.12}$ & $0.25^{+0.12}_{-0.11}$ & $-18.96^{+0.10}_{-0.09}$ & $-18.93^{+0.17}_{-0.14}$ & $2148^{+560}_{-414}$ & $1187^{+442}_{-323}$ \\
1224627 & $189.42770$ & $62.28884$ & $5.86^{+0.01}_{-0.04}$ & $-0.15^{+0.18}_{-0.16}$ & $25.82^{+0.06}_{-0.07}$ & $-0.15^{+0.15}_{-0.16}$ & $-19.00^{+0.10}_{-0.09}$ & $-18.64^{+0.11}_{-0.11}$ & $2589^{+410}_{-348}$ & $1740^{+338}_{-323}$ \\
452280 & $53.02253$ & $-27.73183$ & $5.22^{+0.00}_{-0.10}$ & $-0.41^{+0.12}_{-0.18}$ & $25.74^{+0.08}_{-0.09}$ & $0.09^{+0.08}_{-0.08}$ & $-18.33^{+0.06}_{-0.09}$ & $-18.05^{+0.15}_{-0.12}$ & $2446^{+658}_{-379}$ & $1431^{+470}_{-308}$ \\
1127101 & $189.36534$ & $62.22234$ & $5.21^{+0.02}_{-0.02}$ & $-0.20^{+0.07}_{-0.07}$ & $25.62^{+0.04}_{-0.04}$ & $-0.02^{+0.05}_{-0.04}$ & $-19.91^{+0.03}_{-0.03}$ & $-19.51^{+0.05}_{-0.05}$ & $2429^{+194}_{-185}$ & $1083^{+132}_{-137}$ \\
1101498 & $189.34090$ & $62.24090$ & $5.20^{+0.01}_{-0.10}$ & $-0.35^{+0.10}_{-0.14}$ & $25.63^{+0.09}_{-0.09}$ & $0.07^{+0.13}_{-0.09}$ & $-18.72^{+0.11}_{-0.09}$ & $-18.26^{+0.13}_{-0.10}$ & $2042^{+441}_{-308}$ & $967^{+294}_{-224}$ \\
1129249 & $189.39822$ & $62.24217$ & $5.19^{+0.01}_{-0.04}$ & $-0.18^{+0.14}_{-0.12}$ & $25.80^{+0.09}_{-0.16}$ & $0.21^{+0.16}_{-0.21}$ & $-18.43^{+0.15}_{-0.14}$ & $-18.68^{+0.08}_{-0.11}$ & $1465^{+208}_{-276}$ & $913^{+207}_{-325}$ \\
468645 & $53.00482$ & $-27.85365$ & $5.18^{+0.01}_{-0.03}$ & $-0.16^{+0.09}_{-0.11}$ & $25.62^{+0.08}_{-0.09}$ & $0.05^{+0.12}_{-0.12}$ & $-18.58^{+0.11}_{-0.09}$ & $-18.51^{+0.10}_{-0.08}$ & $2082^{+290}_{-251}$ & $929^{+221}_{-200}$ \\
1029047 & $189.23103$ & $62.25913$ & $5.18^{+0.01}_{-0.02}$ & $-0.41^{+0.07}_{-0.12}$ & $25.80^{+0.04}_{-0.03}$ & $0.12^{+0.08}_{-0.07}$ & $-20.01^{+0.05}_{-0.06}$ & $-19.81^{+0.11}_{-0.07}$ & $3345^{+479}_{-277}$ & $1791^{+338}_{-239}$ \\
479824 & $52.96712$ & $-27.78782$ & $5.15^{+0.02}_{-0.02}$ & $-0.51^{+0.15}_{-0.17}$ & $25.91^{+0.05}_{-0.05}$ & $-0.28^{+0.11}_{-0.11}$ & $-17.95^{+0.09}_{-0.07}$ & $-17.16^{+0.16}_{-0.15}$ & $3743^{+752}_{-646}$ & $3336^{+782}_{-716}$ \\
133802 & $53.18044$ & $-27.77066$ & $5.06^{+0.02}_{-0.02}$ & $-0.60^{+0.10}_{-0.15}$ & $25.74^{+0.05}_{-0.04}$ & $-0.17^{+0.08}_{-0.08}$ & $-18.16^{+0.08}_{-0.06}$ & $-17.32^{+0.12}_{-0.09}$ & $3089^{+547}_{-353}$ & $2210^{+549}_{-293}$ \\
422136 & $52.97557$ & $-27.77382$ & $4.97^{+0.03}_{-0.04}$ & $-0.33^{+0.17}_{-0.24}$ & $25.71^{+0.08}_{-0.08}$ & $-0.07^{+0.14}_{-0.13}$ & $-17.83^{+0.08}_{-0.10}$ & $-17.42^{+0.21}_{-0.15}$ & $2045^{+759}_{-445}$ & $1478^{+568}_{-392}$ \\
43118 & $53.08044$ & $-27.86800$ & $4.93^{+0.02}_{-0.01}$ & $-0.53^{+0.08}_{-0.13}$ & $25.65^{+0.05}_{-0.05}$ & $-0.27^{+0.11}_{-0.10}$ & $-17.86^{+0.07}_{-0.07}$ & $-17.01^{+0.11}_{-0.09}$ & $2540^{+470}_{-317}$ & $1928^{+420}_{-278}$ \\
1147129 & $189.25684$ & $62.32236$ & $4.92^{+0.03}_{-0.01}$ & $-0.23^{+0.08}_{-0.10}$ & $25.68^{+0.05}_{-0.07}$ & $0.01^{+0.09}_{-0.08}$ & $-18.93^{+0.07}_{-0.06}$ & $-18.90^{+0.10}_{-0.09}$ & $2455^{+350}_{-304}$ & $1161^{+283}_{-207}$ \\
1026211 & $189.11840$ & $62.23785$ & $4.91^{+0.01}_{-0.02}$ & $-0.62^{+0.14}_{-0.14}$ & $25.64^{+0.04}_{-0.04}$ & $-0.20^{+0.06}_{-0.07}$ & $-19.45^{+0.03}_{-0.03}$ & $-18.87^{+0.12}_{-0.10}$ & $3503^{+565}_{-451}$ & $1807^{+394}_{-301}$ \\
1073514 & $189.29785$ & $62.17750$ & $4.90^{+0.06}_{-0.01}$ & $-0.29^{+0.11}_{-0.14}$ & $25.79^{+0.05}_{-0.05}$ & $0.16^{+0.10}_{-0.07}$ & $-18.39^{+0.05}_{-0.04}$ & $-18.10^{+0.10}_{-0.08}$ & $1757^{+264}_{-219}$ & $1376^{+278}_{-221}$ \\
411246 & $53.00674$ & $-27.83412$ & $4.84^{+0.01}_{-0.02}$ & $-0.39^{+0.08}_{-0.08}$ & $25.63^{+0.04}_{-0.04}$ & $-0.15^{+0.05}_{-0.05}$ & $-18.45^{+0.03}_{-0.04}$ & $-18.05^{+0.07}_{-0.07}$ & $2657^{+264}_{-234}$ & $1408^{+189}_{-172}$ \\
439263 & $53.19770$ & $-27.76088$ & $4.84^{+-0.01}_{-0.24}$ & $-0.25^{+0.18}_{-0.18}$ & $25.66^{+0.11}_{-0.09}$ & $-0.18^{+0.16}_{-0.18}$ & $-17.36^{+0.13}_{-0.14}$ & $-17.09^{+0.21}_{-0.17}$ & $1930^{+680}_{-494}$ & $1349^{+593}_{-368}$ \\
1132908 & $189.39218$ & $62.26171$ & $4.83^{+0.01}_{-0.05}$ & $-0.37^{+0.08}_{-0.11}$ & $25.69^{+0.06}_{-0.04}$ & $-0.03^{+0.09}_{-0.07}$ & $-19.52^{+0.07}_{-0.05}$ & $-19.02^{+0.08}_{-0.07}$ & $2905^{+347}_{-277}$ & $1407^{+232}_{-209}$ \\
236504 & $53.09761$ & $-27.73729$ & $4.82^{+-0.00}_{-0.23}$ & $-0.32^{+0.25}_{-0.24}$ & $25.76^{+0.11}_{-0.14}$ & $0.17^{+0.22}_{-0.28}$ & $-17.28^{+0.16}_{-0.18}$ & $-17.13^{+0.19}_{-0.18}$ & $1864^{+663}_{-476}$ & $1312^{+665}_{-451}$ \\
1235827 & $189.42667$ & $62.28803$ & $4.80^{+-0.00}_{-3.71}$ & $-0.50^{+0.29}_{-0.27}$ & $25.72^{+0.18}_{-0.22}$ & $0.48^{+0.29}_{-0.19}$ & $-17.51^{+0.23}_{-0.17}$ & $-17.46^{+0.22}_{-0.24}$ & $883^{+560}_{-454}$ & $1066^{+678}_{-571}$ \\
105405 & $53.11830$ & $-27.81016$ & $4.79^{+0.01}_{-0.13}$ & $-0.65^{+0.11}_{-0.12}$ & $25.64^{+0.05}_{-0.06}$ & $0.01^{+0.09}_{-0.11}$ & $-18.01^{+0.07}_{-0.09}$ & $-17.37^{+0.09}_{-0.09}$ & $2354^{+329}_{-258}$ & $1566^{+329}_{-259}$ \\
449873 & $52.97821$ & $-27.78016$ & $4.75^{+0.04}_{-0.02}$ & $-0.59^{+0.14}_{-0.23}$ & $25.75^{+0.05}_{-0.05}$ & $-0.11^{+0.09}_{-0.08}$ & $-17.80^{+0.06}_{-0.07}$ & $-17.32^{+0.17}_{-0.13}$ & $2950^{+758}_{-497}$ & $2110^{+623}_{-433}$ \\
307749 & $53.21204$ & $-27.80678$ & $4.74^{+0.03}_{-0.06}$ & $-0.59^{+0.12}_{-0.11}$ & $25.63^{+0.04}_{-0.04}$ & $-0.29^{+0.08}_{-0.09}$ & $-19.65^{+0.03}_{-0.05}$ & $-19.01^{+0.07}_{-0.09}$ & $3210^{+393}_{-341}$ & $1899^{+271}_{-265}$ \\
446124 & $53.01990$ & $-27.81525$ & $4.74^{+0.04}_{-0.02}$ & $-0.42^{+0.03}_{-0.03}$ & $25.61^{+0.02}_{-0.02}$ & $-0.11^{+0.02}_{-0.02}$ & $-19.34^{+0.01}_{-0.01}$ & $-18.90^{+0.03}_{-0.03}$ & $2496^{+124}_{-95}$ & $1308^{+81}_{-71}$ \\
1022528 & $189.12163$ & $62.30072$ & $4.69^{+0.06}_{-0.10}$ & $-0.21^{+0.18}_{-0.16}$ & $25.64^{+0.10}_{-0.12}$ & $-0.15^{+0.18}_{-0.13}$ & $-18.70^{+0.12}_{-0.12}$ & $-18.38^{+0.11}_{-0.15}$ & $1783^{+370}_{-381}$ & $1191^{+342}_{-358}$ \\
431900 & $52.96811$ & $-27.78685$ & $4.40^{+0.01}_{-0.01}$ & $-0.23^{+0.11}_{-0.13}$ & $26.01^{+0.03}_{-0.04}$ & $0.14^{+0.09}_{-0.13}$ & $-17.73^{+0.06}_{-0.10}$ & $-17.55^{+0.10}_{-0.10}$ & $3385^{+492}_{-440}$ & $2210^{+416}_{-353}$ \\
430168 & $53.00624$ & $-27.82228$ & $4.36^{+-0.00}_{-0.07}$ & $-0.80^{+0.10}_{-0.10}$ & $25.64^{+0.04}_{-0.03}$ & $-0.18^{+0.04}_{-0.02}$ & $-18.98^{+0.03}_{-0.02}$ & $-18.08^{+0.10}_{-0.09}$ & $3811^{+504}_{-413}$ & $2135^{+406}_{-260}$ \\
27706 & $53.07759$ & $-27.87929$ & $4.34^{+0.00}_{-0.02}$ & $-0.47^{+0.21}_{-0.16}$ & $25.77^{+0.08}_{-0.10}$ & $0.05^{+0.17}_{-0.13}$ & $-17.15^{+0.15}_{-0.09}$ & $-16.59^{+0.15}_{-0.16}$ & $1915^{+491}_{-457}$ & $1708^{+472}_{-477}$ \\
445996 & $52.98156$ & $-27.79756$ & $4.30^{+0.03}_{-0.14}$ & $-0.15^{+0.09}_{-0.06}$ & $26.01^{+0.02}_{-0.03}$ & $0.08^{+0.05}_{-0.06}$ & $-18.19^{+0.04}_{-0.04}$ & $-18.55^{+0.06}_{-0.07}$ & $1320^{+162}_{-170}$ & $2172^{+229}_{-256}$ \\
394267 & $53.00657$ & $-27.79110$ & $4.28^{+0.03}_{-0.19}$ & $-0.24^{+0.20}_{-0.21}$ & $25.70^{+0.11}_{-0.12}$ & $-0.08^{+0.16}_{-0.15}$ & $-17.25^{+0.13}_{-0.12}$ & $-17.20^{+0.18}_{-0.18}$ & $1034^{+467}_{-382}$ & $1320^{+551}_{-513}$ \\
188471 & $53.09414$ & $-27.84379$ & $3.99^{+0.07}_{-0.01}$ & $-0.48^{+0.12}_{-0.13}$ & $25.84^{+0.10}_{-0.09}$ & $-0.17^{+0.18}_{-0.21}$ & $-17.29^{+0.17}_{-0.21}$ & $-16.89^{+0.13}_{-0.10}$ & $2087^{+443}_{-341}$ & $1954^{+414}_{-387}$ \\
173329 & $53.05283$ & $-27.87058$ & $3.75^{+0.02}_{-0.07}$ & $-0.18^{+0.11}_{-0.10}$ & $25.68^{+0.11}_{-0.13}$ & $-0.09^{+0.15}_{-0.21}$ & $-18.03^{+0.14}_{-0.14}$ & $-17.63^{+0.10}_{-0.10}$ & - & $1101^{+387}_{-296}$ \\
75595 & $53.07788$ & $-27.84648$ & $3.71^{+0.01}_{-0.04}$ & $-0.51^{+0.12}_{-0.11}$ & $25.77^{+0.05}_{-0.05}$ & $0.05^{+0.12}_{-0.12}$ & $-17.73^{+0.12}_{-0.12}$ & $-17.05^{+0.11}_{-0.09}$ & - & $1935^{+365}_{-292}$ \\
50234 & $53.10296$ & $-27.86323$ & $3.70$ &$-0.25^{+0.01}_{-0.01}$ & $25.85^{+0.01}_{-0.02}$ & $0.04^{+0.04}_{-0.03}$ & $-19.75^{+0.03}_{-0.04}$ & $-19.49^{+0.01}_{-0.01}$ & - & $1639^{+31}_{-35}$ \\
8370 & $53.03412$ & $-27.89744$ & $3.13^{+0.01}_{-0.01}$ & $-0.41^{+0.04}_{-0.04}$ & $25.60^{+0.04}_{-0.03}$ & $-0.12^{+0.10}_{-0.09}$ & $-18.55^{+0.09}_{-0.08}$ & $-17.95^{+0.03}_{-0.03}$ & $1998^{+85}_{-100}$ & $1283^{+102}_{-102}$ \\
171933 & $53.05193$ & $-27.87326$ & $2.98^{+0.01}_{-0.00}$ & $-0.31^{+0.07}_{-0.06}$ & $25.70^{+0.07}_{-0.05}$ & $-0.03^{+0.14}_{-0.11}$ & $-17.34^{+0.17}_{-0.13}$ & $-17.23^{+0.05}_{-0.05}$ & $2256^{+191}_{-164}$ & $1372^{+140}_{-160}$ \\
47159 & $53.12835$ & $-27.86528$ & $2.79$ &$-0.73^{+0.12}_{-0.17}$ & $25.74^{+0.05}_{-0.04}$ & $-0.12^{+0.07}_{-0.04}$ & $-18.56^{+0.04}_{-0.03}$ & $-17.83^{+0.18}_{-0.10}$ & $3229^{+776}_{-378}$ & $2327^{+736}_{-391}$ \\
200199 & $53.11812$ & $-27.80730$ & $2.78^{+0.04}_{-0.01}$ & $-0.30^{+0.04}_{-0.07}$ & $25.82^{+0.05}_{-0.05}$ & $0.14^{+0.10}_{-0.10}$ & $-17.95^{+0.10}_{-0.11}$ & $-17.79^{+0.07}_{-0.04}$ & $2703^{+253}_{-148}$ & $1351^{+192}_{-90}$ \\
206039 & $53.13583$ & $-27.78654$ & $2.66$ &$-0.40^{+0.08}_{-0.07}$ & - & $-0.24^{+0.04}_{-0.04}$ & $-17.91^{+0.02}_{-0.02}$ & $-17.24^{+0.06}_{-0.07}$ & $2497^{+189}_{-213}$ & - \\
1204376 & $189.37799$ & $62.28063$ & $2.64^{+0.01}_{-0.04}$ & $-0.30^{+0.07}_{-0.09}$ & - & $0.06^{+0.09}_{-0.07}$ & $-18.34^{+0.07}_{-0.06}$ & $-18.43^{+0.08}_{-0.06}$ & $2060^{+246}_{-219}$ & - \\
1008622 & $189.19771$ & $62.25095$ & $2.31^{+-0.00}_{-0.26}$ & $-0.20^{+0.01}_{-0.02}$ & $25.66^{+0.01}_{-0.01}$ & $-0.03^{+0.01}_{-0.01}$ & $-19.94^{+0.01}_{-0.01}$ & $-19.76^{+0.02}_{-0.01}$ & $1722^{+51}_{-34}$ & $1112^{+53}_{-34}$ \\
\hline
\end{tabular}
\caption{Properties of a randomly-selected subsample of 50 nebular-dominated galaxy candidates, ordered by decreasing redshift. 
Listed are the JADES DR5 photometric catalog ID, right ascension, declination, redshift, Balmer jump $\Delta m_\mathrm{jump}$ (Eq.~\ref{eq:jump}), observed ionising photon production efficiency $\log\, \xi_\mathrm{ion, obs}$ (cf.\@ Eq.~\ref{eq:xi_ion}), UV colour $m_\mathrm{UV,1} - m_\mathrm{UV,3}$ (cf.\@ Eq.~\ref{eq:uv_colour}), absolute magnitude at ${\sim}1500$~\AA\ rest-frame $M_\mathrm{UV}$, rest-frame optical continuum absolute magnitude $M_\mathrm{opt}$, \OIII\ + \Hb\ rest-frame equivalent width, \Ha\ rest-frame equivalent width. Photometric redshifts are denoted with (16, 84) percentile errors with respect to the best-fit EAZY redshift, while JADES DR4 spectroscopic redshifts are denoted without errors. Empty entries (denoted by -) indicate the parameter could not be determined, because the appropriate emission line (\Ha\ or \OIII\ $\lambda5007$) resides in the gap between wide-band filters, or is redshifted out of the NIRCam spectral range.}
\label{tab:ndg_candidates}
\end{center}
\end{table*}

\begin{figure*}
\centering
\includegraphics[width=\linewidth]{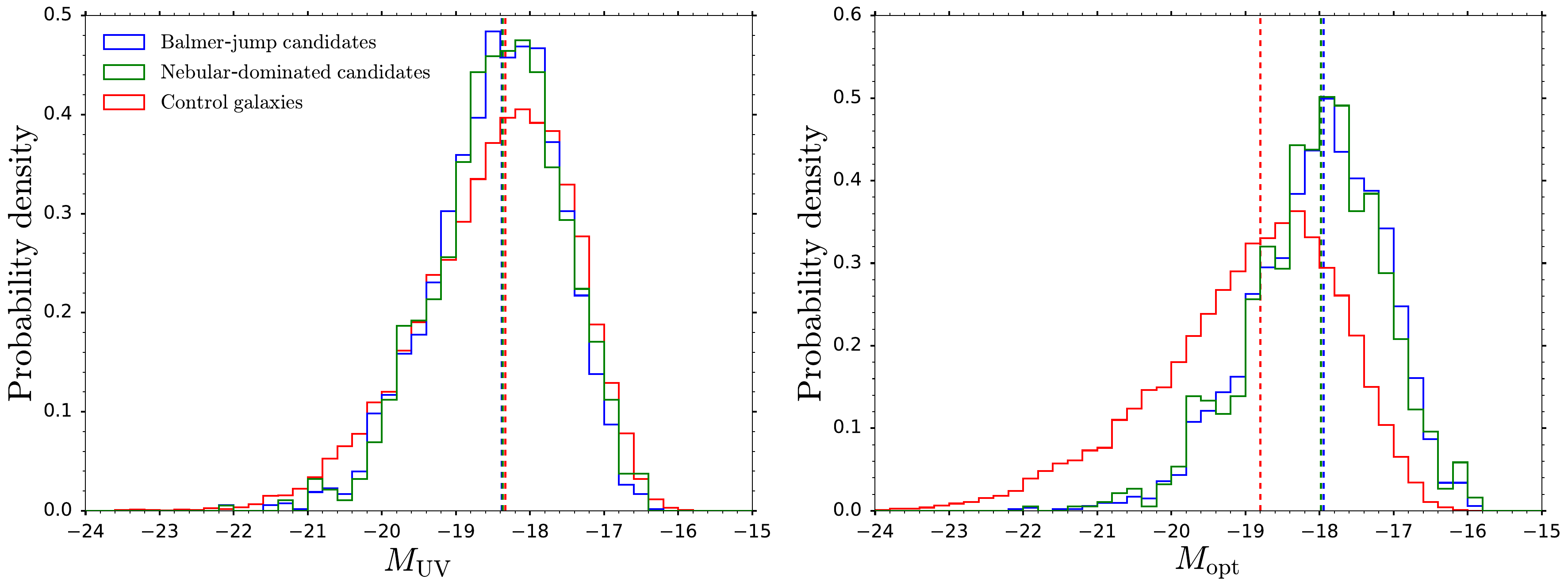}
\caption{Normalised histograms of absolute magnitudes for Balmer-jump candidates (blue), nebular-dominated candidates (green) and control galaxies with $\Delta m_\mathrm{jump} > -0.15$ (i.e.\@ which fail the Balmer-jump selection, red). Left panel: The median (indicated by the vertical dashed lines) absolute magnitude in the first wide-band fully redward of Ly$\alpha$ $M_\mathrm{UV} \approx -18.35$ is similar for the three samples. Right panel: Owing to the redder UV--optical colours for the control sample, their median absolute magnitude in the medium-band filter tracing the rest-frame optical continuum $M_\mathrm{opt} = -18.78$ is brighter than the median $M_\mathrm{opt} = -17.95$ for the Balmer-jump and nebular-dominated candidates.}
\label{fig:continuum_histograms}
\end{figure*}

Furthermore, we see that the standard Bagpipes templates provide adequate fits to the photometry of the nebular-dominated candidates displayed in Fig.~\ref{fig:nd_candidates}. Thus one does not have to invoke star formation following a top-heavy IMF with nebular-dominated emission in all cases to explain their photometry, since regular star formation with a standard IMF \citep[e.g.\@][]{Salpeter1955, Kroupa2001, Chabrier2003} may suffice. Here the large $\log\, (\xi_\mathrm{ion, obs} /\mathrm{(Hz\ erg^{-1})})$ values (25.91, 26.00), exceeding what is possible for a starburst following a normal IMF, are attributable to dust attenuation. This attenuates the rest-frame UV ($L_{\nu, 1500}$) more than the rest-frame optical (\Ha), thus inflating $\xi_\mathrm{ion,obs}$. Assuming the dust attenuation around \Ha\ is approximately given by $A_\mathrm{V}$, and the attenuation around 1500~\AA\ is $A_\mathrm{UV}$, then the shift in $\xi_\mathrm{ion,obs}$ is approximately $\Delta \log \xi_\mathrm{ion,obs} \approx (A_\mathrm{UV} - A_\mathrm{V})/2.5 \approx 0.6 A_\mathrm{V}$, where $A_\mathrm{UV} \approx 2.5A_\mathrm{V}$ for the \citet{Calzetti2000} law. So one should not have the expectation that nebular-dominated galaxies will exhibit unique photometry. Rather the very fact that the photometry of a galaxy is consistent with nebular-dominated templates (or nebular-dominated cuts) makes a compelling case for the source to actually be nebular-dominated. Ultimately, deep follow-up continuum spectroscopy with the NIRSpec PRISM is needed to establish the presence or absence of the two-photon continuum turnover in the nebular-dominated candidate spectra, thus determining whether they are nebular-dominated galaxies or dust-reddened starbursts.

We list the properties of a randomly-selected subsample of 50 of our nebular-dominated galaxy candidates in Table~\ref{tab:ndg_candidates}, arranged in order of decreasing redshift. 
The listed Balmer jump $\Delta m_\mathrm{jump}$, observed ionising photon production efficiency $\xi_\mathrm{ion, obs}$, UV colour $m_\mathrm{UV,1} - m_\mathrm{UV,3}$, absolute UV magnitude at ${\sim}1500$~\AA\ $M_\mathrm{UV}$, the absolute magnitude in the medium-band filter tracing the rest-frame optical continuum $M_\mathrm{opt}$, \OIII\ + \Hb\ equivalent width, \Ha\ equivalent width correspond to the values measured from the actual photometry. The (16, 84) percentiles for each parameter were determined by randomly shifting each photometric data point according to its error (assuming a normal distribution), measuring the subsequent parameter values for the new realisation of the photometry, and repeating this procedure for 100 randomly-shifted samples to establish the parameter errors.

Across $1.5 < z < 8.5$, we identify 2646 Balmer-jump galaxy candidates and 937 nebular-dominated galaxy candidates out of the total sample of 15259 galaxies (see Fig.~\ref{fig:jump_redshift}). We note that some of the nebular-dominated galaxy candidates are at photometric redshifts where \Ha\ is at the edge of a wide-band filter. Given the lower filter throughput (minimally 50~per~cent of maximum, given our redshift cuts), the inferred \Ha\ emission is strong, despite the relatively small observed photometric boost by \Ha. Additionally, due to the more strongly varying throughput at the filter edge, the inferred \Ha\ equivalent width (and resulting $\xi_\mathrm{ion,obs}$) is particularly sensitive to uncertainties in the redshift. Hence it is possible that these nebular-dominated galaxy candidates are less reliable than the candidates at redshifts away from the filter edge. 

\section{Balmer-jump and nebular-dominated statistics} \label{sec:statistics}

Having outlined our selection procedures, we now discuss the general properties of our Balmer-jump and nebular-dominated galaxy candidate samples. We focus here on empirical, model-independent properties. Physical properties, such as stellar mass $M_*$ and star formation rate are of course of interest, though depend on the choice of IMF adopted to convert the observables (photometry) into physically-inferred quantities. Since we believe some of our sources may exhibit nebular-dominated emission, which demands a top-heavy IMF if powered by star formation, physical parameters derived using standard templates incorporating normal IMFs will be inaccurate.

\subsection{Continuum and line statistics}

\begin{figure*}
\centering
\includegraphics[width=\linewidth]{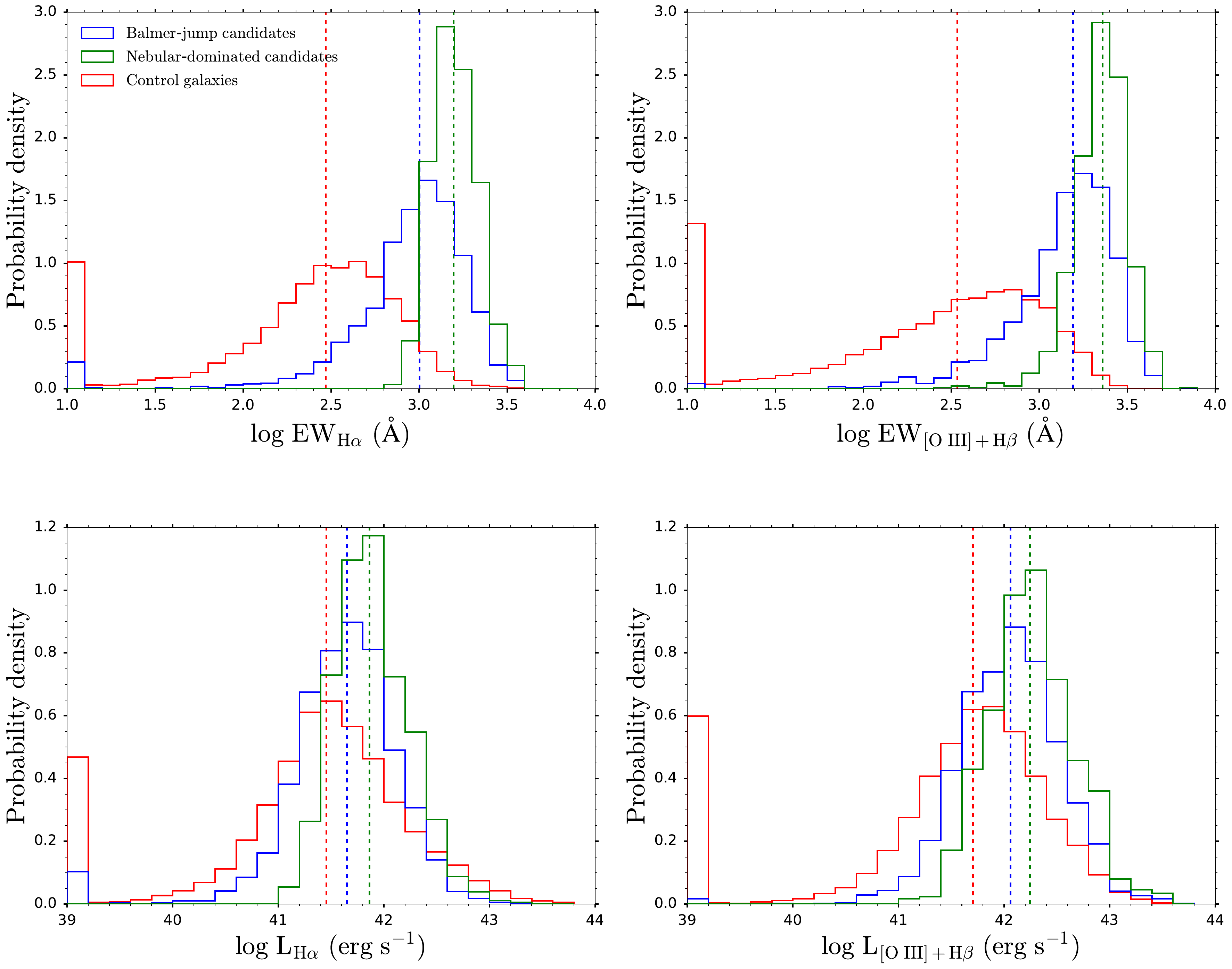}
\caption{Top panel: Similar to Fig.~\ref{fig:continuum_histograms}, now showing normalised histograms for the \Ha\ rest-frame equivalent width $\mathrm{EW_\mathrm{H\alpha}}$ (left) and \OIII\ $\lambda\lambda 4959,5007$ + \Hb\ rest-frame equivalent width $\mathrm{EW_{[O\ III] + H\beta}}$ (right). The median equivalent widths for nebular-dominated and Balmer-jump candidates are substantially larger than for control galaxies, at 1567~\AA, 1007~\AA, 294~\AA\ for \Ha, and 2292~\AA, 1555  ~\AA, 342~\AA\ for \OIII\ + \Hb, respectively. Bottom panel: Normalised histograms for the \Ha\ luminosity $L_{\mathrm{H}\alpha}$ (left) and \OIII\ + \Hb\ luminosity $L_\mathrm{[O\ III] + H\beta}$ (right). The median $\log\, (L_\mathrm{H\alpha}/\mathrm{erg\ s^{-1}})$ is 41.86, 41.65 and 41.46, for nebular-dominated candidates, Balmer-jump candidates and control galaxies, respectively. For $\log\, (L_\mathrm{[O\ III] + H\beta}/\mathrm{erg\ s^{-1}})$, the medians are 42.24, 42.06 and 41.71, respectively.}
\label{fig:line_histograms}
\end{figure*}

We show normalised histograms of continuum properties in Fig.~\ref{fig:continuum_histograms}, for our Balmer-jump candidates (satisfying the Balmer-jump criterion: Eq.~\ref{eq:jump_cut}, blue), the subset of nebular-dominated candidates (further satisfying Equations~\ref{eq:xi_ion}/\ref{eq:ew} and \ref{eq:uv_colour}, green), and control galaxies which do not satisfy the Balmer-jump criterion (i.e.\@ which have $\Delta m _\mathrm{jump} > -0.15$, red). As a reminder, we apply no apparent/absolute magnitude cut or stellar mass limit cut to our samples. The Balmer-jump and nebular-dominated candidates, together with the control sample, are all drawn from the same GOODS-S and GOODS-N fields which typically vary in depth from 28.5--31.0~AB~mag in 0.2~arcsec diameter circular apertures \citep{Johnson2026, Robertson2026}. We find that applying apparent magnitude cuts $m_\mathrm{UV}$ (mimicking the depths in shallower JWST imaging programs) qualitatively preserves the relative trends between the Balmer-jump, nebular-dominated and control samples, yet biases the median absolute magnitudes and luminosities for these samples to brighter values. The absolute magnitudes displayed in Fig.~\ref{fig:continuum_histograms}correspond to total flux densities, derived from scaling by the flux density ratio between the Kron aperture and a 0.3~arcsec diameter circular aperture. From the left panel we see that these systems exhibit relatively similar distributions in $M_\mathrm{UV}$, the absolute magnitude in the first wide-band filter fully redward of Ly$\alpha$. The median $M_\mathrm{UV}$ value (indicated by the vertical dashed lines) for the three samples is $\approx -18.35$. Given (by definition of the Balmer jump selection) the redder UV--optical colours for the control sample, these tend to exhibit a brighter $M_\mathrm{opt}$, the absolute magnitude in the medium-band filter tracing the rest-frame optical continuum, with a median $M_\mathrm{opt}=-18.80$ compared to $-17.95$ for the Balmer-jump and nebular-dominated candidate samples (see right panel). \citet{Roberts-Borsani2024} stack NIRSpec PRISM spectra for Ly$\alpha$ emitters (LAEs, $\mathrm{EW_{Ly\alpha}} > 30$~\AA) and non-LAEs ($\mathrm{EW_{Ly\alpha}} < 10$~\AA) at $5 < z < 8$, finding similar qualitative results. Whilst the LAE and non-LAE stacks have comparable $M_\mathrm{UV} \approx -18.6$, the LAE stack exhibits a Balmer jump, which the non-LAE stack lacks. Thus, the stronger line-emitting LAE stack (somewhat analogous to our Balmer-jump and nebular-dominated candidate samples), has a fainter optical continuum compared to the non-LAE stack (somewhat analogous to our control sample), despite the similar $M_\mathrm{UV}$.

We show normalised histograms of the emission-line properties in Fig.~\ref{fig:line_histograms}. As expected from our Balmer-jump cut, which aims to select young starbursts, our Balmer-jump candidates generally have larger \Ha\ equivalent widths (top-left panel) than the control sample, with medians of 1007~\AA\ and 294~\AA, respectively. There is a low EW tail of Balmer-jump candidates, which could correspond to sources with blue slopes (rather than an actual Balmer jump) that satisfy our UV--optical colour cut, or sources which spuriously satisfy the cut due to photometric uncertainties (systematic/statistical), which do not have particularly strong line emission. Given the high $\xi_\mathrm{ion,obs}$ of our nebular-dominated candidates (Eq.~\ref{eq:xi_ion}), and the correlation between $\xi_\mathrm{ion,obs}$ and $\mathrm{EW_{H\alpha}}$ (Fig.~\ref{fig:xi_ion_ew}), these sources generally have even larger \Ha\ equivalent widths, with a median of 1567~\AA, with our additional cuts mostly eliminating the low EW tail from the sample. Similar trends are seen for the \OIII\ + \Hb\ equivalent widths (top-right panel), with median equivalent widths of 342~\AA, 1555~\AA\ and 2292~\AA\ for the control galaxies, the Balmer-jump candidates, and the nebular-dominated candidates, respectively. This mirrors the correlation seen between the Balmer jump $\Delta m_\mathrm{jump}$ and $\mathrm{EW_{[O\ III]}}$ from the NIRSpec PRISM stacking analysis of $4 < z < 10$ galaxies by \citet{Hayes2025}. Thus our continuum-selected Balmer-jump candidates can generally be classified as extreme emission-line galaxies, \citep[$\mathrm{EW_{rest}} \gtrsim 750$~\AA,][]{Withers2023, Boyett2024, Davis2024, Llerena2024}.

\begin{figure*}
\centering
\includegraphics[width=\linewidth]{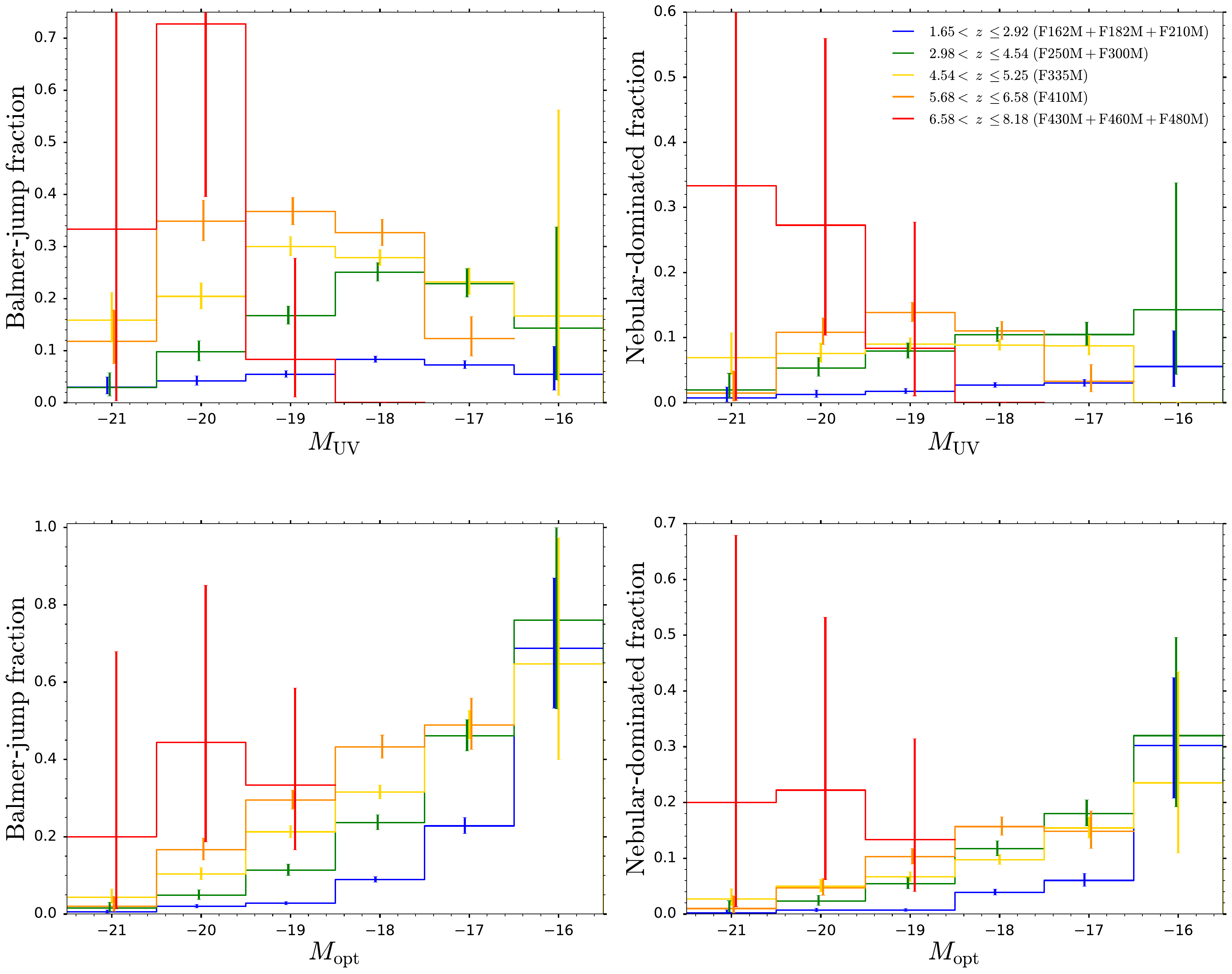}
\caption{Left panels: Balmer-jump fractions, defined as the number of Balmer-jump candidates divided by the total number of galaxies in absolute magnitude bins of $\Delta M = 1$ width, for $M_\mathrm{UV}$ (top) and $M_\mathrm{opt}$ (bottom). Error bars denote the errors propagated in quadrature, assuming Poisson errors on the counts. Galaxies are split in redshift (colour coding), according to the medium-band filter(s) used to trace the rest-frame optical continuum. The Balmer-jump fraction tends to increase with increasing redshift, is relatively flat with $M_\mathrm{UV}$, and is strongly increasing with increasing $M_\mathrm{opt}$ (which may be driven by selection bias). Right panels: Nebular-dominated fractions, which exhibit similar qualitative trends to the Balmer-jump fractions. Note that the vertical axis limits are different in the various panels.}
\label{fig:jump_nd_fractions}
\end{figure*}

Despite their fainter optical continua $M_\mathrm{opt}$, Balmer-jump candidates and nebular-dominated candidates tend to have brighter \Ha\ (bottom-left panel) and \OIII\ + \Hb\ luminosities (bottom right), due to their higher line equivalent widths. The median $\log\, (L_\mathrm{H\alpha}/\mathrm{erg\ s^{-1}})$ is 41.46, 41.65 and 41.86, for control galaxies, Balmer-jump candidates and nebular-dominated candidates, respectively. For $\log\, (L_\mathrm{[O\ III] + H\beta}/\mathrm{erg\ s^{-1}})$, the medians are 41.71, 42.06 and 42.24, respectively. Thus the \OIII\ + \Hb\ emission is generally $\approx$2--2.5$\times$  brighter than \Ha\ across our samples. The brighter line luminosities for our Balmer-jump and nebular-dominated candidates relative to the control sample, is qualitatively consistent with the results of \citet{Llerena2024}, who found that extreme emission-line galaxies colour-selected via their emission line photometric excess tended to have brighter \Ha\ and \OIII\ + \Hb\ fluxes than their control sample.

We note a subset of sources, especially control galaxies, with very small ($<10$~\AA) EWs and low line luminosities ($<10^{39}$~erg~s$^{-1}$). These need not all be quiescent galaxies. From our colour-based approach for determining the line equivalent widths, if the wide-band $f_\nu$ measurement covering the emission line happens to be fainter than the rest-frame optical continuum measurement, then the inferred equivalent width is negative, which we set to 10~\AA\ for display purposes in Fig.~\ref{fig:line_histograms}. This situation can occur due to photometric noise on the flux density measurements and/or a blue (red) optical slope in the case of \Ha\ (\OIII\ + \Hb), coupled with relatively weak emission lines. Furthermore, our colour-based approach assumes a flat optical $f_\nu$ slope. In the case of very red control galaxies, e.g.\@ dusty galaxies, quiescent galaxies, Little Red Dots \citep{Kokorev2024, Kocevski2025}, this could cause the \Ha\ EW (\OIII\ + \Hb\ EW) to be overestimated (underestimated). 

\subsection{Redshift evolution}

We further investigate the incidence rate of Balmer jumps and possible nebular-dominated emission in the galaxy population. We count the number of Balmer-jump candidates and nebular-dominated candidates in absolute magnitude bins of width $\Delta M = 1$, and divide this by the total number of galaxies (= Balmer-jump candidates + control galaxies) within that bin. We split galaxies in redshift, according to the medium band used to trace the rest-frame optical continuum level. In the case of low statistics, we group multiple medium bands (i.e.\@ redshift intervals) together. The resulting Balmer-jump fractions and nebular-dominated fractions are shown in the left and right panels of Fig.~\ref{fig:jump_nd_fractions}, respectively. 

The Balmer-jump fractions clearly increase with increasing redshift. When binning by $M_\mathrm{UV}$ (top panels), the Balmer-jump fraction grows from ${\sim}7$ per cent at $1.65 < z \leq 2.92$ (using the F162M+F182M+F210M filters, blue) to ${\sim}19$ per cent at $2.92 < z \leq 4.54$ (F250M+F300M, green) and ${\sim}32$ per cent at $5.68 < z \leq 6.58$ (F410M, orange). This rise in Balmer-jump fraction (a continuum-based tracer of starburst activity) with increasing redshift is qualitatively similar to the rise in extreme ($\mathrm{EW_{rest}}>750$~\AA) emission-line galaxy fractions (line-based tracer) with redshift \citep{Boyett2024}. 

Across the different redshift bins, the Balmer-jump fractions are generally relatively flat with $M_\mathrm{UV}$, with perhaps a tentative deficit at the brightest magnitudes. On the other hand, the Balmer-jump fraction exhibits a clearly increasing trend with increasing $M_\mathrm{opt}$ (bottom-left panel), i.e.\@ Balmer-jump galaxies dominate at fainter optical magnitudes. For example, at $4.54 < z \leq 5.25$ (F335M, yellow), the Balmer-jump fraction rises from ${\sim}4$ per cent at $M_\mathrm{opt}=-21$ to ${\sim}21$ per cent at $M_\mathrm{opt}=-19$ and ${\sim}49$ per cent at $M_\mathrm{opt}=-17$. This might suggest that while galaxies can be UV-faint for a variety of reasons (low-mass starburst, inactive/dusty massive galaxy), galaxies tend to be optically-faint if they are low-mass starbursts. However, the rising Balmer-jump fraction with increasing $M_\mathrm{opt}$ may also (partly) reflect a selection bias. By definition, Balmer-jump galaxies will be brighter in the UV than in the optical, so they should remain detectable (passing our $5\sigma$ SNR cut) in the rest-frame UV, even at faint $M_\mathrm{opt}$. On the other hand, control galaxies, with their red UV--optical colours \citep[possibly Balmer breaks, e.g.\@][]{Kuruvanthodi2024, Langeroodi2024, Looser2024, Trussler2024, Trussler2025, Baker2025, Witten2025b, CoveloPaz2026} will be fainter in the UV than in the optical, hindering their rest-frame UV detection at fainter $M_\mathrm{opt}$, so they may not be selected. The fainter the $M_\mathrm{opt}$ bin, the smaller the maximal red UV--optical colour that will still pass our $5\sigma$ initial selection procedure, increasingly biasing the sample to Balmer-jump galaxies.

The nebular-dominated fractions follow similar qualitative trends. Binning by $M_\mathrm{UV}$ (top-right panel), the nebular-dominated fraction increases from ${\sim}2$ per cent at $1.65 < z \leq 2.92$ to ${\sim}11$ per cent at $5.68 < z \leq 6.58$. While a $M_\mathrm{UV}$ trend is not immediately evident, the nebular-dominated fraction clearly increases with increasing $M_\mathrm{opt}$. 

\begin{figure}
\centering
\includegraphics[width=\linewidth]{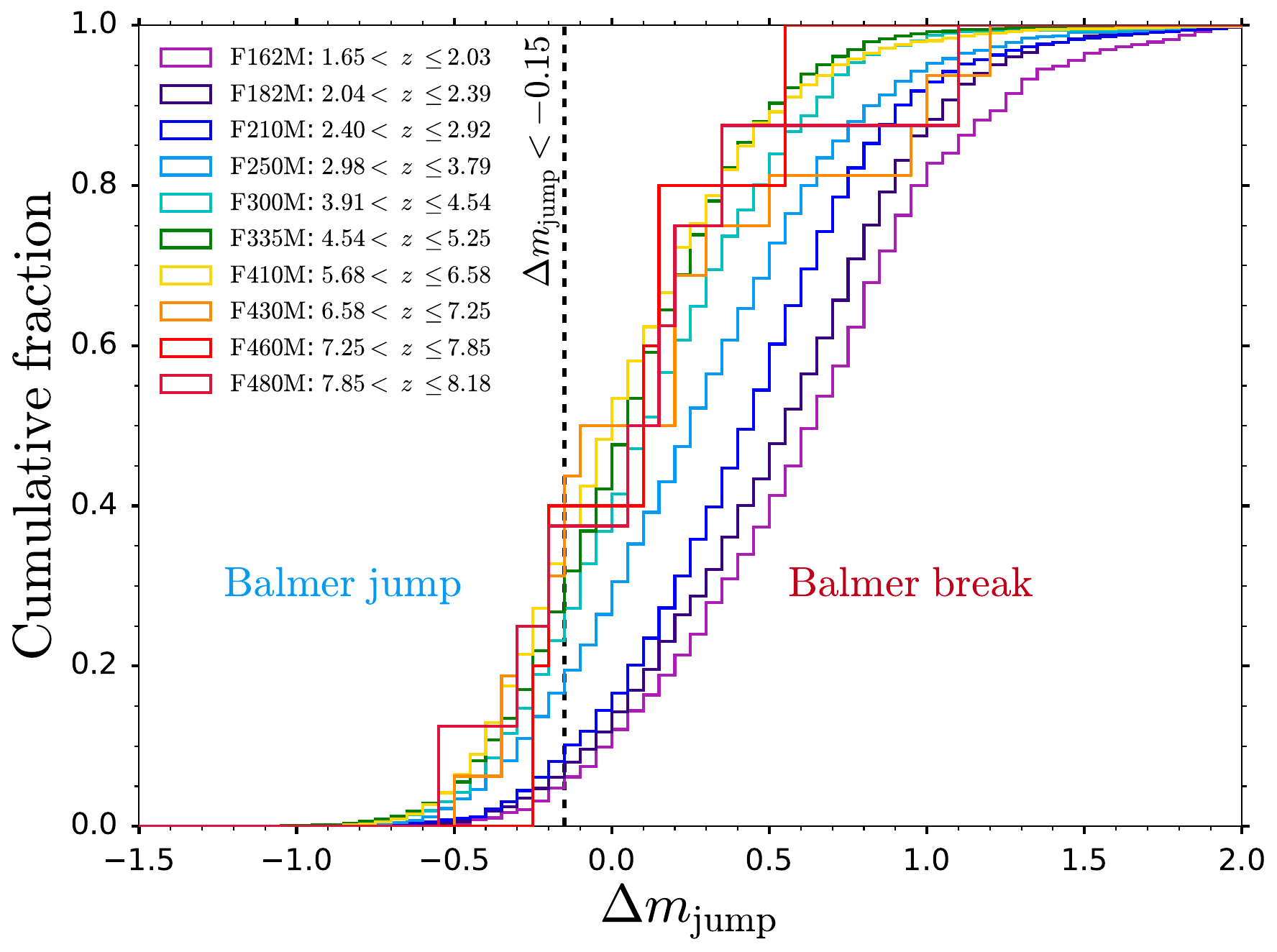}
\caption{The redshift-evolution (colour coding) of the distribution of $\Delta m_\mathrm{jump} = m_\mathrm{b, UV} - m_\mathrm{r,opt}$, the gap in flux density between the UV and optical (Eq.~\ref{eq:jump}). Negative values correspond to Balmer jumps and/or blue UV--optical slopes. Positive values correspond to Balmer breaks and/or red UV--optical slopes. The vertical dashed line at $\Delta m_\mathrm{jump} = -0.15$ corresponds to our Balmer-jump selection criterion. Balmer jumps become increasingly more common at higher redshift, indicating that the Universe is dominated by young starbursts at high redshift, and transitions to being dominated by evolved galaxies with sizable Balmer breaks at lower redshift.}
\label{fig:jump_distribution}
\end{figure}

Finally, we show the redshift-evolution of the distribution of Balmer jumps $\Delta m_\mathrm{jump}$ in Fig.~\ref{fig:jump_distribution}. Negative $\Delta m_\mathrm{jump}$ values correspond to Balmer jumps and/or blue UV--optical slopes, while positive values correspond to Balmer breaks and/or red UV--optical slopes. The vertical dashed line at $\Delta m_\mathrm{jump} = -0.15$ denotes our Balmer-jump candidate selection criterion. As already indicated by Fig.~\ref{fig:jump_nd_fractions}, Balmer jumps become increasingly common at higher redshift. For example, the cumulative fraction of galaxies with $\Delta m_\mathrm{jump} < -0.15$ is only 5~per~cent at $1.65 < z \leq 2.03$ (F162M, purple) and rises to 32~per~cent at $5.68 < z \leq 6.58$ (F410M, yellow). Moreover, while the median $\Delta m_\mathrm{jump} = 0.02$ at $5.68 < z \leq 6.58$ indicates that Balmer jumps are common, the median $\Delta m_\mathrm{jump} = 0.67$ at $1.65 < z \leq 2.03$ indicates that Balmer breaks dominate in the galaxy population. Thus Fig.~\ref{fig:jump_distribution} traces the transition of a Universe dominated by young starbursts with prominent Balmer jumps at high redshift, to a Universe dominated by evolved galaxies with sizable Balmer breaks at lower redshift. Similarly, \citet{Roberts-Borsani2024} and \citet{Langeroodi2024} both find, through stacking analyses of NIRSpec PRISM spectra at $5 < z < 10$ and $3 < z < 10$ respectively, that the Balmer break in spectra stacked according to redshift tends to increase with decreasing redshift. Moreover, a Balmer jump emerges in the stacked spectra (analogous to our median $\Delta m_\mathrm{jump} \leq 0$ measurement) for redshift stacks with $z > 6$. 

\section{Conclusions} \label{sec:conclusions}

Motivated by the serendipitous spectroscopic discovery of high-redshift galaxies exhibiting the UV downturn associated with two-photon continuum emission (implying nebular-dominated emission, or possibly a downturn due to damped Ly$\alpha$ absorption), we develop a photometric search method to identify further nebular-dominated galaxy candidates. Key to this is the extensive medium-band imaging offered by JADES, which enables the identification of Balmer-jump galaxy candidates across a wide range of redshifts ($1.5 < z < 8.5)$.

We outline how the profile of the nebular continuum depends on \ion{H}{II} region temperature and collisionally-enhanced two-photon continuum emission, which requires high temperatures and non-negligible neutral \ion{H}{I} fractions in the \ion{H}{II} region. The collisional enhancement of two-photon emission raises the nebular continuum (reducing $\xi_\mathrm{ion, neb}$), which favours the detection of the two-photon downturn, makes the UV slope less red, and affects the magnitude shift associated with the Balmer jump. As the Balmer jump decreases with increasing temperature, we show that nebular-dominated emission (possibly powered by top-heavy, metal-poor star formation) does not necessarily have the largest Balmer jumps due to the likely higher \ion{H}{II} region temperatures, despite the relative lack of stellar continuum. Normal starbursts \citep[following a regular IMF, e.g.][]{Salpeter1955, Kroupa2001, Chabrier2003} lack the stellar ionising photon production efficiency $\xi_\mathrm{ion, *}$ for the UV downturn associated with two-photon continuum emission to be evident in the spectra, being hidden underneath the much brighter stellar continuum. It is only with the increased normalisation of the nebular continuum powered by star formation following a top-heavy IMF with much higher $\xi_\mathrm{ion, *}$, that the nebular continuum dominates over the incident starlight and the two-photon downturn becomes prominent: the galaxy is nebular-dominated.

We identify Balmer-jump galaxy candidates through their deficit in rest-frame optical continuum level (relative to the UV), using medium-band filters that reside in the emission line gap between the strong \OIII\ $\lambda 5007$ and \Ha\ lines. Motivated by the smallest Balmer jumps seen in the nebular-dominated galaxy candidates identified spectroscopically thus far, and in current Pop III models, we set a relatively modest Balmer jump cut $\Delta m_\mathrm{jump} < -0.15$ for our Balmer-jump selection. As Balmer jumps are a general recombination feature of all young starbursts, and thus do not necessarily demand star formation following a top-heavy IMF, we apply additional selection criteria to identify possible nebular-dominated galaxy candidates. We demand a high $\log \, (\xi_\mathrm{ion, obs} /\mathrm{(Hz\ erg^{-1})}) > 25.60$ to power the strong nebular continuum, as well as a relatively non-blue UV colour (e.g.\@ $\mathrm{F115W} - \mathrm{F200W} > -0.4$) indicating a lack of stellar continuum emission. However, the same star formation following a top-heavy IMF that would power the strong nebular continuum, possibly also powers strong UV lines and significant collisionally-enhanced two-photon continuum emission, which reduce the otherwise distinct gap in $\xi_\mathrm{ion, obs}$ and UV colour between normal starbursts and nebular-dominated emission. Hence our nebular-dominated selection criteria are still quite modest, and dust-reddened starbursts may form part of our nebular-dominated candidate list.

We investigate the continuum and emission-line properties of our starburst candidates. We find that nebular-dominated candidates, Balmer-jump candidates and control galaxies (with $\Delta m_\mathrm{jump} > -0.15$) are similarly distributed in absolute UV magnitude $M_\mathrm{UV}$, with a similar median $M_\mathrm{UV} \approx -18.35$ for the three samples. Owing to their redder UV--optical colours, the control sample exhibits brighter rest-frame optical continuum absolute magnitudes, with a median $M_\mathrm{opt} = -18.80$, compared to the median $M_\mathrm{opt} = -17.95$ for the Balmer-jump and nebular-dominated candidates. As expected from the Balmer-jump selection and subsequent $\xi_\mathrm{ion, obs}$ cut, Balmer-jump candidates and nebular-dominated candidates are very strong line emitters. The median rest-frame equivalent widths for nebular-dominated candidates, Balmer-jump candidates and control galaxies are 1567~\AA, 1007~\AA, 294~\AA\ for \Ha, and 2292~\AA, 1555~\AA, 342~\AA\ for \OIII\ + \Hb, respectively. The median \Ha\ luminosities $\log\, (L_\mathrm{H\alpha}/\mathrm{erg\ s^{-1}})$ are 41.86, 41.65 and 41.46, while the median \OIII\ + \Hb\ luminosities $\log\, (L_\mathrm{[O\ III] + H\beta}/\mathrm{erg\ s^{-1}})$ are 42.24, 42.06 and 41.71, respectively.

Enabled by the plethora of medium bands (F162M, F182M, F210M, F250M, F300M, F335M, F410M, F430M, F460M, F480M), we trace the evolution of Balmer jumps across cosmic time. We find that the fraction of galaxies exhibiting Balmer jumps tends to increase with increasing redshift. The Balmer-jump fraction grows from ${\sim}7$ per cent at $1.65 < z \leq 2.92$ to ${\sim}19$ per cent at $2.92 < z \leq 4.54$ and ${\sim}32$ per cent at $5.68 < z \leq 6.58$. Balmer-jump fractions are relatively flat with $M_\mathrm{UV}$, though strongly increase with increasing $M_\mathrm{opt}$, i.e.\@ Balmer-jump galaxies dominate at fainter optical magnitudes. However, this $M_\mathrm{opt}$ trend may (partly) be driven by a selection bias, that disfavours the rest-frame UV detection of red UV--optical galaxies at faint optical magnitudes. Nebular-dominated fractions follow similar qualitative trends. The nebular-dominated fraction increases from ${\sim}2$ per cent at $1.65 < z \leq 2.92$ to ${\sim}11$ per cent at $5.68 < z \leq 6.58)$. While Balmer jumps are common at high redshift, with a median $\Delta m_\mathrm{jump} = 0.02$ at $5.68 < z \leq 6.58$, the median $\Delta m_\mathrm{jump} = 0.67$ at $1.65 < z \leq 2.03$, indicating that Balmer breaks dominate in the galaxy population. Thus the Universe is dominated by young starbursts with prominent Balmer jumps at early cosmic epochs, and transitions to being dominated by evolved galaxies with sizable Balmer breaks at lower redshift. 

Finally, we close by noting that our nebular-dominated candidates were identified using the select few filters that comprise our Balmer jump, $\xi_\mathrm{ion, obs}$ and UV colour selections. These measurements are further affected by emission lines which reduce the inferred Balmer jump (\ion{He}{I} $\lambda 5876$) and reduce $\xi_\mathrm{ion, obs}$ plus make the UV slope more blue (\ion{C}{IV}, \ion{He}{II}, \ion{O}{III}], \ion{C}{III}]). Through the development of nebular-dominated galaxy templates, this selection procedure can be refined. By utilising the full set of available photometry via SED-fitting, which naturally accounts for the previously mentioned emission line effects, an even stronger case can be made for possible nebular-dominated candidates. Even then however, normal starbursts can exhibit comparable photometry to starbursts following a top-heavy IMF with nebular-dominated emission. Hence deep follow-up continuum spectroscopy with the NIRSpec PRISM is ultimately needed to establish the presence or absence of the two-photon continuum turnover in the spectra, thus determining whether the galaxy is truly nebular-dominated or rather a dust-reddened starburst. 

\section*{Acknowledgements}

We thank the referee for their useful comments which helped to improve this article. JAAT thanks Lisa Kewley for valuable discussions that greatly aided in the interpretation of results. JAAT acknowledges support from the Simons Foundation and \emph{JWST} program 3215. Support for program 3215 was provided by NASA through a grant from the Space Telescope Science Institute, which is operated by the Association of Universities for Research in Astronomy, Inc., under NASA contract NAS 5-03127. AJB and AJC acknowledge funding from the "FirstGalaxies" Advanced Grant from the European Research Council (ERC) under the European Union’s Horizon 2020 research and innovation programme (Grant agreement No.\@ 789056). AJC also gratefully acknowledges support from the Cosmic Dawn Center through the DAWN Fellowship. The Cosmic Dawn Center (DAWN) is funded by the Danish National Research Foundation under grant No. 140. DJE is supported as a Simons Investigator and by JWST/NIRCam contract to the University of Arizona, NAS5-02105. SC acknowledges support by European Union’s HE ERC Starting Grant No.\@ 101040227 - WINGS. CJC, NJA and QL acknowledge support from the ERC Advanced Investigator Grant EPOCHS (788113). DA and TH acknowledge support from STFC in the form of PhD studentships. ECL acknowledges support of an STFC Webb Fellowship (ST/W001438/1). BDJ acknowledges support from \emph{JWST}/NIRCam contract to the University of Arizona, NAS5-02105. TJL gratefully acknowledges support from the Swiss National Science Foundation through a Postdoc Mobility Fellowship and from the \emph{JWST} Program 5997. BER acknowledges support from the NIRCam Science Team contract to the University of Arizona, NAS5-02105, and JWST Program 3215. ST acknowledges support by the Royal Society Research Grant G125142. FS acknowledges funding from \emph{JWST}/NIRCam contract to the University of Arizona, NAS5-02105 and support for \emph{JWST} program \#2883, 4924, 5105, 6434 provided by NASA through grants from the Space Telescope Science Institute, which is operated by the Association of Universities for Research in Astronomy, Inc., under NASA contract NAS 5-03127. The research of CCW is supported by NOIRLab, which is managed by the Association of Universities for Research in Astronomy (AURA) under a cooperative agreement with the National Science Foundation. CNAW acknowledges support from the JWST/NIRCam contract to the University of Arizona NAS5-02105.

This work is based on observations made with the NASA/ESA \emph{Hubble Space Telescope} (\emph{HST}) and NASA/ESA/CSA \emph{James Webb Space Telescope} (\emph{JWST}) obtained from the Mikulski Archive for Space Telescopes (MAST) at the Space Telescope Science Institute (STScI), which is operated by the Association of Universities for Research in Astronomy, Inc., under NASA contract NAS 5-03127 for \emph{JWST}, and NAS 5–26555 for \emph{HST}. The observations used in this work are associated with \emph{JWST} programs 1180, 1181, 1210, 1286, 1963, 2514, 3215, 3990, 5997. We also utilise public imaging datasets from \emph{JWST} programs 2561 (UNCOVER) and 4111 (Mega Science), as well as spectra from \emph{JWST} programs 2198 and 2756. Some of the spectroscopic data products presented herein were retrieved from the Dawn JWST Archive (DJA). These DJA spectroscopic products were produced using msaexp \citep{Brammer2022}. DJA is an initiative of the Cosmic Dawn Center (DAWN), which is funded by the Danish National Research Foundation under grant DNRF140. The authors acknowledge the FRESCO team (PID 1895, PI: Oesch) for developing their observing program with a zero-exclusive-access period.

This research made use of Astropy,\footnote{http://www.astropy.org} a community-developed core Python package for Astronomy \citep{astropy2013, astropy2018}.

\section*{Data Availability}

The \emph{JWST} imaging data and photometry used in this work are publicly available through JADES Data Release 5 (DR5), see \citet{Johnson2026} and \citet{Robertson2026}. Any remaining data underlying the analysis in this article will be shared on reasonable request to the first author.



\bibliographystyle{mnras}
\bibliography{main.bib} 




\appendix

\section{Two-photon downturn: dropout photometric deficit} \label{app:deficit}

The steep downturn associated with the two-photon continuum represents a significant departure from the generally rather smooth rest-frame UV slopes. It is for this reason that the two-photon (or possibly DLA) downturn is such a notable feature in rest-frame UV spectroscopy. In this section we discuss the challenges in identifying the imprint of the lack of flux density associated with the two-photon downturn in rest-frame UV photometry. 

\begin{figure}
\centering
\includegraphics[width=\linewidth] {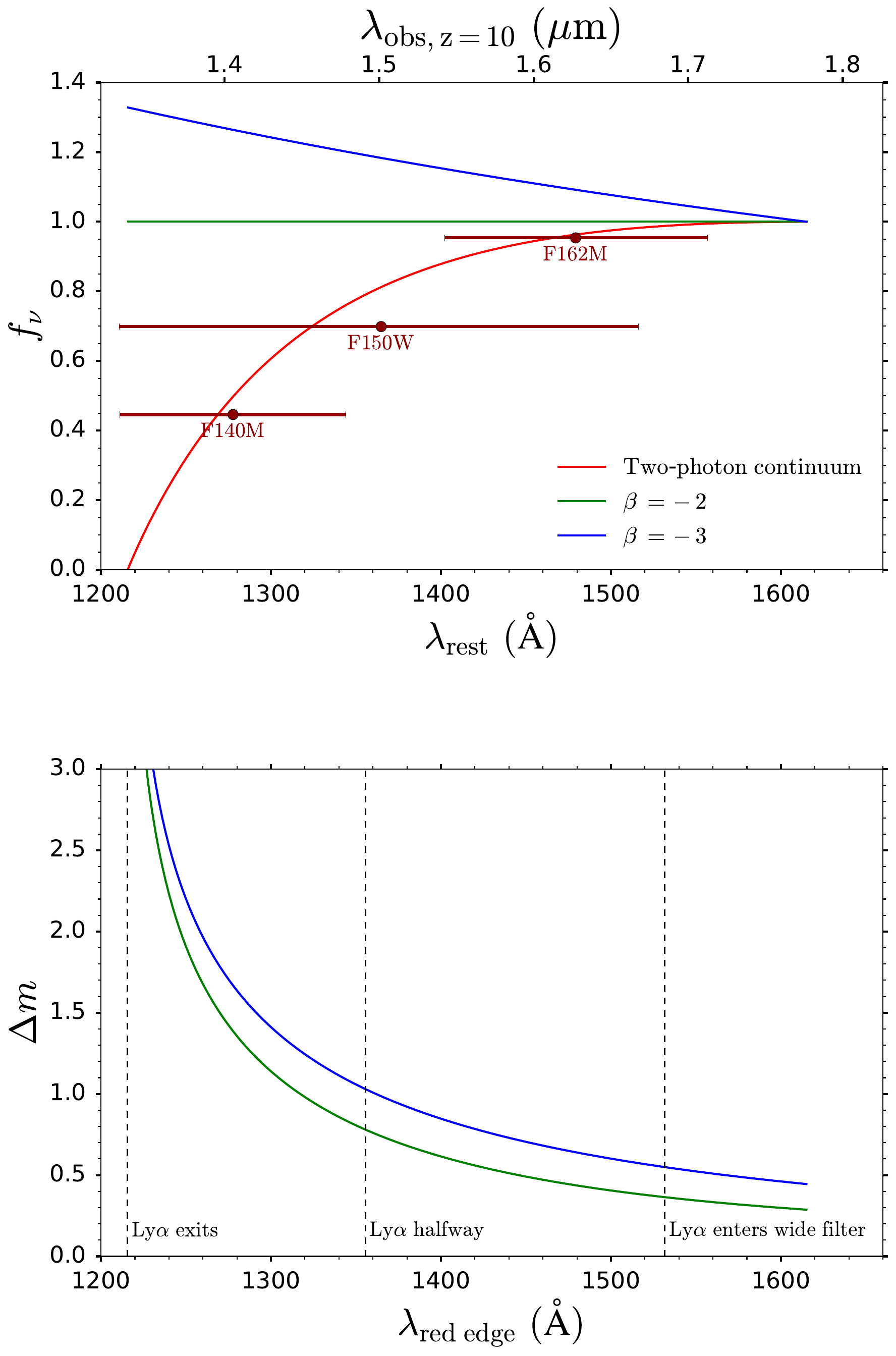}
\caption{Top panel: The steep downturn associated with the two-photon continuum (red) results in a lack of bandpass-averaged flux density in dropout filters (here F140M and F150W at $z=10$) compared to regular smooth UV slopes, e.g.\@ $\beta = -2, -3$ (green and blue, respectively), causing the source to appear fainter in the dropout filter than it otherwise should be given its redshift. Bottom panel: The resulting photometric deficit $\Delta m$, i.e.\@ the magnitude difference between the bandpass-averaged flux density measured in the wide-band dropout filter in the case of two-photon continuum emission vs.\@ $\beta = -2$ (green) and $\beta = -3$ (blue), increases as Ly$\alpha$ shifts through the dropout filter (with the rest-frame wavelength at the red edge of the filter $\lambda_\mathrm{red\ edge}$ decreasing), due to the widening gap in flux density between the two-photon continuum and regular UV slopes. As the photometric deficit caused by the two-photon downturn is degenerate with the lack of flux density due to the Ly$\alpha$ break, precise redshift constraints are required to distinguish between these two cases. Thus, nebular-dominated candidates (as well as galaxies with DLA absorption) could be identified in this way, through the imprint of the two-photon downturn (or DLA downturn) on dropout photometry, though any Ly$\alpha$ emission would reduce the photometric deficit, complicating the search.} 
\label{fig:twophoton_deficit}
\end{figure}

We begin by noting that even NIRCam medium-band filters (with $R{\sim}10$) are too wide to adequately sample the decline in flux density (towards shorter wavelengths) associated with the two-photon downturn. As shown in the top panel of Fig.~\ref{fig:twophoton_deficit}, although the two-photon continuum reaches its peak value (in $f_\nu$) at 1615~\AA, it only begins to appreciably decline at $\lambda \lesssim 1450$~\AA, so there is only really a single medium-band filter (and one wide-band filter) tracing the lack of flux density associated with the downturn. Better sampling would demand 3--4 filters with $R{\sim}20$ to reveal the steep drop in the continuum that is evident in spectra.

Rather one has to rely on the lack of flux density in one wide band (and one medium band, if available), specifically the dropout filter containing the Ly$\alpha$ break, as the gap in flux density (and thus bandpass-averaged flux density) between the steeply declining two-photon continuum (red) and regular UV slopes, e.g.\@ $\beta = -2$ (green) and $\beta = -3$ (blue), is greatest closer to Ly$\alpha$. We show the magnitude difference between the bandpass-averaged flux density that would be measured in the wide-band dropout filter if two-photon continuum is present vs.\@ a $\beta = -2$ slope (green) and a $\beta = -3$ slope (blue) in the bottom panel of Fig.~\ref{fig:twophoton_deficit}. Opposite to an emission line causing a photometric excess, the steep decline in the two-photon continuum causes a photometric deficit in the dropout filter, causing the source to appear much fainter than it otherwise should be (given normal UV slopes) for its redshift. As Ly$\alpha$ begins to enter the wide-band filter (the scenario in the top panel of Fig.~\ref{fig:twophoton_deficit}), the photometric deficit $\Delta m$ is already 0.37~mag, for $\beta = -2$. By the time Ly$\alpha$ is located halfway through the dropout filter, the photometric deficit is 0.78~mag, continuing to grow as Ly$\alpha$ passes through the dropout filter, as the bandpass-averaged gap in flux density between the two-photon continuum and regular UV slopes increases. To be general, we have assumed a top-hat filter with $R=4.5$ (typical for NIRCam) when generating the photometric deficit curves, plotted against the rest-frame wavelength at the red edge of the dropout filter, which decreases with increasing redshift as Ly$\alpha$ passes through the filter. 

\begin{figure*}
\centering
\includegraphics[width=\linewidth]{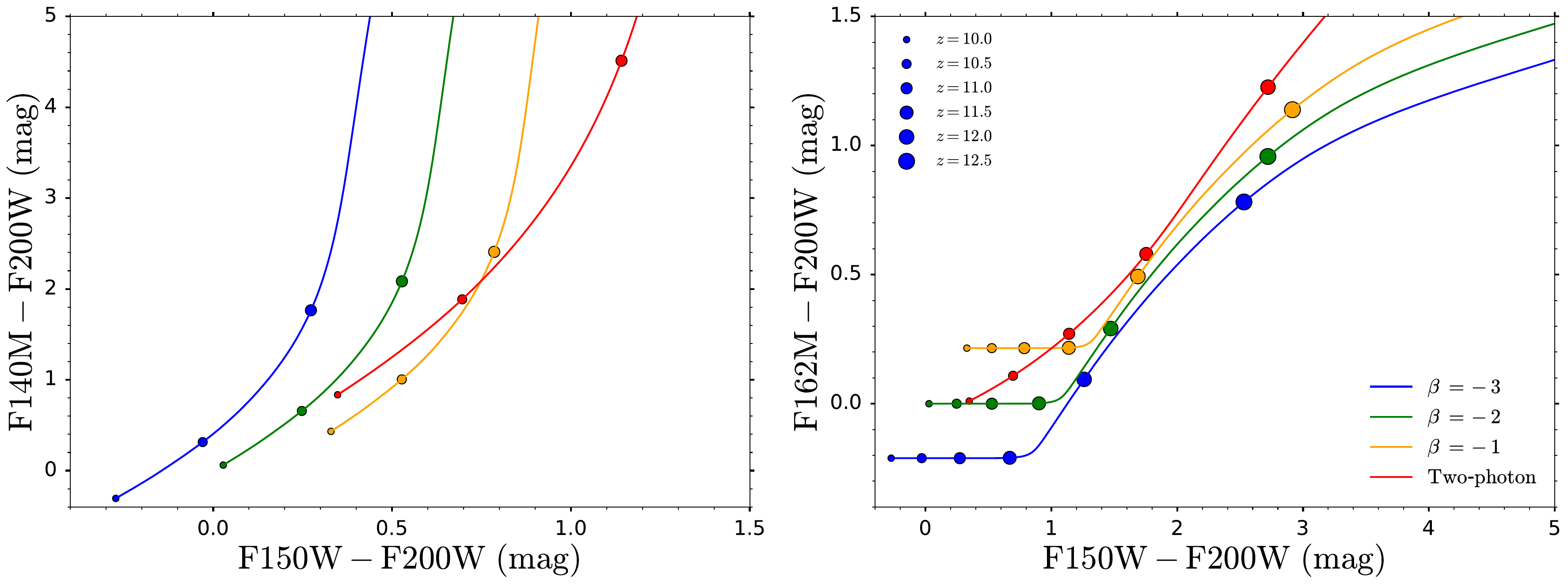}
\caption{At $z > 10$, the Ly$\alpha$ break starts passing through the NIRCam medium-band filters. The dual medium/wide dropout measurement helps to lift the degeneracy between the two-photon downturn photometric deficit and the Ly$\alpha$ break, without precise redshift constraints. The steep two-photon downturn results in a significant departure (red) from the medium/wide dropout colours for regular UV slopes: $\beta = -3$ (blue), $\beta = -2$ (green), $\beta = -1$ (orange). Left panel: Dropout colours at $10 < z < 13$ (circles denote $\Delta z =0.5$ steps, with circle size increasing with redshift) with $\mathrm{F140M} - \mathrm{F200W}$, $\mathrm{F150W} - \mathrm{F200W}$. Right panel: Dropout colours with $\mathrm{F162M} - \mathrm{F200W}$, $\mathrm{F150W} - \mathrm{F200W}$.}
\label{fig:dropout_colours}
\end{figure*}

The photometric deficit caused by the two-photon downturn is reminiscent of the lack of flux density \citep[e.g.\@][]{Heintz2024, Heintz2025} in the dropout filter caused by damped Ly$\alpha$ absorption \citep[see e.g.\@ the photometric deficit in F115W for members of a $z=7.88$ protocluster,][]{Witten2025}. Indeed, it has been demonstrated that at high-redshift, photometric redshifts tend to overestimate the true (spectroscopic) redshifts of galaxies \citep{ArrabalHaro2023, Fujimoto2023, Finkelstein2024, Hainline2024, Asada2025, Heintz2025}. Damped Ly$\alpha$ absorption is widespread and tends to be strong in the Epoch of Reionisation \citep{Asada2025, Heintz2025}, causing a prominent UV downturn resulting in a lack of flux density in the dropout filter \citep{Witten2025}, in addition to the usual deficit caused by the Ly$\alpha$ break. The Ly$\alpha$ damping wings from a highly neutral IGM cause a weaker softening of the Ly$\alpha$ break \citep{Curtis-Lake2023}, resulting in a subtler lack of flux density in the dropout filter. Photometric redshift fitting codes which do not include this additional DLA absorption solely attribute this lack of flux density to the Ly$\alpha$ break, causing the redshift to be overestimated \citep{Asada2025, Heintz2025, Witstok2026}. Hence the photometric deficit associated with typical high-redshift damped Ly$\alpha$ absorption \citep[$N_\mathrm{HI} \approx 10^{20.5\textrm{--}21.5}~\mathrm{cm}^{-2}$,][]{Asada2025} is appreciable enough to bias the photometric redshifts in this way. Given that the spectral profile of the two-photon continuum of GS-9422 can closely resemble that of an extreme DLA system \citep[with $N_\mathrm{HI} \approx 10^{23}~\mathrm{cm}^{-2}$,][]{Cameron2024} it should have a significant effect on the dropout photometry, as indicated in the bottom panel of Fig.~\ref{fig:twophoton_deficit}.

However, one of the main challenges in identifying nebular-dominated candidates through the two-photon downturn photometric deficit is distinguishing this from the Ly$\alpha$ break, as both cause a lack of flux density in the dropout filter. This requires precise redshift constraints, offered by e.g.\@ slitless spectroscopy \citep[e.g.\@][]{Sun2022, Sun2023, Helton2024, Fudamoto2025}, and possibly extensive medium-band photometry \citep[][]{Suess2024, Muzzin2025, Sarrouh2026}. If the redshift of the source is known, then the contribution of the Ly$\alpha$ break can be precisely determined. Sources that appear appreciably fainter in the dropout filter than they should be given their redshift \citep{Witten2025}, likely have damped Ly$\alpha$ absorption and/or strong two-photon continuum emission.

At $z > 10$ the Ly$\alpha$ break starts passing through the NIRCam medium-band filters. This offers an additional avenue to search for the photometric deficit associated with the two-photon downturn, perhaps not dependent on spectroscopic redshift constraints. Generally, the strengths of the Ly$\alpha$ break in a wide-band filter and its accompanying medium-band filters are closely related, determined by their relative spectral ranges, and further affected by the UV slope. We show this at $10 < z < 13$ using the F140M, F150W, F162M dropout filters, and F200W detection filter, in Fig.~\ref{fig:dropout_colours}. In the left panel, F140M starts dropping out at the same redshift as F150W, with its dropout strength $\mathrm{F140M} - \mathrm{F200W}$ evolving more quickly with redshift than $\mathrm{F150W} - \mathrm{F200W}$, due to its narrower spectral range. On the other hand (right panel), F162M starts dropping out when the Ly$\alpha$ break is already halfway through F150W, with its dropout strength $\mathrm{F162M} - \mathrm{F200W}$ evolving more quickly with redshift, with both the F162M and F150W filters fully dropping out at the same redshift. Thus galaxies follow well-defined trajectories in their medium- and wide-band dropout colours, according to their UV slopes ($\beta = -3, -2, -1$ are shown in blue, green and orange, respectively). The two-photon continuum (red), with its strong downturn, further affects the dropout strength, causing a significant departure from the general medium- and wide-band dropout trends. In principle, nebular-dominated candidates could be identified through their unique medium/wide dropout colours in this way, taking into account also the UV slopes measured using the rest-frame UV photometry. In practice, due to the small colour offsets in the dropout--dropout curves, it will be difficult as dropout photometry is generally low SNR, unless candidates are especially bright or the medium-band imaging is particularly deep.

Finally, we close by noting that we have only been considering the contribution of two-photon continuum emission to the dropout photometry. However, if Ly$\alpha$ emission is present it will reduce the photometric deficit caused by the two-photon downturn, making this nebular-dominated search method even more challenging. Indeed, this is the case for GS-9422 (Fig.~\ref{fig:GS-9422}), with the Ly$\alpha$ photometric boost mostly hiding the two-photon photometric deficit, meaning that nebular-dominated candidates cannot always be identified in this manner. Although we would expect nebular-dominated galaxies to have intrinsically very strong Ly$\alpha$ emission \citep[e.g.\@][]{Zackrisson2011, Nakajima2022, Trussler2023}, it could effectively appear very weak, if the system does not reside in a substantially ionised bubble \citep[e.g.\@][]{Trussler2023, Fujimoto2024, Witstok2024, Witstok2025, Witstok2025b}. Hence it should still be possible to identify some (not all) nebular-dominated galaxies via this photometric deficit procedure. 

\section{Indirect Balmer-jump selection} \label{app:indirect}

\begin{figure*}
\centering
\includegraphics[width=.475\linewidth] {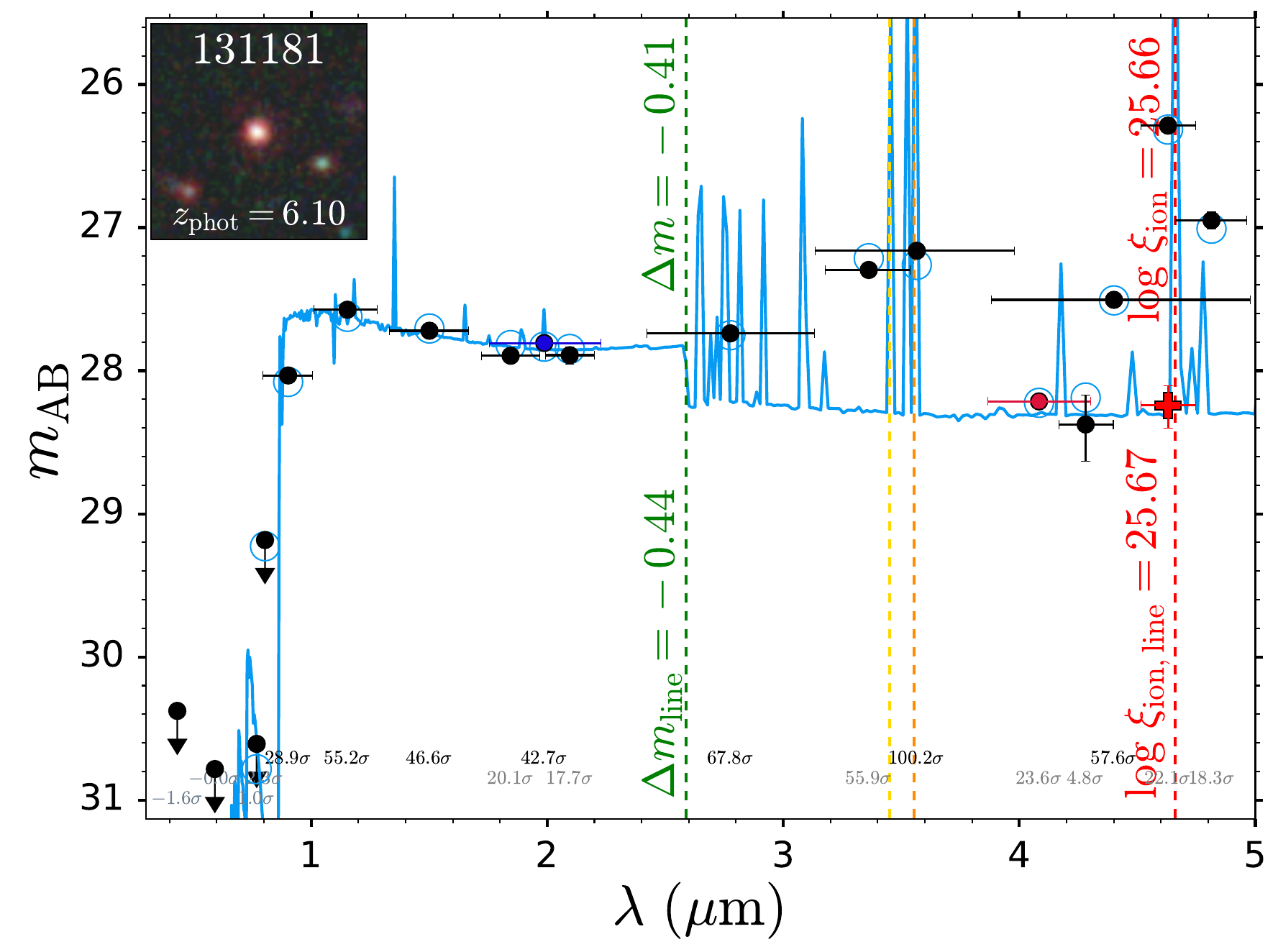} \hfill
\includegraphics[width=.475\linewidth]{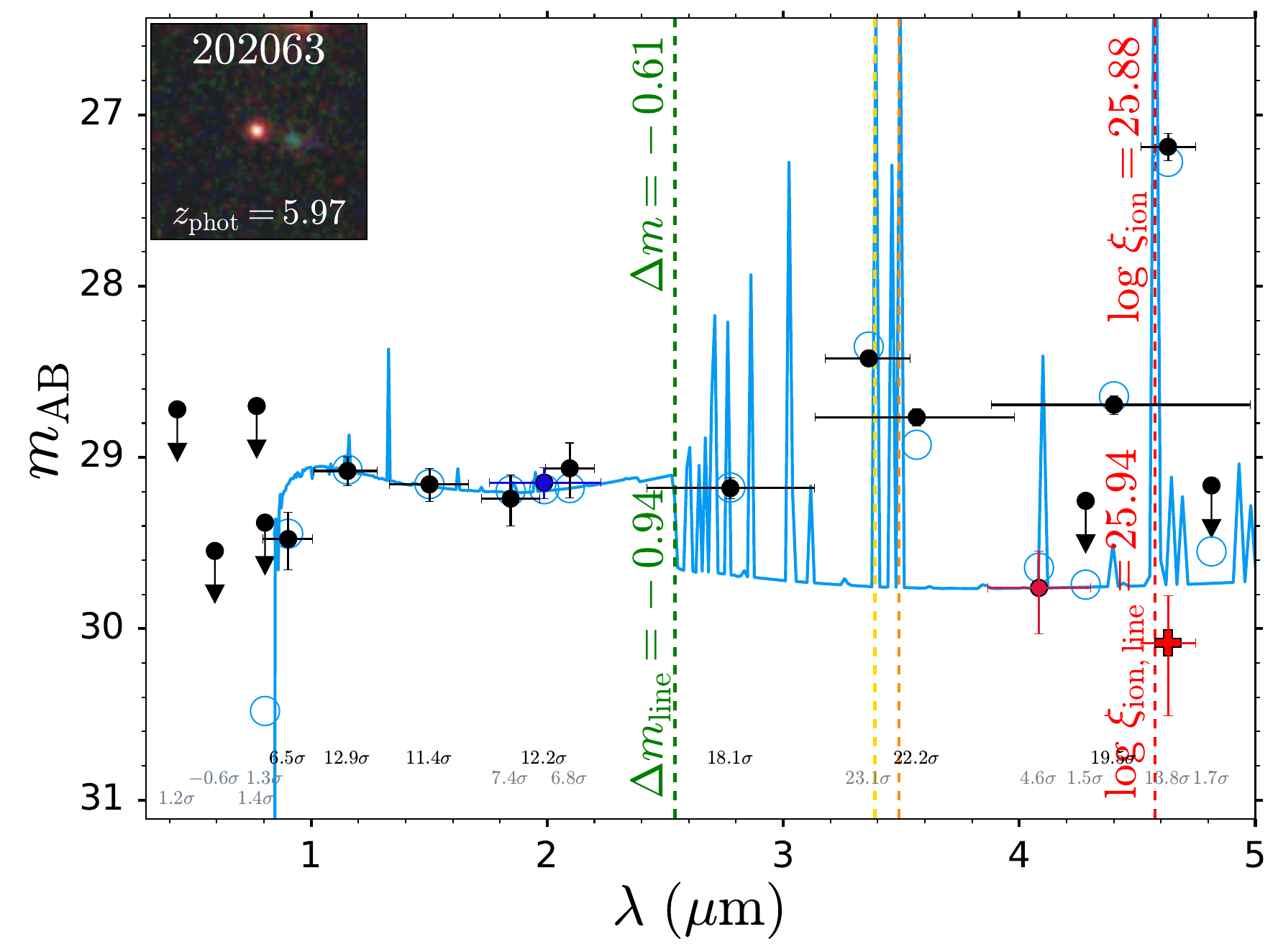}
\caption{Balmer-jump galaxy candidates identified indirectly, using the photometric boost by \Ha\ in both a medium- and wide-band filter to indirectly determine the rest-frame optical continuum level (red plus symbol). Left panel: A Balmer-jump candidate where the rest-frame optical continuum level determined indirectly is comparable to that determined directly (red circle). Right panel: A Balmer-jump candidate where the indirect approach has vastly overestimated the Balmer jump, due to the large uncertainty on the rest-frame optical continuum level for a given uncertainty on the differential photometric boost between the medium and wide bands, in the limit of high equivalent width emission lines.} 
\label{fig:indirect_candidates}
\end{figure*}

In the main body of the article we determined the Balmer jump directly by probing the deficit in the rest-frame optical continuum directly with a medium-band filter. Here we discuss the prospects of determining the Balmer jump indirectly using the emission line photometric excess in a medium-band filter. The benefit of this approach is that it relies on an excess (emission line boost), rather than a deficit (drop in continuum), so the SNR is greater and thus can be pushed to fainter systems. However, this approach also has its challenges, which we discuss. 

Owing to their narrower spectral range, the bandpass-averaged flux density is boosted more strongly above the continuum level by a given emission line in a medium band than in a wide band. It is this differential photometric boost $\mathrm{d}m = m_\mathrm{med} - m_\mathrm{wide}$ between the medium- $m_\mathrm{med}$ and wide-band measurements $m_\mathrm{wide}$ that encodes information about the emission line equivalent width, and thus the rest-frame optical continuum level. In the limit of very weak (i.e.\@ low equivalent width) lines, there is no photometric boost and the medium and wide bands trace the same underlying continuum and so are equal (if the continuum is flat in $f_\nu$). In the limit of very strong emission lines (i.e.\@ high equivalent width), the differential photometric boost approaches an asymptotic limit, set by the relative widths of the medium- and wide-band filters. Assuming the wide-band spectral resolution $R = 4.5$ (typical for most NIRCam wide bands) and the medium-band spectral resolution $R = 9$ (typical for most NIRCam medium bands), then the limiting differential photometric boost $\mathrm{d}m_\mathrm{lim} \approx -2.5\log _{10} (9/4.5) \approx -0.75$. This limiting behaviour can be seen using the approximate formula for the photometric boost $\Delta m = -2.5\log_{10}(1 + \mathrm{EW_{rest}(1+z)/\Delta \lambda})$, as $\mathrm{EW_{rest}}$ becomes substantially large. For emission lines of intermediate strength, the differential photometric boost $\mathrm{d}m$ is somewhere between these two extremes, increasing in absolute value with increasing emission line equivalent width. Hence the differential photometric boost can be used to determine the line equivalent width, and thus the underlying rest-frame optical continuum level.

We take the measured differential photometric boost for the medium and wide bands probing \Ha\, together with the best-fit photometric redshift. Iterating over different line equivalent widths, we determine the bandpass-averaged flux densities in the medium and wide bands, assuming a flat $f_\nu$ spectrum with a single emission line (\Ha) at the redshift of the source. We find the line equivalent width that yields a differential photometric boost that matches the observed value. The continuum level of the flat $f_\nu$ spectrum is scaled so that the model bandpass-averaged flux densities match the observed flux densities in the medium and wide bands. This establishes the rest-frame optical continuum level. 

We show two examples of Balmer-jump candidates identified following this indirect approach in Fig.~\ref{fig:indirect_candidates}. In the left panel, the rest-frame optical continuum level determined indirectly through the F460M filter (red plus symbol) is comparable to that determined directly through the F410M filter (red circle). Hence, the Balmer jump determined indirectly $\Delta m_\mathrm{line} = -0.44$ is very similar to the value determined directly $\Delta m_\mathrm{jump} = -0.41$. However, in the right panel, the indirect approach has vastly overestimated the Balmer jump, with $\Delta m_\mathrm{line} = -0.94$ vs.\@ $\Delta m_\mathrm{jump} = -0.61$. 

The issue with this indirect approach is that in the limit of large emission line equivalent widths (relevant for nebular-dominated galaxies), the inferred rest-frame optical continuum level is very sensitive to slight shifts in the measured differential photometric boost. Approaching the asymptotic value, large changes in equivalent width (and thus continuum level) only translate into small changes in the differential photometric boost. So any error on the medium- and wide-band measurements (statistical or systematic) can cause a very inaccurate determination of the Balmer jump (and to a lesser degree $\xi_\mathrm{ion, obs}$). It is for this reason that we do not utilise this indirect approach in our colour-based analysis. However, we note that the photometric excess in medium bands can certainly be useful for deriving continuum constraints when performing SED-fitting of the full photometry, increasing the redshift range over which the continuum level can be inferred for a given medium-band filter. 


\bsp	
\label{lastpage}
\end{document}